\documentclass[longauth]{aa}

\usepackage[utf8]{inputenc}
\usepackage{xcolor}
\usepackage{graphicx}
\usepackage[varg]{txfonts}
\usepackage{afterpage}
\usepackage{CJK}
\usepackage{placeins}
\usepackage{longtable}
\usepackage{amsmath} 
\usepackage{footnote}
\usepackage{rotating}
\usepackage{booktabs}
\usepackage{siunitx}
\usepackage{graphicx}
\usepackage{float} 
\usepackage{fancyhdr}
\usepackage{threeparttable}
\usepackage{hyperref} 

\usepackage{silence}
\newcommand{\ha} {\mbox{H$\alpha$}\,}
\newcommand{\hb} {\mbox{H$\beta$}\,}
\newcommand{\hg} {\mbox{H$\gamma$}\,}

\newcommand{\Ha} {\mbox{H$\alpha$}\,}
\newcommand{\Hb} {\mbox{H$\beta$}\,}

\newcommand{\Feii} {\ion{Fe}{ii}\,}

\newcommand{\Caii} {\ion{Ca}{ii}\,}

\newcommand{\Hei} {\ion{He}{i}\,}

\newcommand{\Nii} {\ion{N}{ii}\,}

\newcommand{\Oi} {\ion{O}{i}\,}
\newcommand{\Oii} {\ion{O}{ii}\,}

\newcommand{\Mgi} {\ion{Mg}{i}\,}

\newcommand{\msun}{\mbox{M$_{\odot}$\,}}

\begin{document}

   \title{SN 2017ati: A luminous type IIb explosion from a massive progenitor}

   \subtitle{}
\author{
    Z.-H. Peng \begin{CJK}{UTF8}{gbsn} (彭泽辉)\end{CJK}\orcid{0009-0000-7773-553X}\inst{\ref{inst1}} 
    \and S. Benetti \orcid{0000-0002-3256-0016}\inst{\ref{inst2}}
    \and Y.-Z. Cai \begin{CJK}{UTF8}{gbsn} (蔡永志)\end{CJK}\orcid{0000-0002-7714-493X}\inst{\ref{inst3}, \ref{inst4}, \ref{inst2}}\thanks{Corresponding authors: yongzhi.cai@inaf.it (CYZ)} 
    \and A. Pastorello \orcid{0000-0002-7259-4624}\inst{\ref{inst2}}
    \and J.-W. Zhao \orcid{0009-0009-8633-8582}\inst{\ref{inst5},\ref{inst23}\thanks{meow.jiewei.zhao@gmail.com (ZJW)}} 
    \and A. Reguitti\orcid{0000-0003-4254-2724}\inst{\ref{inst2},\ref{inst19}}
    \and Z.-Y. Wang \orcid{0000-0002-0025-0179}\inst{\ref{inst27},\ref{inst28}}
    \and E. Cappellaro \orcid{0000-0001-5008-8619}\inst{\ref{inst2}}
    \and N.~Elias-Rosa \orcid{0000-0002-1381-9125}\inst{\ref{inst2},\ref{inst22}}
    \and Q.-L. Fang \orcid{0000-0002-1161-9592}\inst{\ref{inst7}} 
    \and M. Fraser \orcid{0000-0003-2191-1674}\inst{\ref{inst12}} 
    \and T. Kangas \orcid{0000-0002-5477-0217}\inst{\ref{inst10}, \ref{inst8}} 
    \and E. Kankare \orcid{0000-0001-8257-3512}\inst{\ref{inst8}} 
    \and Z.~Kostrzewa-Rutkowska \inst{\ref{inst15}} 
    \and P. Lundqvist \orcid{0000-0002-3664-8082}\inst{\ref{inst18}}
    \and S. Mattila\orcid{0000-0001-7497-2994}\inst{\ref{inst8}, \ref{inst21}}
    \and T.~M.~Reynolds \orcid{0000-0002-1022-6463}\inst{\ref{inst8}, \ref{inst13}, \ref{inst14}}
    \and M.~D.~Stritzinger \orcid{0000-0002-5571-1833}\inst{\ref{inst9}} 
    \and A.~Somero \orcid{0000-0001-6566-9192}\inst{\ref{inst8}}
    \and L. Tomasella \orcid{0000-0002-3697-2616}\inst{\ref{inst2}}
    \and S.-P.~Pei \orcid{0000-0002-0851-8045}\inst{\ref{inst33}}
    \and Y.-J.~Yang \orcid{0000-0003-3017-352X}\inst{\ref{inst34}}
    \and J.-J. Zhang \orcid{0000-0002-8296-2590}\inst{\ref{inst3},\ref{inst4}}
    \and Y. Pan \begin{CJK}{UTF8}{gbsn} (潘宇)\end{CJK}\orcid{0000-0001-7261-8297}\inst{\ref{inst1}\thanks{panyu@cqupt.edu.cn (PY)}}
}

\institute{
\label{inst1}School of Electronic Science and Engineering, Chongqing University of Posts and Telecommunications, Chongqing 400065, P.R. China \and
\label{inst2}INAF - Osservatorio Astronomico di Padova, Vicolo dell'Osservatorio 5, 35122 Padova, Italy \and
\label{inst3}Yunnan Observatories, Chinese Academy of Sciences, Kunming 650216, P.R. China \and
\label{inst4}International Centre of Supernovae, Yunnan Key Laboratory, Kunming 650216, P.R. China   \and
\label{inst5}South-Western Institute for Astronomy Research, Yunnan Key Laboratory of Survey Science, Yunnan University, Kunming, Yunnan 650500, P.R. China \and
\label{inst23}Yunnan Key Laboratory of Survey Science, Yunnan University, Kunming, Yunnan 650500, P.R. China \and
\label{inst19}INAF - Osservatorio Astronomico di Brera, Via E. Bianchi 46, 23807 Merate (LC), Italy \and
\label{inst27}School of Astronomy and Space Science, University of Chinese Academy of Sciences, Beijing 100049, P.R. China   \and
\label{inst28}National Astronomical Observatories, Chinese Academy of Sciences, Beijing 100101, P.R. China\and
\label{inst22}Institute of Space Sciences (ICE, CSIC), Campus UAB, Carrer de Can Magrans, s/n, E-08193 Barcelona, Spain \and
\label{inst7}National Astronomical Observatory of Japan, National Institutes of Natural Sciences, 2-21-1 Osawa, Mitaka, Tokyo 181-8588, Japan \and
\label{inst12}School of Physics, O'Brien Centre for Science North, University College Dublin, Belfield, Dublin 4, Ireland \and
\label{inst10}innish Centre for Astronomy with ESO (FINCA), FI-20014 University of Turku, Finland \and
\label{inst8}Department of Physics and Astronomy, University of Turku, FI-20014 Turku, Finland \and
\label{inst15}Kapteyn Astronomical Institute, University of Groningen, 9700 AV Groningen, The Netherlands \and
\label{inst18}The Oskar Klein Centre, Department of Astronomy, Stockholm University, AlbaNova, SE-10691 Stockholm, Sweden \and
\label{inst21}School of Sciences, European University Cyprus, Diogenes Street, Engomi, 1516 Nicosia, Cyprus \and
\label{inst13}Cosmic Dawn Center (DAWN) \and
\label{inst14}Niels Bohr Institute, University of Copenhagen, Jagtvej 128, DK-2200, Copenhagen N, Denmark \and
\label{inst9}Department of Physics and Astronomy, Aarhus University, Ny Munkegade 120, DK-8000 Aarhus C, Denmark \and
\label{inst33}School of Physics and Electrical Engineering, Liupanshui Normal University, Liupanshui, Guizhou, 553004, P.R. China \and
\label{inst34}Department of Mathematics and Physics, School of Biomedical Engineering, Southern Medical University, Guangzhou 510515, P.R. China
}
   \authorrunning{Z.-H. Peng et al.} 
   \titlerunning{SN 2017ati}

   \date{Received ; accepted }
 
  \abstract
  {
  We present optical photometric and spectroscopic observations of the Type~IIb supernova (SN)~2017ati. 
  It reached the maximum light at about 27~d after the explosion and the light curve shows a broad, luminous peak with an absolute $r$-band magnitude of $M_{r} = -18.48 \pm 0.16$~mag. 
  At about 50~d after maximum light, SN~2017ati exhibits a decline rate close to that expected from the $^{56}$Co $\rightarrow$ $^{56}$Fe radioactive decay, at 0.98 mag per 100 days, as usually observed in SNe IIb. However, it remains systematically brighter at late times by about 1--2~mag, exceeding the usual upper luminosity range of this class. 
  As a result, modelling the light curve of SN~2017ati with a standard $^{56}$Ni decay scenario requires a large nickel mass of up to $\sim0.37\,M_{\odot}$ and still fails to reproduce the early-time light curve adequately. 
  In contrast, incorporating additional energy input from a magnetar yields a significantly improved fit to the light curve of SN~2017ati, which would reduce the nickel mass to $\sim0.21\,M_{\odot}$, still close to the upper end of the range typically inferred for SNe~IIb. Comparing the fitted results of SN~2017ati with the known sample of SNe~IIb indicates that its luminosity evolution is best explained by a combination of neutron star spin-down energy and radioactive nickel deposition.
  From late-time nebular spectra of SN~2017ati, the luminosity of the [\Oi]~$\lambda\lambda6300,6364$ doublet implies an oxygen mass of $\sim1.82-3.34\,M_{\odot}$, and the combination of a [\Caii]/[\Oi] flux ratio of $\sim0.5$ with nebular spectral model comparisons favours a progenitor zero-age main-sequence mass of $\geq17\,M_{\odot}$.
  }

   \keywords{stars: mass-loss -- supernovae: general -- supernovae:  individual:   SN\,2017ati}

   \maketitle

\nolinenumbers
\section{Introduction}

Core-collapse supernovae (CCSNe) constitute a broad and heterogeneous class of stellar explosions, exhibiting a wide range of observational behaviours that reflect the diversity of progenitor systems and explosion mechanisms.
Within this family, Type~II SNe are defined by the presence of hydrogen features in photospheric spectra.
SNe~Ib are distinguished by the absence of hydrogen signatures while displaying strong helium lines, whereas Type~Ic events show neither hydrogen nor helium features.
SNe~IIb occupy a transitional regime, in which hydrogen features are visible at early epochs but gradually fade as helium lines strengthen, leading to spectral characteristics that increasingly resemble those of SNe~Ib at later phases \citep{Filippenko1997ARA&A..35..309F, Gal-Yam2017hsn..book..195G, Modjaz2019NatAs...3..717M}.

In addition to the classical spectroscopic categories, several additional sub-classes of CCSNe have been firmly established over the past two decades, including Types~Ibn and Icn. 
Collectively, SNe~Ib, Ic, IIb, Ibn, and Icn are commonly grouped under the class of stripped-envelope SNe \citep[SE-SNe,][]{Clocchiatti1996ApJ...459..547C, Matheson2001AJ....121.1648M, Gangopadhyay2025arXiv251204010G}, reflecting progenitor stars that experienced partial or near-complete removal of hydrogen and/or helium layers prior to explosion. 
Within this population, the majority of SNe~IIb display peak absolute magnitudes fainter than $-18.0$~mag. 
However, a small fraction of SNe~IIb are markedly more luminous \citep[e.g. SN~2018gk, DES14X2fna;][]{Bose2021MNRAS.503.3472B, Grayling2021MNRAS.505.3950G, Gomez2022ApJ...941..107G}. 
These overluminous events occupy an intermediate luminosity regime between normal CCSNe and superluminous SNe \citep[$M_{\rm peak}\lesssim-21$~mag;][]{Gal2019ARA&A..57..305G}.

In the standard framework of SNe~IIb, the optical light curve is predominantly powered by the radioactive decay of $^{56}$Ni produced during the explosion into $^{56}$Co, followed by its subsequent decay to stable $^{56}$Fe. 
Analytical and semi-analytical formulations of this radioactive-decay scenario, such as the Arnett model \citep{Arnett1982ApJ...253..785A} and the more recent treatment by \citet{Khatami2019ApJ...878...56K}, enable estimates of key explosion parameters from the observed light curves \citep{Taddia2018A&A...609A.136T, Stritzinger2018A&A...609A.135S,Bersten2018Natur.554..497B}. 
By contrast, several other classes of CCSNe are dominated by alternative energy sources, for example strong interaction between the ejecta and circumstellar material in SNe~IIn \citep{Chevalier1982ApJ...259..302C, Chugai1991MNRAS.250..513C, Moriya2013MNRAS.435.1520M}. 
Even in such cases, a $^{56}$Ni-based framework can still provide useful constraints on explosion properties \citep[e.g.][]{Prentice2016MNRAS.458.2973P, Meza2020A&A...641A.177M}. 
For events whose luminosity evolution is primarily governed by radioactive decay, including SNe~IIb, a higher peak brightness generally implies a larger synthesized mass of $^{56}$Ni required to power the light-curve maximum.

More recently, magnetar-based scenarios have been shown to successfully reproduce the observed light curves of  Broad-Line Type Ic SNe (SNe Ic-BL).
Within this framework, the luminosity evolution is sustained by a combination of radioactive energy input and additional power supplied by a central engine, arising from the spin-down of a rapidly rotating neutron star \citep{Kasen2010ApJ...717..245K, Woosley2010ApJ...719L.204W}. In particular, \cite{Wang2017ApJ...837..128W} (but see \citealt{Dessart2017A&A...603A..51D}) showed that a magnetar model can successfully reproduce the light curves of the SNe~Ic-BL SN~1998bw and SN~2002ap, with magnetar energy dominating within the first $\sim50$~d and again at very late phases, thereby explaining the deviations from a pure $^{56}$Ni-powered evolution. 
By contrast, for a subset of X-ray transients linked to SNe~Ic-BL, such as EP~250108a/SN~2025kg, one could expect that the early emission post explosion originates from magnetar energy deposition, and not solely from shock-cooling \citep{Eyles2025ApJ...988L..14E,Rastinejad2025ApJ...988L..13R,Srinivasaragavan2025ApJ...988L..60S, Zhu2025MNRAS.544L.139Z, Li2025arXiv250417034L}. In these events, multi-wavelength observations show clear evidence for circumstellar interaction, indicating that both magnetar input and ejecta-CSM interaction contribute to the luminosity evolution. These studies provide a direct demonstration that hybrid energy sources may operate in the earliest phases of some broad-line Type Ic SNe, complementing the classical magnetar-powered light curve framework.

In addition, a subset of SNe~IIb, such as SNe~1993J \citep{Richmond1994AJ....107.1022R}, 2011fu \citep{Morales2015MNRAS.454...95M},  2016gkg \citep{Arcavi2017ApJ...837L...2A, Bersten2018Natur.554..497B}, and 2024iss \citep{Chen2025arXiv251022997C}, display an early-time excess in their light curves prior to the main maximum. 
This feature is commonly interpreted as emission from post-shock-breakout cooling associated with a progenitor consisting of a compact core embedded within an extended, low-mass envelope \citep{Bersten2012ApJ...757...31B, Nakar2014ApJ...788..193N, 2025A&A...698A.306M,2025A&A...698A.305M}. 
The duration of this phase is typically limited to a few days and is not universally detected among SNe~IIb, as described in the case of  SN~2008ax \citep{Pastorello2008MNRAS.389..955P, Tsvetkov2009PZ.....29....2T, Taubenberger2011MNRAS.413.2140T}.
When observed, this early bump provides valuable constraints on progenitor properties such as the stellar radius and the role of binary interaction through comparison with hydrodynamic models \citep{Sapir2017ApJ...838..130S}.

In this work, we present optical photometric and spectroscopic observations of SN~2017ati, a comparatively luminous member ($M_{r} = -18.48 \pm 0.16$~mag) of the SN~IIb population.
This study is structured as follows: Section~\ref{section:Basic_information} introduces SN~2017ati. 
Section~\ref{section:photometry} examines the light-curve evolution and colour behaviour, and explores a range of semi-analytic models to interpret the luminosity and temporal evolution of SN~2017ati.
Section~\ref{section:spect} is devoted to spectroscopic analysis, including a comparison of the spectral characteristics with those of other Type~IIb SNe. 
A detailed discussion is given in Section~\ref{section:sum}, followed by a summary of the main results in Section~\ref{section:conclusion}.
In addition, the data reduction procedures are described in Appendix~\ref{section:data}, the supplementary data tables are compiled in Appendix~\ref{SpecInfo}, and the supplementary figures are presented in Appendix~\ref{sect:supfigures}. 

\section{Basic information for SN 2017ati}
\label{section:Basic_information}

SN~2017ati (also known as ATLAS17era\footnote{\url{https://atlas.fallingstar.com/?utm_source=chatgpt.com}}, CSS170303-094957+671059\footnote{\url{http://nesssi.cacr.caltech.edu/catalina/20170303/1703031670144113388.html}}, and Gaia17aiq\footnote{\url{http://gsaweb.ast.cam.ac.uk/alerts/alert/Gaia17aiq/}}) was discovered on 2017 February 6 (MJD= 57790.4) with the Gaia photometric instrument on board the Gaia spacecraft \citep{Delgado2017TNSTR.184....1D}, and was subsequently classified as a Type~IIb SN by \citet{Benetti2017TNSCR.258....1B}.
The first detection on 2017 February 5 (MJD = 57789.5) yielded an apparent magnitude of  $m_V = 17.62 \pm 0.16$~mag (Vega system) on ASAS-SN, slightly earlier than the reported epoch, while the last non-detection on 2017 January 5 (MJD = 57758.5) provided a limiting apparent magnitude of $m_c = 20.89$~mag in the \textit{cyan} filter (AB system) on ATLAS.

The coordinates of SN~2017ati are RA = $09^\mathrm{h}49^\mathrm{m}56^\mathrm{s}.70$ and Dec = $+67^\circ10'59''.56$. The location of SN 2017ati is shown in Fig.~\ref{fig:location}.
SN~2017ati was initially reported to the Transient Name Server (TNS) as a hostless SN.
However, a wider-field inspection revealed that SN~2017ati lies between two galaxies, at projected distances of $36^{\prime\prime}$ (KUG~0946+674) and $76^{\prime\prime}$ (WISEA~J094948.98+671113.0) from their respective nuclei \citep{Balakina2019mmag.conf...32B, Taggart2021MNRAS.503.3931T}. Following the method of \citet{Sako2018PASP..130f4002S}, we compute the normalized projected distance between the SN and each galaxy using the so-called directional light radius (DLR), defined as
\begin{equation}
d_{\mathrm{DLR}} = \frac{\text{SN--galaxy angular separation (arcsec)}}{\text{DLR (arcsec)}}.
\end{equation}
Here, SN--galaxy angular separation denotes the angular distance between the SN and the geometric center of the galaxy, measured in arcseconds on the celestial sphere, while the DLR represents the effective radius of the galaxy in the direction toward the SN. Assuming that the galaxies KUG~0946+674 (DLR = $36.60^{\prime\prime}$) and WISEA~J094948.98+671113.0 (DLR = $31.41^{\prime\prime}$) are circular, we obtain $d_{\mathrm{DLR}} \sim 0.98$ for KUG~0946+674 and $d_{\mathrm{DLR}} \sim 2.42$ for WISEA~J094948.98+671113.0.
This places SN 2017ati about 10~kpc (36$^{\prime\prime}$) from the centre of the nearest galaxy,  KUG~0946+674.
The two galaxies appear to be interacting, and such interaction could have triggered a region of enhanced star formation, which may account for the occurrence of a CCSN at such a remote location from the host galaxy disc.

We adopt a recessional velocity for SN~2017ati of $v = 5084 \pm 13\,\mathrm{km\,s^{-1}}$ \footnote{\url{https://ned.ipac.caltech.edu/byname?objname=SN+2017ati&hconst=67.8&omegam=0.308&omegav=0.692&wmap=4&corr_z=1}}\citep{Mould2000ApJ...529..786M}, corrected for the effects of the Virgo Cluster, the Great Attractor, and the Shapley Supercluster, corresponding to a redshift of $z = v/c = 0.01696$.
These corrections account for the peculiar motions of the Milky Way and other nearby large-scale structures, which would otherwise affect the derived recessional velocity \citep{Marinoni1998ApJ...505..484M}. 
Assuming a standard cosmological model with \( H_0 = 73 \pm 5 \, \mathrm{km \, s^{-1} \, Mpc^{-1}} \), \( \Omega_M = 0.27 \), and \( \Omega_\Lambda = 0.73 \) \citep{Spergel2007ApJS..170..377S}, we obtain a luminosity distance of \( d_L = 70.80 \pm 5.20 \, \mathrm{Mpc} \), corresponding to a distance modulus of \( \mu = 34.25 \pm 0.16 \, \mathrm{mag} \) for SN~2017ati.

For the interstellar reddening, we use \( E(B-V)_{\text{Gal}} = 0.104 \, \mathrm{mag} \) \citep[$A_V^{\text{Gal}}$ = 0.322 mag]{Schlafly2011ApJ...737..103S} for the Galactic reddening component, as derived from the NASA/IPAC Extragalactic Database (NED)\footnote{\url{https://ned.ipac.caltech.edu}}, and assume a reddening law with \( R_V = 3.1 \) \citep{Cardelli1989ApJ...345..245C}. However, the extinction contribution from the host galaxy remains uncertain, but likely negligible given the outer location of the object.
We therefore adopt a total reddening of \( E(B-V)_{\text{tot}} = 0.104 \, \mathrm{mag} \) towards SN~2017ati, acknowledging that this value likely represents a lower limit to the true extinction.

\section{Photometry}
\label{section:photometry}

\subsection{Apparent magnitude light curves}
\label{sect:decline}

We followed the photometric evolution of SN~2017ati for approximately 300 days post-discovery. The resulting optical light curves are shown in Fig.~\ref{fig:light_curve}. We estimated the explosion epoch of SN~2017ati using the fireball expansion method. A 2nd-order polynomial fit, Gaussian Process (GP), and Artificial Neural Network (ANN) models were applied to the data obtained within a 20-day interval before and around the $V$-band maximum in flux space \citep[e.g.][]{Gonzalez-Gaitan2015MNRAS.451.2212G, Wang2024A&A...691A.156W}, as shown in the top panel of Fig.~\ref{fig:explosion+peak}.
Following this approach, the explosion epoch of SN~2017ati is calculated to be $\mathrm{MJD}=57784.5^{+1.4}_{-2.3}$, derived as the average of three independent estimates based on the epoch at which the flux extrapolates to zero. This epoch is adopted as the reference time throughout this paper.
To estimate the peak brightness of SN~2017ati, we fitted the $o$-band light curve over a two week interval around maximum using a 2nd-order polynomial fit, GP, and ANN, and the mean of these fits gives $o = 16.17 \pm 0.01$~mag at MJD~$57812.0 \pm 0.8$, as shown in the bottom panel of Fig.~\ref{fig:explosion+peak}. 

After the maximum light, the multi-band light curves of SN\,2017ati display an approximately linear decline, particularly pronounced in the $roiz$ bands\footnote{The \textit{Gaia} $G$-band photometry, which broadly corresponds to the $r$ band, was used in the construction of the $r-i$ colour evolution over the range $0-0.6$~mag (see \url{https://gea.esac.esa.int/archive/documentation/GDR2/Data_processing/chap_cu5pho/sec_cu5pho_calibr/ssec_cu5pho_PhotTransf.html}).}.  Notably, variations in the decline rates appear around 50~days, especially in the $BgV$ bands, while the $u$- and $c$-band data are too sparse to provide a reliable estimate. We estimated the post-peak decline slopes in each band through linear fits to the data after maximum, with the results summarized in Table~\ref{tab:decline_rate}.
Owing to the noticeable change in the light-curve slope around 50\,d in the $BgV$ bands, we derived decline rates between peak and 50\,d ($\gamma_{0-50}$) and between 50\,d and 160\,d ($\gamma_{50-160}$) for these bands, while a single decline rate was measured for the $oiz$ bands. In addition, we measured a late-time decline rate for the $r/G$ band between 160 and 300\,d ($\gamma_{160-300}$).
During the early phase, the decline proceeds more rapidly in the bluer bands than in the redder ones. For example, during the early phase, the $g$ band declines rapidly at $\sim$ 4.49 mag 100~\ d $^{-1}$, whereas the $r$ band fades more slowly at only $\sim1.37$ mag 100~\ d $^{-1}$.  At later phases, the light-curve decline becomes significantly slower, with decline rates approaching that expected from the $^{56}$Co$\rightarrow$$^{56}$Fe radioactive decay of 0.98~mag~100~d$^{-1}$ (e.g. $\gamma_{50\text{--}160}(B) \sim$ 0.97~ mag 100~\ d $^{-1}$).

\begin{table}
    \centering
    \caption{Decline rates (in $\rm mag\, /100\,d$) of the multi-band light curves of SN\,2017ati.}
    \renewcommand{\arraystretch}{1.2}
    \setlength{\tabcolsep}{15pt}
    \begin{tabular}{ccc}
        \hline\hline
        Filter & $\gamma_{0-50}$ & $\gamma_{50-160}$ \\
        \hline
        $B$ & 4.72$\pm$0.59 & 0.97$\pm$0.06 \\
        $g$ & 4.49$\pm$0.29 & 0.96$\pm$0.04 \\
        $V$ & 4.53$\pm$0.21 & 1.31$\pm$0.03 \\
        \hline
        Filter & $\gamma_{0-160}$& $\gamma_{160-200}$ \\
        \hline
        $r/G$ & 1.37$\pm$0.01 & 1.46$\pm$0.01 \\
        $o$ & 2.76$\pm$0.04 & -- \\
        $i$ & 1.81$\pm$0.02 & -- \\
        $z$ & 1.54$\pm$0.02 & -- \\
        \hline\hline
    \end{tabular}
    \label{tab:decline_rate}
\end{table}

\subsection{Absolute magnitude light curves}
\begin{figure}[htbp]
\centering
\includegraphics[width=1.0\linewidth]{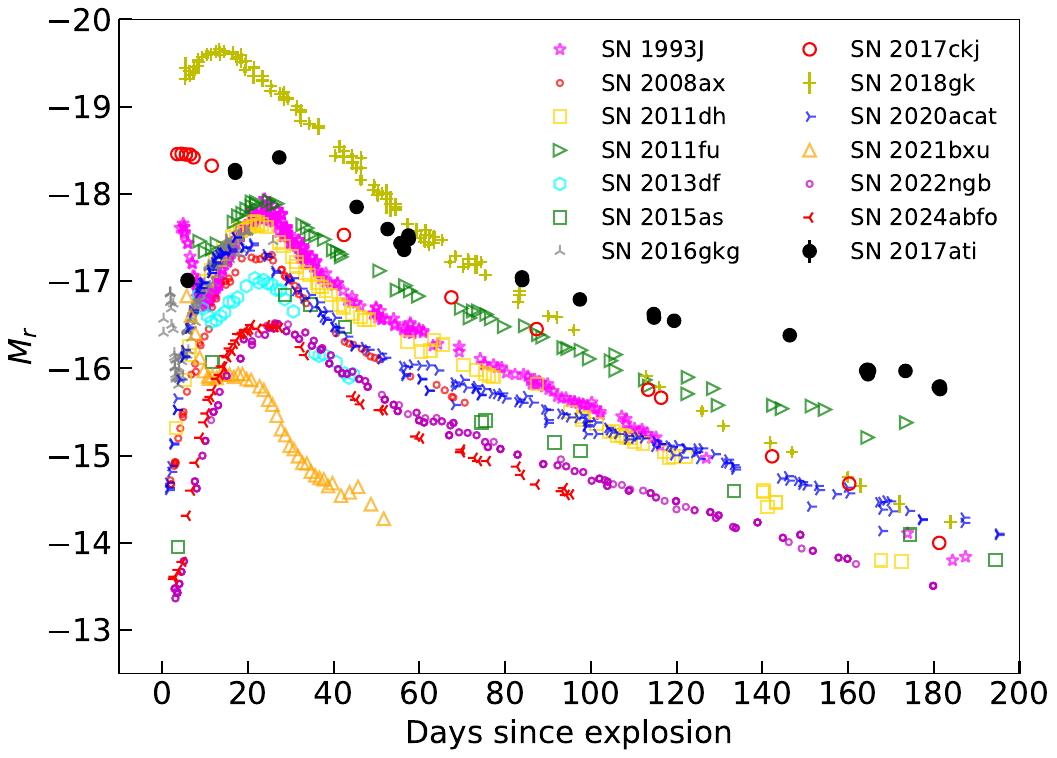} 
    \caption{Absolute \textit{r/G}-band light curve of SN~2017ati compared with other SNe IIb in \textit{r/R}-band.  All light curves have been corrected for reddening and shifted according to the distances listed in Table \ref{tab:SNe_IIb}. }
    \label{fig:Absolute_magnitude}
\end{figure}

In this work, a comparison sample of Type~IIb SNe is compiled to serve as a reference for the photometric and spectroscopic analysis of SN~2017ati. 
The objects included in Appendix~Table~\ref{tab:SNe_IIb} are all classified as Type~IIb SNe and generally have well-sampled $r$-band light curves within the first $\sim$200~d after the explosion. The only exception is SN~2021bxu, which does not have complete coverage within the first 200~d after the explosion, although its light-curve evolution shows some differences compared with the other objects in the sample.
Within this sample, SN~2008ax exhibits a spectral evolution closely matching that of SN~2017ati  (see detailed comparison in Sect. \ref{Spectra}), while SN~2017ckj shows a peak luminosity comparable to SN~2017ati but reaches maximum light at an earlier epoch. 

Adopting the distance and extinction values from Sect.~\ref{section:Basic_information}, a second-order polynomial fit to the $r$-band light curve of SN~2017ati yields a peak absolute magnitude of $M_{r} = -18.48 \pm 0.16$~mag at $\mathrm{MJD} = 57811.1 \pm 0.7$, placing it toward the luminous end of our selected comparison sample. 
Type~IIb SNe generally display comparable luminosities near the radioactive-decay-powered maximum, typically spanning \mbox{$-16.5$ to $-18$}~mag \citep{Taddia2018A&A...609A.136T, Stritzinger2018A&A}, with the $r$/$R$-band peak absolute magnitudes of the selected comparison sample provided in Table~\ref{tab:SNe_IIb}. 
A comparison of the absolute $r$-band magnitudes for SN~2017ati and the reference Type~IIb SNe is shown in Fig.~\ref{fig:Absolute_magnitude}.
Within this context, SN~2017ati lies near the upper boundary of the typical luminosity range. 
SN~2018gk \citep[$M_{r} = -19.64 \pm 0.24$~mag;][]{Bose2021MNRAS.503.3472B} attains an even higher peak luminosity, while SN~2017ckj \citep[$M_{r} = -18.46 \pm 0.07$~mag;][]{Li2025A&A...704A.233L} exhibits a peak magnitude comparable to that of SN~2017ati.
However, neither SN~2018gk nor SN~2017ckj exhibits a light curve shape similar to that of SN~2017ati, whereas SN~2011fu does (apart from the initial bump) but is slightly fainter.
In contrast, SN~2021bxu does not show a distinct peak but exhibits a plateau at a mostly comparable epoch \citep[$M_{\mathrm{r}} \sim -15.93$\,mag;][]{Desai2023MNRAS.524..767D}, appearing significantly fainter.

In contrast, the standard radioactive-decay framework successfully accounts for the light curves of typical Type~IIb SNe, but it fails to reproduce the observed properties of energetic SNe~Ic-BL. Specifically, it cannot simultaneously explain their high peak luminosities and their late-time photometric evolution. 
For luminous events such as SN~2017ati, which exceed the brightness of ordinary Type~IIb SNe, we therefore consider the possibility of an additional energy source contributing to the observed luminosity.

\subsection{Colour evolution}
\label{section:Colour_evolution}

\begin{figure}[htbp]
\centering
\includegraphics[width=1.0\linewidth]{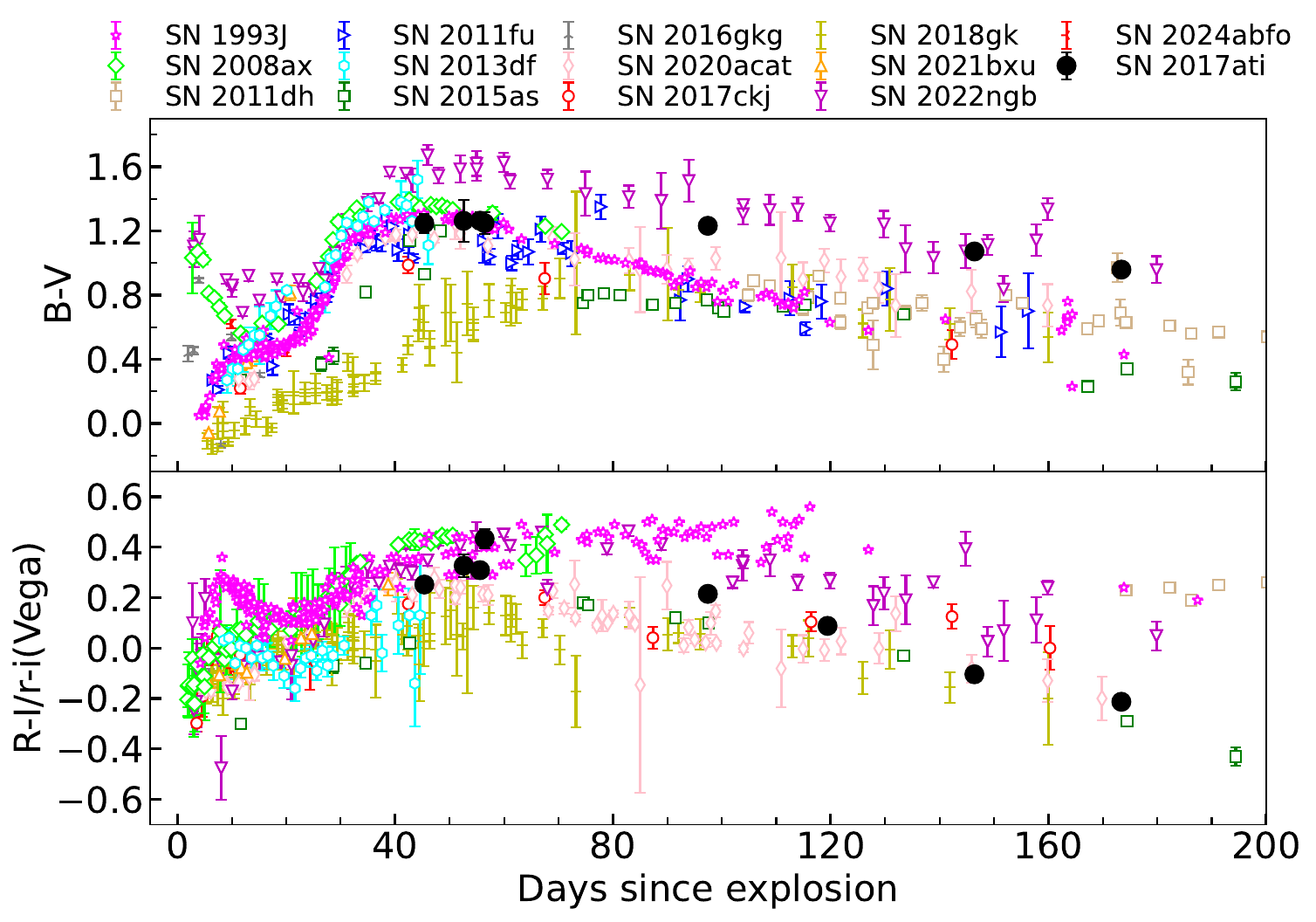} 
    \caption{Colour evolution of SN\,2017ati, compared to a sample of SNe~IIb.
    \emph{Upper panel}: $B~-~V$ colour evolution. \emph{Lower panel}: ($R~-~I$) or ($r~-~i$) colour evolution.
    The colour curves are corrected for both Galactic and host galaxy extinction.}
    \label{fig:colour_evolution}
\end{figure}

The intrinsic colour evolution of SN~2017ati is presented together with the comparison Type~IIb sample, with all colour curves corrected for reddening using the parameters provided in Table~\ref{tab:SNe_IIb}. During the early phases after the explosion, the $B-V$ colour of SN~2017ati is missing, whereas the comparison Type~IIb sample exhibits a rising trend. Between 40 and 60 days after the explosion, the $B-V$ colour of SN~2017ati remains at around 1.26~mag. Most comparison Type~IIb SNe reach their peak colour in this interval, except SN~2018gk, which continues to redden. From 90 to 180 days after the explosion, the $B-V$ colour of SN~2017ati gradually decreases from $1.23 \pm 0.04$~mag to $0.96 \pm 0.03$~mag.
Overall, the $B-V$ colour evolution of SN~2017ati broadly traces the behaviour seen in previously studied Type~IIb SNe (see the top panel in Fig.~\ref{fig:colour_evolution}).

For SN~2017ati, the $r-i$ colour evolution closely resembles that observed in SNe~2015as, 2018gk, and 2020acat, showing an initial rise to a well-defined peak followed by a rapid decline, suggesting a broadly comparable physical evolution  (see the bottom panel in Fig.~\ref{fig:colour_evolution}).
Owing to the lack of very early-time observations, the initial shock-cooling phase cannot be directly constrained. During the photospheric phase, between $\sim$40 and 60~d after the explosion, the $r-i$ colour becomes progressively redder and reaches a maximum of $0.43 \pm 0.04$~mag at about 56~d, which is consistent with continued ejecta expansion and photospheric cooling, causing the spectral energy distribution to shift toward longer wavelengths. At later epochs, between approximately 90 and 180~d after the explosion, the $r-i$ colour of SN~2017ati declines rapidly, showing behaviour similar to that observed in SNe~2015as, 2020acat, and 2018gk. In contrast, SNe~1993J, 2011dh, and 2017ckj do not display a comparable decrease at similar phases, but instead maintain an approximately constant $R-I/r-i$ colour.

\subsection{Pseudo-bolometric light curves} 
\begin{figure}[htbp]
\centering
\includegraphics[width=1.0\linewidth]{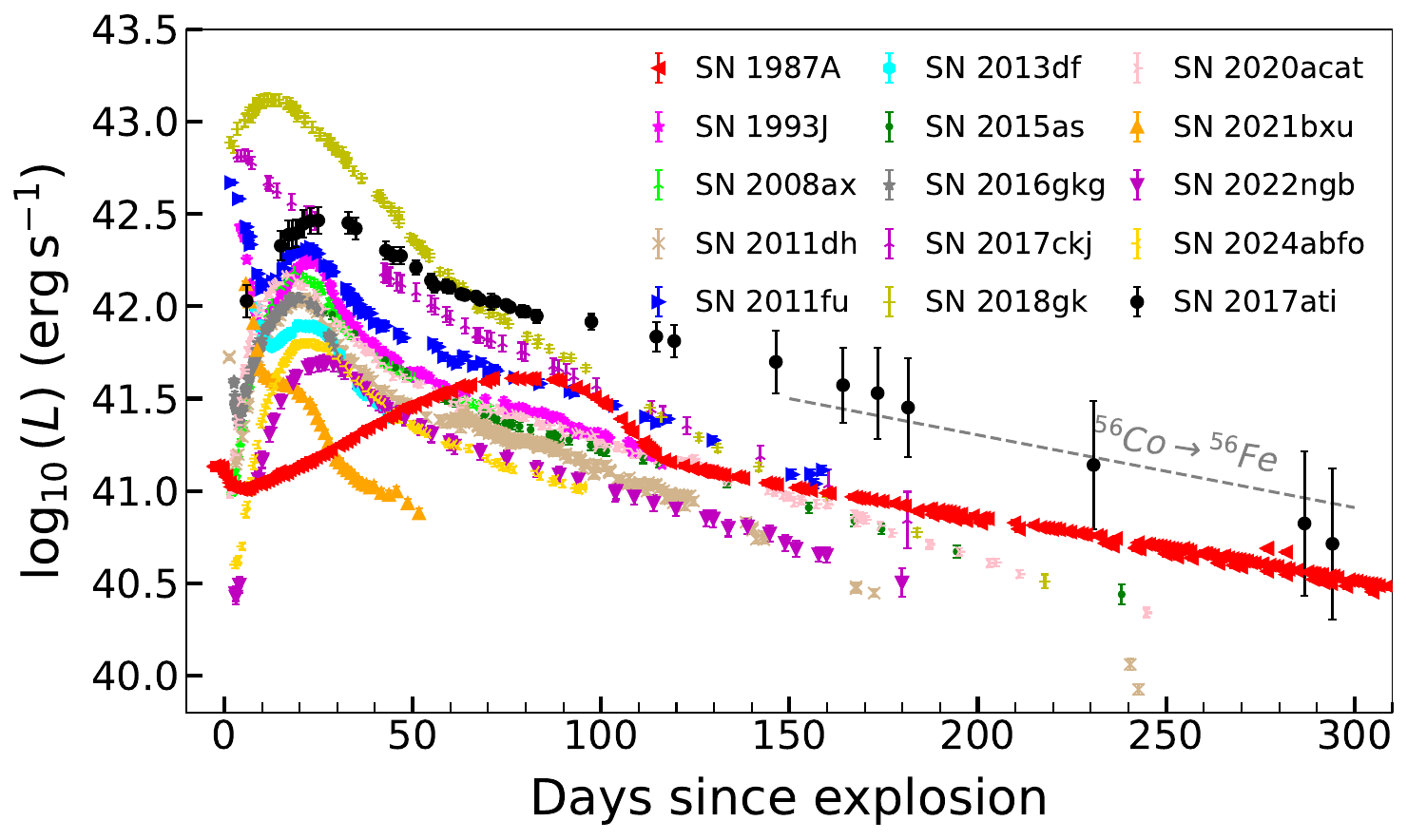} 
    \caption{Comparison of the pseudo-bolometric light curve of SN~2017ati with those of SN~1987A and other Type IIb SNe. The grey dashed line illustrates the expected slope of the light curve under the assumption that all energy from $^{56}$Co decay is fully thermalized by the ejecta.}
    \label{fig:Lbol}
\end{figure}

Using the available optical photometry and the \texttt{SuperBol}\footnote{\url{https://github.com/mnicholl/superbol}} code \citep{Nicholl_2018}, a pseudo-bolometric light curve of SN~2017ati was constructed from the optical bands. For epochs with missing data in certain filters, values were estimated by extrapolating from the available measurements. For a consistent comparison, pseudo-bolometric light curves of the other SNe were constructed using the $U/u$ to $I/i$ bands, with filters lacking sufficient data not included in the fitting procedure.

The pseudo-bolometric light curve of SN~2017ati, derived through blackbody fits, is compared with those of other Type~IIb SNe summarised in Table~\ref{tab:SNe_IIb} and is shown in Fig.~\ref{fig:Lbol}.
SN~2017ati exhibits a peak luminosity of around \(L \sim 3.00 \times 10^{42} \, \mathrm{erg\,s^{-1}}\), which is lower than those of SNe~2017ckj (\(L \sim 6.65 \times 10^{42} \, \mathrm{erg\,s^{-1}}\)) and 2018gk (\(L \sim 1.43 \times 10^{43} \, \mathrm{erg\,s^{-1}}\)), yet still relatively bright compared with other SNe IIb in the sample. The rise time of SN~2017ati is approximately 29.1~d, in close agreement with the values determined for SNe~2008ax, 2022ngb, and 2024abfo \citep{Pastorello2008MNRAS.389..955P, Zhao2025arXiv251209384Z, Reguitti2025A&A...698A.129R}.
In contrast, events classified as Type~IIb such as SNe~1993J, 2011fu, 2013df, and 2016gkg present pronounced double-peaked profiles, indicating that contribution from an expanded progenitor envelope is more significant in these cases than in SN~2017ati.

Based on the bolometric luminosity, an estimate of the \({}^{56}\mathrm{Ni}\) mass synthesised in SN~2017ati is derived. For hydrogen-rich SNe, including SN~2017ati, determining the fraction of luminosity contributed by \({}^{56}\mathrm{Ni}\) near peak is challenging, as hydrogen recombination can lead to uncertain estimates \citep{Charalampopoulos2025A&A...700A.138C}. Using the peak luminosity of SN~2017ati and adopting the Arnett rule \citep{Arnett1982ApJ...253..785A}, the \({}^{56}\mathrm{Ni}\) mass is evaluated to be $\sim 0.40 M_{\odot}$.  This value should be treated as an upper limit. 

For SE-SNe, \citet{Clocchiatti1997ApJ...491..375C} introduced a parametrised description of the luminosity decline at late phases:
\begin{equation}
L(t)=L_0(t) \times\left[1-e^{-\left(T_0 / t\right)^2}\right],
\end{equation}
where \(T_{0}\) denotes the characteristic timescale governing the transition from complete to incomplete trapping of the radioactive decay energy.
Assuming homologous expansion with spherical geometry and radioactive material located centrally, \citet{Jerkstrand2012A&A...546A..28J} derived the theoretical luminosity for the case in which the energy released by the decay of \({}^{56}\mathrm{Co}\) is fully trapped, expressed as
\begin{equation}
L_{0}(t) = 9.92 \times 10^{41} \frac{M_{\mathrm{Ni}}}{0.07\,\mathrm{M}_{\odot}}
\left(e^{-t/111.4} - e^{-t/8.8}\right)\,\mathrm{erg\,s^{-1}},
\end{equation}
where \(M_{\mathrm{Ni}}\) denotes the ejected mass of \({}^{56}\mathrm{Ni}\).
Using this formulation and performing an Markov chain Monte Carlo (MCMC) fit, the \({}^{56}\mathrm{Ni}\) mass for SN~2017ati is determined to be \(0.29^{+0.05}_{-0.04}\,\mathrm{M_{\odot}}\), along with a characteristic timescale of \(T_{0}=225.8^{+58.4}_{-45.9}\,\mathrm{d}\).
For SE SNe, the characteristic timescale is typically in the range 
\(T_{0} \approx 80-140~\mathrm{d}\) \citep{Sharon2020MNRAS.496.4517S}. 
The unusually large value of \(T_{0}\) derived for SN~2017ati may reflect the presence of additional energy sources beyond radioactive decay or a relatively massive progenitor star.

A comparison with the tail luminosity of SN~1987A, for which
$M(^{56}\text{Ni})_{\text{SN~1987A}} = 0.075 \pm 0.005 M_{\odot}$ \citep{Arnett1996SupernovaeAN}, yields a value of \(0.47 \pm 0.13\,\mathrm{M}_{\odot}\) for SN~2017ati. These estimates of the \({}^{56}\mathrm{Ni}\) mass for SN~2017ati are relatively high compared with values usually found for SNe IIb, suggesting that additional energy input beyond radioactive heating may contribute to powering the light curve.  Using the equation presented in \citet{Osmar2023ApJ...955...71R}, we estimate $M_{\mathrm{Ni}} = 0.18^{+0.05}_{-0.04}\,M_{\odot}$ for SN~2017ati.
All estimated values of the \({}^{56}\mathrm{Ni}\) mass, together with the \({}^{56}\mathrm{Ni}\) mass inferred from the models applied in Sect. \ref{sec:MOSFiT}, are summarised in Table \ref{tab:Ni}.

\subsection{Modelling the multi-band light curves with MOSFiT}
\label{sec:MOSFiT}

Given that SN~2017ati exhibits luminosity differences compared with previously studied SNe IIb, we employed the publicly available Modular Open-Source Fitter for Transients (\texttt{MOSFiT}\footnote{\url{https://mosfit.readthedocs.io/en/latest/index.html}}; \citealt{Guillochon2018ApJS..236....6G}) to model the multi-band light curves and explore potential powering mechanisms. This tool takes multi-band photometry as input and enables the use of prior distributions for model parameters, providing constraints on alternative energy contributions.
A MCMC approach is used in \texttt{MOSFiT} to explore the parameter space associated with each built-in model, resulting in a best-fitting light curve and corresponding parameter constraints. 

In this work, three energy-input scenarios available in \texttt{MOSFiT} are examined to interpret the observed luminosity evolution of SN~2017ati.
\begin{itemize}
\item \({}^{56}\mathrm{Ni}\) decay: luminosity arises solely from the decay chain of \({}^{56}\mathrm{Ni}\) and \({}^{56}\mathrm{Co}\), adopting the standard treatment for Type IIb events \citep{Nadyozhin1994ApJS...92..527N}.
\item \({}^{56}\mathrm{Ni}\) decay plus CSM interaction: luminosity is powered both by radioactive decay and  ejecta interaction with circumstellar material, following the model of \citet{Chatzopoulos2013ApJ...773...76C}.
\item \({}^{56}\mathrm{Ni}\) decay plus magnetar: this model adopts the radioactive framework above and introduces an additional contribution from a magnetar central engine, following the formulation of \citet{Nicholl2017ApJ...850...55N}.
\end{itemize}
For every scenario, \texttt{MOSFiT} is applied with an MCMC inference framework, adopting the \texttt{PHOTOSPHERES.TEMPERATURE\_FLOOR} prescription for the photospheric evolution, following the treatment in \citet{Nicholl2017ApJ...850...55N}. 
Throughout the fitting process, a uniform opacity for UVOIR emission of \(\kappa = 0.195~\mathrm{cm^{2}\,g^{-1}}\) is utilised, following the value recommended by \citet{Nagy2018ApJ...862..143N}.

The results of the \({}^{56}\mathrm{Ni}\) decay model, together with the corresponding corner plot, are displayed in Fig.~\ref{fig:MOSFIT-Ni} of Appendix~\ref{sect:supfigures}. The fitted parameters are listed in Table~\ref{tab:MOSFiT}.
For the \({}^{56}\mathrm{Ni}\) decay model, seven free parameters are employed, yielding an ejecta mass of \(M_{\rm ej}=1.70^{+0.44}_{-0.38}\,\mathrm{M_{\odot}}\), a \({}^{56}\mathrm{Ni}\) mass of \(M_{\mathrm{^{56}Ni}} = 0.37^{+0.07}_{-0.10}\,\mathrm{M_{\odot}}\), a gamma-ray opacity of \(\kappa_{\gamma}=0.06^{+0.03}_{-0.01}\,\mathrm{cm^{2}\,g^{-1}}\), a temperature floor of \(T_{\rm min}=4.47^{+0.21}_{-0.20}\times10^{3}\,\mathrm{K}\), an ejecta velocity of \(v_{\rm ej}=3.63^{+0.26}_{-0.24}\times10^{3}\,\mathrm{km\,s^{-1}}\), and an additional uncertainty term of \(\sigma = 0.37^{+0.03}_{-0.02}\).
The inferred explosion epoch from the model is \(t_{\rm exp}=-33.97^{+1.55}_{-1.03}\,\mathrm{days}\) (measured relative to the first detection), which is earlier than the explosion time estimated from our analysis by \(28.97\) days. 

It is evident that a framework based solely on \({}^{56}\mathrm{Ni}\) radioactive decay fails to reproduce the multi-band light curves of SN~2017ati. Significant discrepancies emerge at early phases, where the observed flux cannot be matched by the model. Although the corner plot suggests statistical convergence, the inferred explosion epoch is markedly earlier than our estimated value, the \({}^{56}\mathrm{Ni}\) mass is higher than that commonly deduced for SNe IIb, and an extra uncertainty of \(\sigma = 0.37\) is required to maintain the fitting quality. Furthermore, the reduced $\chi^{2}$ of the best-fitting model is $ 3.95$, indicating a poor overall fit to the light curves. These limitations indicate that this model does not offer a physically reliable representation of the radiative evolution of SN~2017ati.

We subsequently examined a model combining \({}^{56}\mathrm{Ni}\) radioactive heating with a contribution from CSM interaction. The resulting fit, along with the corresponding corner plot, are shown in Fig.~\ref{fig:CSMNI}, and the fitted parameters are summarised in Table~\ref{tab:MOSFiT}.
Although clear signatures of CSM interaction are not evident in either the spectra or light curves of SN~2017ati, the possibility of interaction with surrounding material cannot be entirely excluded, particularly in the case of a low-density CSM \citep{Luc2022, Peng2026A&A...705A.104P} and in the absence of ultraviolet observations.
This composite framework provides an improved reproduction of the multi-band light curves of SN~2017ati, particularly at early phases and near the peak luminosity. 
However, the fit requires a markedly elevated $^{56}\mathrm{Ni}$ mass of $0.49^{+0.19}_{-0.12}\,M_{\odot}$. The  reduced $\chi^{2}$ of the best-fitting $^{56}\mathrm{Ni}$ plus CSM model is 1.61, representing a substantial improvement over the pure $^{56}\mathrm{Ni}$ decay model. Nevertheless, the inclusion of such a CSM component would produce a rapidly evolving light curve similar to that of SN~2016iog, rather than the evolution rate observed in SN~2017ati, which is consistent with typical SNe~IIb \citep{Peng2026A&A...705A.104P}. Moreover, the corresponding H-line absorption features in the spectra would disappear, a behavior not seen in SN~2017ati (see Fig. \ref{fig:sp_ev}).
Overall, the reduced best-fitting values of these parameters deviate significantly from realistic expectations, making  difficult to regard the model as a credible description of SN~2017ati.

We further explored a model in which the \({}^{56}\mathrm{Ni}\) radioactive heating is supplemented by energy input from a magnetar. This scenario provides a markedly improved reproduction of the multi-band light curves of SN~2017ati, with the scatter reduced to \(\sigma = 0.17^{+0.01}_{-0.02}\). The resulting parameters are also more consistent with those commonly inferred for SNe IIb. In this framework, nine free parameters are employed. 
The best-fitting model and the corresponding corner plot are shown in Fig.~\ref{fig:mosfit}, and the inferred parameters are listed in Table~\ref{tab:MOSFiT}.
The \({}^{56}\mathrm{Ni}\) decay plus magnetar fit yields a magnetic field strength of \(B=13.2^{+4.6}_{-4.3}\times10^{14}~\mathrm{G}\), together with an initial spin period of \(P_{\rm spin}=28.2^{+2.7}_{-6.3}~\mathrm{ms}\). The inferred ejecta mass is \(M_{\rm ej}=1.82^{+0.53}_{-0.59}~\mathrm{M_{\odot}}\) and is accompanied by a \({}^{56}\mathrm{Ni}\) mass of \(M_{\mathrm{^{56}Ni}}=0.21^{+0.08}_{-0.12}~\mathrm{M_{\odot}}\). The fit additionally requires a gamma-ray opacity of \(\kappa_{\gamma}=0.33^{+0.12}_{-0.05}~\mathrm{cm^{2}\,g^{-1}}\), a temperature floor of \(T_{\rm min}=4.57^{+0.11}_{-0.10}\times10^{3}~\mathrm{K}\), and an ejecta velocity of \(v_{\rm ej}=7.24^{+0.70}_{-0.64}\times10^{3}~\mathrm{km\,s^{-1}}\). The model implies an explosion epoch of \(t_{\rm exp}=-7.36^{+1.04}_{-0.99}~\mathrm{d}\) relative to the first detection, which is close to the independently estimated value of \(-5.0\)~d, and adopts a neutron star mass of \(M_{\rm ns}=1.5~\mathrm{M_{\odot}}\). The reduced $\chi^{2}$ of the best-fitting $^{56}\mathrm{Ni}$ plus magnetar model is 1.26, representing a substantial improvement over the previous models and indicating the most satisfactory fit among the three scenarios explored. This demonstrates that the inclusion of magnetar energy provides a physically plausible and statistically robust description of the multi-band light curves of SN~2017ati.

These physical parameters derived from the modelling of SN~2017ati are broadly consistent with expectations for SNe~IIb. 
The inferred ejecta mass of $M_{\rm ej}=1.82^{+0.53}_{-0.59}\,\mathrm{M_{\odot}}$ lies within the range of $\sim1.7-5.2\,\mathrm{M_{\odot}}$ obtained for SE-SNe from radiative-transfer model grids \citep{Dessart2016MNRAS.458.1618D}. 
The $^{56}$Ni mass of $M_{\mathrm{^{56}Ni}} = 0.21^{+0.08}_{-0.12}\,\mathrm{M_{\odot}}$ falls within the distribution observed for SNe~IIb, but is close to the upper end of the reported range of $0.018-0.226\,\mathrm{M_{\odot}}$ \citep{Anderson2019A&A...628A...7A, Meza2020A&A...641A.177M, Osmar2023ApJ...955...71R}.
The ejecta velocity of $v_{\rm ej}=7.24^{+0.70}_{-0.64}\times10^{3}\,\mathrm{km\,s^{-1}}$ is consistent with the spectral velocities measured in Sect.~\ref{sec:vel}.
The measured magnetic field strength of $B = 13.2^{+4.6}_{-4.3} \times 10^{14}~\mathrm{G}$ significantly exceeds the typical range of magnetic fields observed in newly born neutron stars and instead falls within the characteristic range for magnetars \citep[$3 \times 10^{14}-10^{15}$~G;][]{Beniamini2019MNRAS.487.1426B}. 
Compared with the other models discussed above, a framework combining $^{56}$Ni radioactive heating with additional energy input from a magnetar yields parameter values that provide a more physically plausible match to the observed properties for SN~2017ati.
As noted by \citet{Osmar2024Natur.628..733R}, SE-SNe may represent progenitors of magnetars. SN~2017ati may therefore provide a possible example of this scenario, in which a magnetar contributes to the observed luminosity.

\section{Spectroscopy}
\label{section:spect}
\subsection{Spectral sequence}
\label{section:spectrasequence}

\begin{figure*}[ht]
    \centering
    \includegraphics[width=\textwidth]{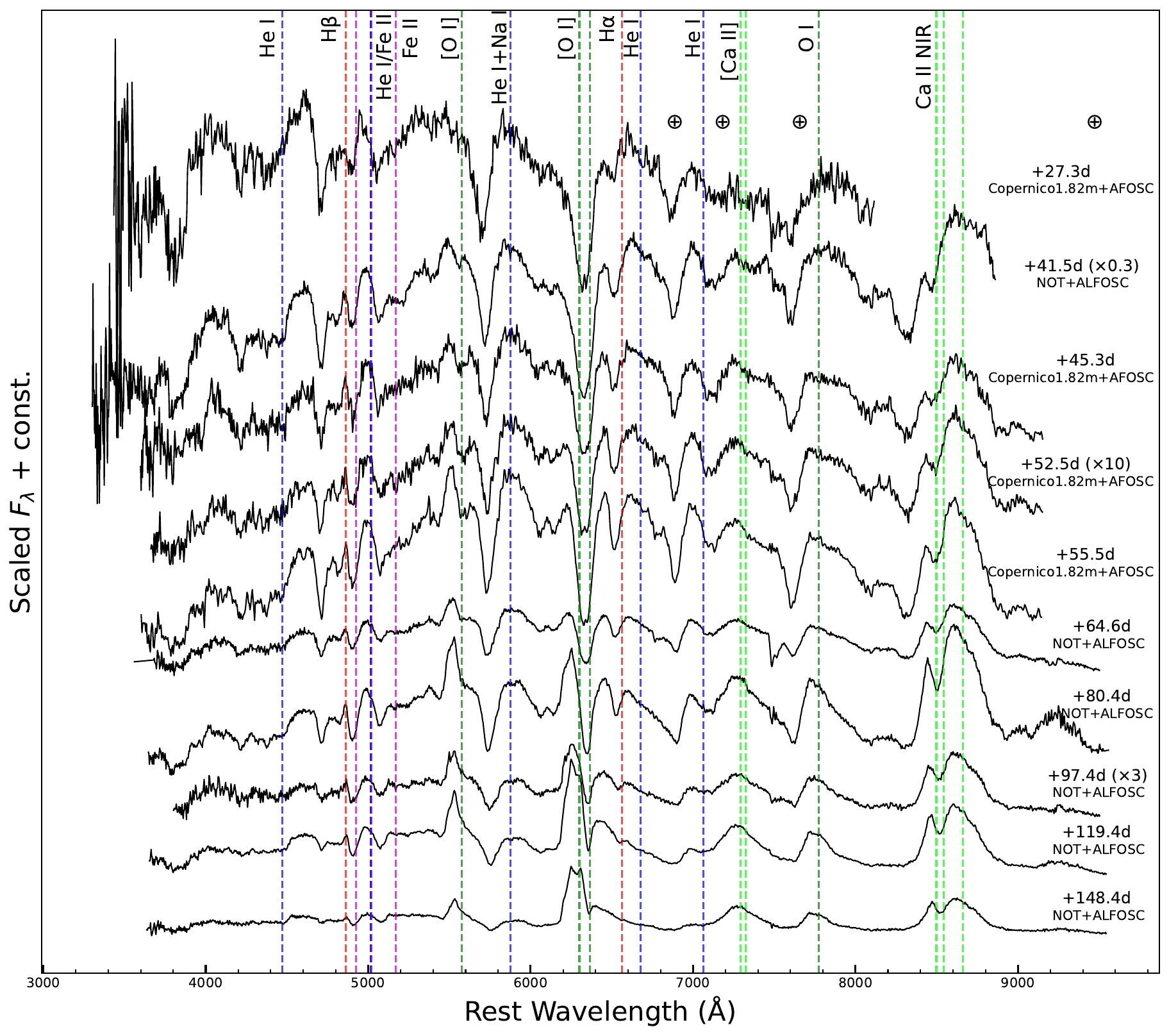} 
    \caption{Optical spectral evolution of SN 2017ati from +27.3 days to +148.4 days since the explosion. The spectra have been corrected for reddening and redshift, and vertically shifted for better visualization.  The epochs are indicated to the right of each spectrum. The positions of major telluric absorption lines are denoted by the $\bigoplus$ symbol.}
    \label{fig:sp_ev}
\end{figure*}

The spectroscopic campaign of SN~2017ati covers phases from +27.3~d to +148.4~d after the explosion, commencing near maximum light and extending well into the nebular regime. The observational log is listed in Table~\ref{tab:spec}, and the spectral sequence is shown in Fig.~\ref{fig:sp_ev}, illustrating the temporal evolution over the photospheric and  the early nebular stages.

The earliest spectrum of SN~2017ati was secured near the optical peak, and the initial five spectra (from +27.3~d to +55.5~d) reveal persistent and pronounced H and He signatures. 
Prominent lines of \ha\,$\lambda6563$, \hb\,$\lambda4861$, \hg\,$\lambda4340$, \Hei\,$\lambda\lambda5876,7065$, and \Oi\,$\lambda7774$ are clearly detected and exhibit a gradual increase in strength as the ejecta cools and recedes.
A detailed analysis of this behaviour is presented in Sect.~\ref{sec:vel}.

As time progresses, nebular signatures such as [\Oi]~$\lambda\lambda6300,6364$ and [\Caii]~$\lambda\lambda7291,7324$, which are characteristic of the nebular phase, become clearly visible from $+64.6$~d. During this phase, the H$\alpha$ emission exhibits a noticeable decline, whereas the [\Oi]~$\lambda\lambda6300,6364$ feature continues to strengthen. 
In the spectrum at $+148.4$~d, SN~2017ati shows a small bump on the redshifted side of the [\Oi] doublet, closely resembling that observed in SN~1993J. This excess may be associated with contributions from H$\alpha$, \Hei, and [\Nii] \citep{Patat1995A&A...299..715P, Houck1996ApJ...456..811H}, is discussed further in Sect.~\ref{Nebularsp-O}.
The \Caii NIR triplet ($\lambda\lambda8498,8542,8662$) is already detected in emission at $+41.5$~d and remains persistently strong at later epochs. 

\subsection{Spectra line profile and velocity evolution}
\label{sec:vel}

\begin{table}[htbp]
\centering
\caption{Velocity measurements for the main absorption minimum with uncertainties (in km/s).}
\label{tab:linev}
\resizebox{\columnwidth}{!}{
\begin{tabular}{c c c c c}
\hline\hline
Phase$^a$ & H$\alpha$    & H$\beta$ & He I $\lambda5876$ & Fe II $\lambda5018$ \\
\text{[days]}    & \text{[km s$^{-1}$]} & \text{[km s$^{-1}$]} & \text{[km s$^{-1}$]} & \text{[km s$^{-1}$]} \\
\hline
+27.3 & 11220(370)   & 9070(300) & 9110(380) & 7680(380) \\
+41.5 & 11030(250)   & 9270(200) & 8110(230) & 7140(160) \\
+45.3 & 10851(240)   & 9290(240) & 7750(270) & 6550(330) \\
+52.5 & 10720(220)   & 9540(200) & 7410(270) & 7050(370) \\
+55.5 & 10520(230)   & 9220(220) & 7280(310) & 6970(470) \\
+64.6 & 10220(180)   & 9140(200) & 7090(240) & 6780(240) \\
+80.4 & 9870(190)    & 9040(210) & 6790(210) & 6790(220) \\
+97.4 & --           & --        & 6360(260) & 6630(230) \\
\hline\hline
\end{tabular}}
\begin{flushleft}
    $^a$ Phases are calculated relative to the explosion epoch (MJD = 57784.5) in the reference frame of the observer. 
    \end{flushleft}
\end{table}

\begin{figure*}[ht]
    \centering
    \includegraphics[width=\textwidth]{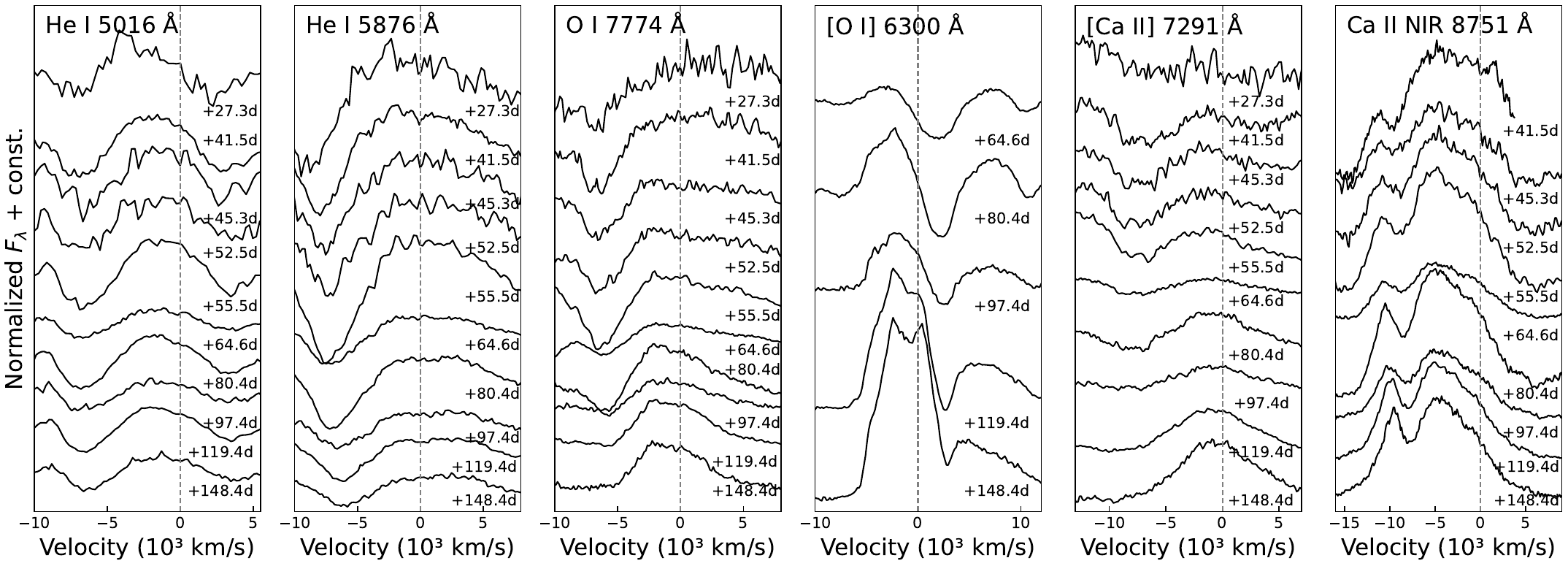} 
    \caption{Line profiles of \Hei, \Oi, $\rm [\Oi]$,   $\rm [\Caii]$ and  $\rm \Caii\ NIR$ within the spectra of SN\,2017ati.
    The dashed lines indicate the velocities corresponding to the rest wavelengths of the emission lines at $\lambda\lambda5016$, 5876, 7774, 6300, 7291, and 8751.
    The epoch of each spectrum is given in the rest frame, relative to the estimated explosion date.}
    \label{fig:line_profiles}
\end{figure*}

\begin{figure}[htbp]
\centering
\includegraphics[width=1.0\linewidth]{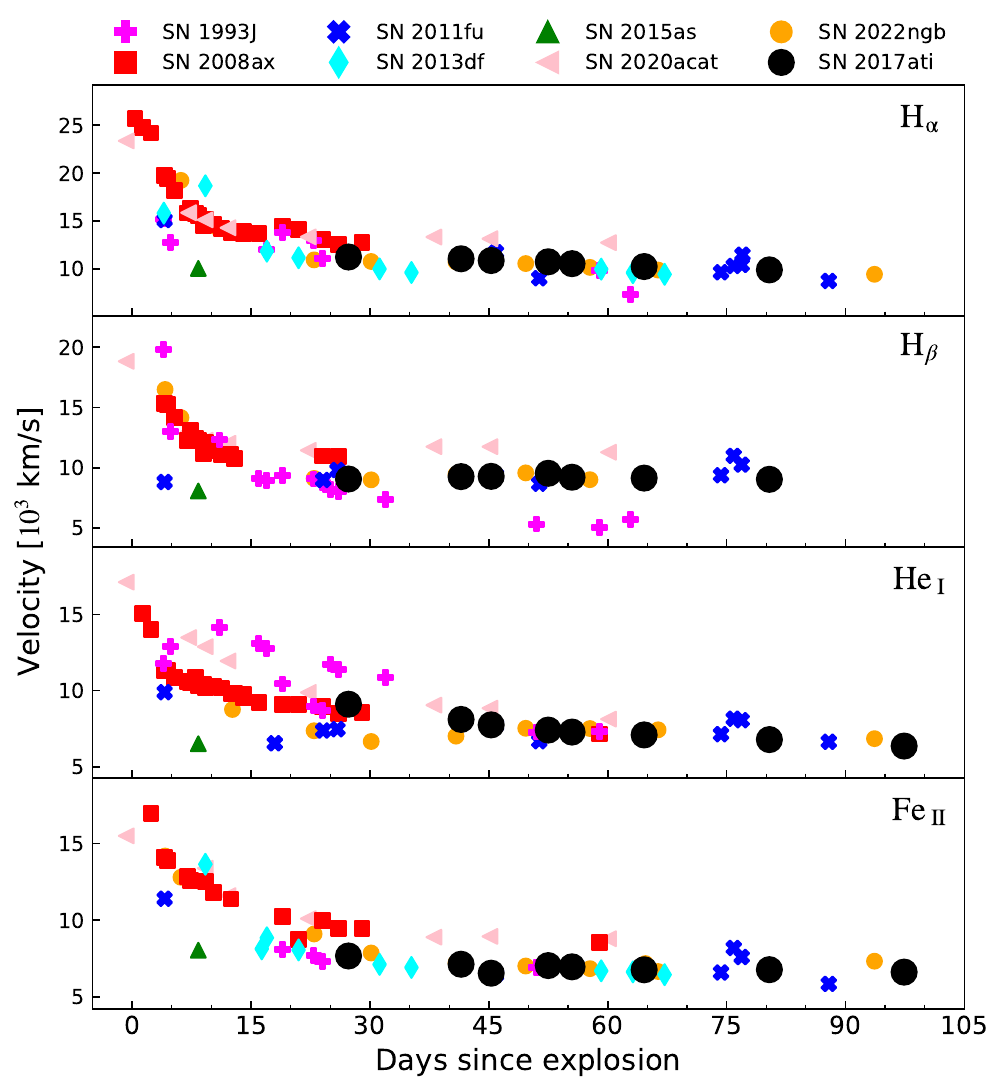}
\caption{Evolution of the line velocities for \Ha $\mathrm{\lambda}6563$ \AA, \Hb $\mathrm{\lambda}4861$ \AA, \Hei\ $\mathrm{\lambda}5876$ \AA, and \Feii\ $\mathrm{\lambda}5018$ \AA. SN~2017ati and other comparisons are marked with different  symbols. The measured values and uncertainties for SN 2017ati are listed in Table~\ref{tab:linev}.}
\label{spec_velocity}
\end{figure}

\begin{figure}
    \includegraphics[width=\columnwidth]{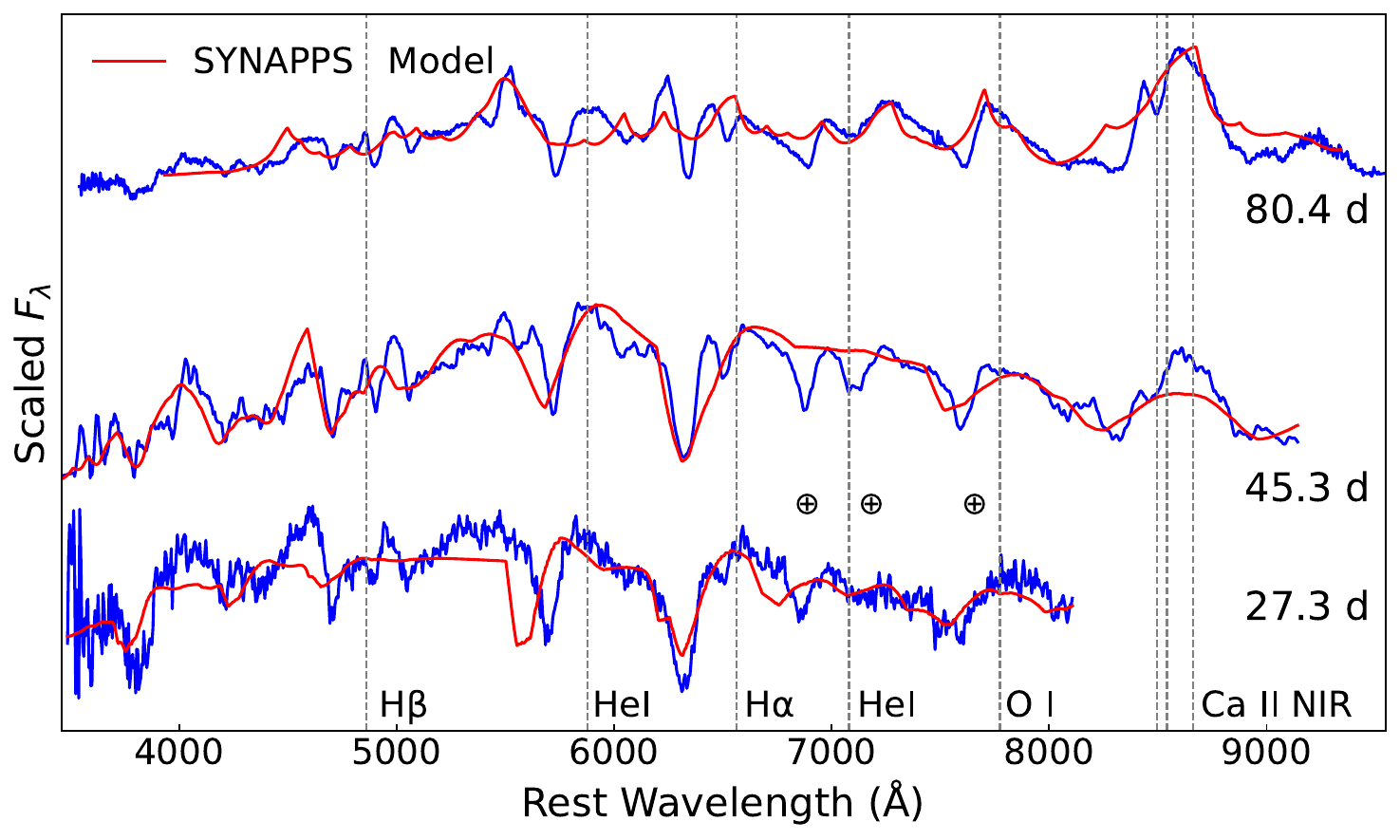}
    \caption{Comparison between observed spectra and SYNAPPS models, together with line identifications, for SN 2017ati at three epochs. The spectra are corrected for extinction and redshift.}
\label{fig:SYN}
\end{figure}

The evolution of line profiles for \Hei\,$\lambda\lambda5016,5876$, \Oi\,$\lambda7774$, [\Oi]~$\lambda6300$, [\Caii]~$\lambda\lambda7291,7324$, and the \Caii\ NIR triplet in spectra of SN\,2017ati is shown in Fig.~\ref{fig:line_profiles}. In the earliest spectrum of SN\,2017ati at $+27.3$~d, the \Hei and \Oi\ lines exhibit clear broad P Cygni profiles, indicating outwardly expanding ejecta and the presence of these elements in rapidly moving material.
With time, the blueshift of the \Hei\ absorption features decreases gradually, from about $-10, 000~\mathrm{km~s^{-1}}$ for \Hei $\lambda5876$ at early phases to roughly $-5,000~\mathrm{km~s^{-1}}$ at $+148.4$~d. 
This behaviour is naturally explained by the recession of the line-forming region from the fast, outer ejecta to progressively deeper layers with lower expansion velocities, as the ejecta expand and become increasingly optically thin \citep{Dessart2005AA}.
It is noteworthy that in the earliest five spectra, the \Hei\ $\lambda\lambda5016,5876$ and \Oi\ $\lambda7774$ lines exhibit a pronounced boxy morphology from 27.3\,d to 52.5\,d, as illustrated in the first three panels of Fig.~\ref{fig:line_profiles}.
This boxy appearance is not seen in the Balmer features, which instead maintain standard P Cygni profiles over the same epochs.

The boxy structure of the \Hei and \Oi lines suggests that their emission originates from a confined region of the ejecta, occupying a narrow velocity interval of approximately $-5,000~\mathrm{km\,s^{-1}}$ to $0~\mathrm{km\,s^{-1}}$. 
Such a configuration implies a stratified distribution in which the bulk of the intermediate-mass elements reside within a well-defined shell rather than being efficiently mixed into deeper layers or the fast outer ejecta.
The absence of a similar morphology in the H features indicates that hydrogen is present at different velocities without forming a comparably sharp shell, consistent with a partially stripped envelope. This stratified structure also provides a natural explanation for the later emergence of the double-peaked [\Oi]~$\lambda\lambda6300,6364$ emission, typically associated with an oxygen-rich, radially confined or asymmetric zone. 

The spectrum of SN~2017ati at $+64.6$~d already exhibits a weak [\Oi]~$\lambda\lambda6300,6364$ emission component, which subsequently increases in strength with time. In SNe~IIb, the [\Oi] emission becomes progressively more prominent toward late phases, as observed in well-studied events such as SNe~1993J, 2008ax, and 2011fu \citep{Richmond1994AJ....107.1022R, Pastorello2008MNRAS.389..955P, Tsvetkov2009PZ.....29....2T, Taubenberger2011MNRAS.413.2140T, Fang2022ApJ...928..151F}. 
This behaviour reflects the increasing contribution of the energy deposited in the oxygen-rich inner ejecta as the photosphere recedes, enabling forbidden-line emission to emerge under low-density, homologously expanding conditions. 
In the final spectrum of SN~2017ati at $+148.4$~d, the [\Oi]~$\lambda\lambda6300,6364$ feature develops a clear double-peaked structure with an additional red-side excess, which is discussed in detail in Sect.~\ref{Nebularsp-O}. 
During this period, the [\Oi]~$\lambda\lambda6300,6364$ exhibits a pronounced blueshift that gradually diminishes, declining from nearly $-5,000~\mathrm{km\,s^{-1}}$ at $+64.6$~d to approximately $-2,500~\mathrm{km\,s^{-1}}$ by $+148.8$~d.

As the spectra evolve, the [\Caii]~$\lambda\lambda7291,7324$ emission lines of SN~2017ati display an approximately symmetric profile and gradually increase in strength with time. By contrast, the near-infrared \Caii\  triplet is already prominent at early epochs and shows a pronounced blueshift. This contrasting behaviour can be understood in terms of the different formation conditions of permitted and forbidden calcium transitions. The \Caii\ near-infrared triplet originates predominantly in the dense, optically thick outer ejecta at early phases, where resonance scattering and line opacity favour blueshifted absorption and emission from the approaching hemisphere. As the ejecta expand and the density decreases, the line-forming region recedes to deeper layers and the forbidden [\Caii] transitions become increasingly efficient, giving rise to stronger and more symmetric emission profiles that trace the more globally distributed calcium-rich material.

To further quantify the kinematic evolution implied by these spectral changes, SN~2017ati is analysed through measurements of spectral line expansion velocities. 
Gaussian profile fitting is applied to the principal absorption minima in the spectra of SN~2017ati to derive the corresponding velocities.
The analysis concentrates on \Ha, \Hb, \Hei\,$\lambda5876$, and \Feii\,$\lambda5018$. The resulting velocity evolution is compared with that of a representative sample of SNe IIb. The temporal behaviour of these line velocities for SN~2017ati and the comparison objects is shown in Fig.~\ref{spec_velocity}, while the measured values and associated uncertainties for SN~2017ati are listed in Table~\ref{tab:linev}.

The velocity evolution of SN~2017ati broadly follows the trends observed in the majority of Type~IIb SNe events included in the comparison sample.
Owing to the absence of early-time spectra, the initial line velocities of SN~2017ati cannot be constrained. 
At intermediate and late phases, the H-line velocity of SN~2017ati occupies an intermediate range among Type~IIb events, with values similar to those measured in SNe~2011fu, 2013df, and 2022ngb, lower than those observed in SN~2020acat, and higher than those inferred for SN~1993J, as illustrated in the two upper panels of Fig.~\ref{spec_velocity}.
For SN~2017ati, the \Ha\ velocity shows a gradual decline from $\sim11{,}220~\mathrm{km\,s^{-1}}$ at $+27.3$~d after the explosion to $\sim9{,}870~\mathrm{km\,s^{-1}}$ at $+80.4$~d. 
In contrast, the \Hb\ velocity in SN~2017ati increases from $\sim9{,}070~\mathrm{km\,s^{-1}}$ at $+27.3$~d to a peak of $\sim9{,}540~\mathrm{km\,s^{-1}}$ at $+52.5$~d, before declining to $\sim9{,}040~\mathrm{km\,s^{-1}}$ by $+80.4$~d. The evolution of H$\beta$ is likely similar to that of H$\alpha$, although H$\beta$ lies in a more crowded spectral region, which results in differences in their measured velocities.
At later epochs, the H features in SN~2017ati fade from the spectra, preventing further reliable velocity measurements.

A comparable decline is seen in the \Hei\ and \Feii\ features of SN~2017ati, with their velocities broadly consistent with those measured in the representative Type~IIb SNe.
The \Hei\,$\lambda5876$ velocity decreases from $\sim9{,}110~\mathrm{km\,s^{-1}}$ at $+27.3$~d to $\sim6{,}360~\mathrm{km\,s^{-1}}$ by $+97.4$~d, while the \Feii\,$\lambda5018$ line exhibits a more gradual evolution, declining from $\sim7{,}680~\mathrm{km\,s^{-1}}$ at $+27.3$~d to $\sim6{,}630~\mathrm{km\,s^{-1}}$ at $+97.4$~d. The velocities inferred from \Hei\ and \Feii\ lines are systematically lower and exhibit a smoother temporal evolution than those of the Balmer features (see the bottom two panels of Fig.~\ref{spec_velocity}), consistent with their formation in deeper ejecta layers as the line-forming region recedes. This behaviour has been previously reported in SE-SNe \citep[e.g.][]{Matheson2000AJ....120.1499M, Branch2002ApJ...566.1005B}.

Because hydrogen and helium lines do not reliably trace the photospheric velocity, the \Feii\,$\lambda5018$ line is commonly adopted as a photospheric proxy, as it forms in deeper layers of the ejecta \citep{Dessart2005AA}.
Following this approach, the photospheric velocity of SN~2017ati is estimated from the \Feii\,$\lambda5018$ absorption measured at $-0.2$~d relative to the \textit{o}-band maximum (corresponding to $+27.3$~d after the explosion), yielding a value of $\sim7{,}680~\mathrm{km\,s^{-1}}$. This estimate is in good agreement with the ejecta velocity inferred in Sect.~\ref{sec:MOSFiT} using \texttt{MOSFiT} within the radioactive-decay plus magnetar central-engine framework, which gives $v_{\rm ej}=7.24^{+0.70}_{-0.64}\times10^{3}~\mathrm{km\,s^{-1}}$. The consistency between these independent estimates supports the physical credibility of the parameters derived from the model.

To identify the ions contributing to the photospheric spectra of SN~2017ati, synthetic spectra were computed with \textsc{SYNAPPS}\footnote{\url{https://github.com/rcthomas/es/}} \citep{SYN, Thomas2013ascl.soft08008T}. 
For SN~2017ati, the synthetic calculations were constructed to reproduce the observed spectra at epochs of $+27.3$, $+45.3$, and $+80.4$~d after the explosion.
The synthetic spectra include contributions from a selected set of atomic species, including \ion{H}{i}, \Hei, \Oii, \ion{Ti}{ii}, \ion{Sc}{ii}, \ion{Ca}{ii}, \ion{Fe}{ii}, and \ion{Ba}{ii}, which collectively reproduce the dominant absorption and emission features observed in the spectra. 
The corresponding SYNAPPS models illustrating the line identifications at three representative epochs for SN~2017ati are shown in Fig.~\ref{fig:SYN}. Although the SYNAPPS model does not reproduce the observed line profiles perfectly, particularly at wavelengths shorter than $5500$~\AA, it robustly identifies the dominant contributing species, including \ha, \Hei, \Oi, and \Caii, while the depression near $7000$~\AA\ can be attributed to residual telluric absorption.

\subsection{Comparison of Type IIb SNe spectra}
\label{Spectra}

\begin{figure}
    \includegraphics[width=\columnwidth]{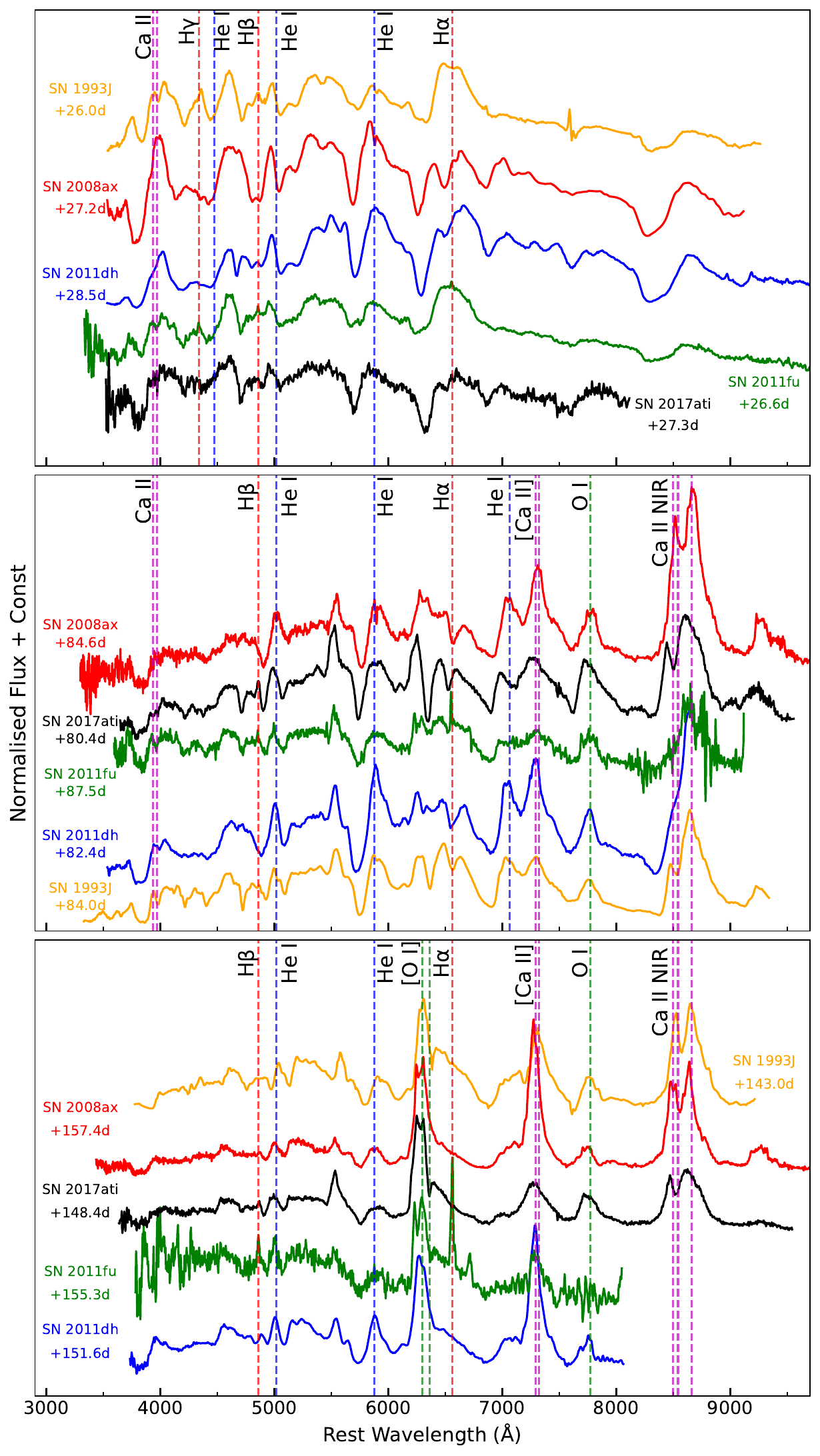}
    \caption{Comparisons of the spectra of SN~2017ati with other Type IIb events. Key spectral lines are highlighted with corresponding labels.}
    \label{fig:sp_com}
\end{figure}

To place SN~2017ati in the context of typical SNe IIb, including SNe~1993J, 2008ax, 2011dh, and 2011fu, its spectra are compared with those of well-studied objects observed at similar phases\footnote{Spectra of the comparison objects were obtained from the WISeRep database \citep[\url{https://www.wiserep.org/},][] {Yaron2012PASP..124..668Y, Goldwasser2022TNSAN.191....1G}.} (see Fig.~\ref{fig:sp_com}). 
Prominent spectral features are identified and labelled in each panel. For consistency and to facilitate a clearer comparison of line profiles and relative strengths, all spectra have been normalized. Each spectrum was corrected for both redshift and line-of-sight extinction. 

The spectrum of SN~2017ati at $27.3$~d after the explosion is compared with those of other SNe IIb observed at a similar epoch, as shown in the upper panel of Fig.~\ref{fig:sp_com}. 
Although P Cygni profiles of \ha\ are present in all spectra at this phase, their overall strengths and velocity extents vary among different SNe IIb. 
The central absorption feature embedded within the \ha\ emission profile of SN~2017ati is clearly detected and closely resembles that seen in SNe 1993J, 2008ax, and 2011dh. This component is likely associated with the P Cygni absorption of the \Hei\,$\lambda6678$ line, while a comparable feature is not evident in the spectra of SN 2011fu. The \Hei\ P Cygni features in SN~2017ati and the other SNe~IIb in the comparison sample show clear blueshifts. At this epoch, the overall spectral morphology of SN~2017ati shows a particularly close resemblance to that of SNe~2008ax and 2011dh.

The spectrum of SN~2017ati at $+80.4$~d, shown in the middle panel of Fig.~\ref{fig:sp_com}, is compared with a subset of SNe~IIb observed at similar phases. 
At this epoch, SN~2017ati shows a clearly detected, single-peaked [\Oi]~$\lambda\lambda6300,6364$ profile together with a still-prominent \Ha\ P Cygni feature. 
SN~2017ati exhibits significantly enhanced [\Caii]~$\lambda\lambda7291,7324$ emission along with the presence of additional metal lines, including \Mgi]~$\lambda4571$, \Feii~$\lambda\lambda4924,5018,5169$, and \Oi~$\lambda\lambda7772,7774$, indicating a progressive transition towards a nebular-dominated spectral regime. 
Overall, the spectrum of SN~2017ati at this phase closely resembles that of SN~1993J and broadly matches the general properties of the other comparison SNe~IIb.

The nebular-phase spectrum of SN~2017ati at  $+148.4$~d post-explosion, shown in the lower panel of Fig.~\ref{fig:sp_com}, is compared with a selection of Type~IIb SNe, highlighting the late-time spectral evolution during the nebular phase.
At this stage, the spectra of SN~2017ati are dominated by prominent nebular emission features, including [\Oi]~$\lambda\lambda6300,6364$, [\Caii]~$\lambda\lambda7291,7324$, and the \Caii\ NIR triplet at $\lambda\lambda8498,8542,8662$, all of which are well developed, while H$\alpha$ has largely faded.
The spectra of SN~2017ati at this phase show a strong similarity to those of SNe~1993J and 2008ax, with the [\Oi]~$\lambda\lambda6300,6364$ feature exhibiting a double-peaked profile, reminiscent of SNe~2008ax and 2011fu.
A detailed discussion of the nebular-phase spectra is provided in Sect.~\ref{Nebularsp}.

\subsection{Nebular spectra} 
\label{Nebularsp}

Nebular-phase spectra of CCSNe are commonly employed to infer progenitor  and  explosion properties. These spectra, obtained once the ejecta become optically thin, provide direct insights into the composition, distribution, and kinematics of the inner layers of the SN. In particular, analysis of nebular emission lines, such as [\Oi] and [\Caii], allows for the estimation of oxygen mass, assessment of line-profile asymmetries, and comparison with theoretical explosion models. Together, these diagnostics can be used to estimate the zero-age main-sequence (ZAMS) mass of the progenitor and to explore details of the explosion geometry and the nucleosynthesis \citep{Fang2024NatAs...8..111F}.

\subsubsection{Nebular [\Oi] line profile} 
\label{Nebularsp-O}

\begin{figure}
    \includegraphics[width=\columnwidth]{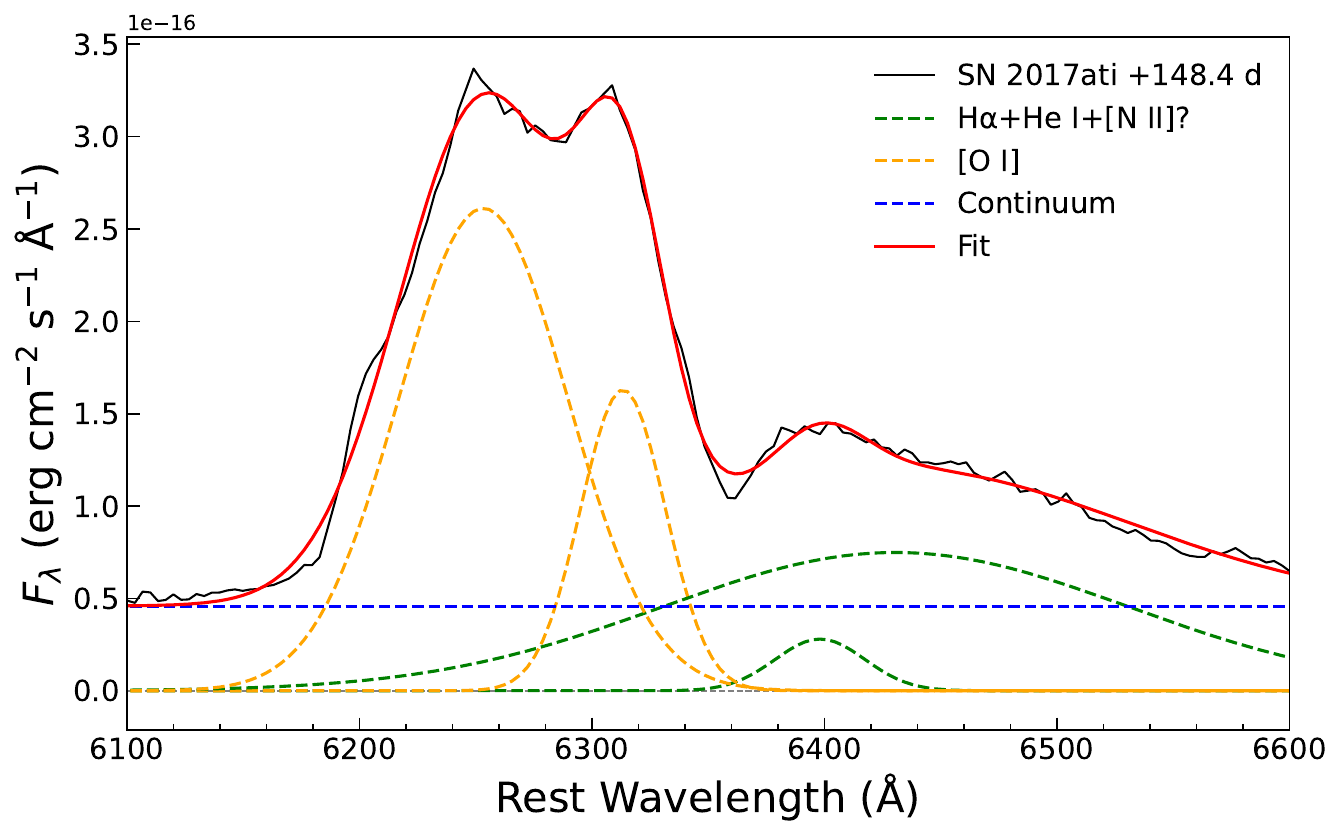}
    \caption{Multi component Gaussian fit to the [\Oi]~$\lambda\lambda6300,6364$ doublet in the nebular-phase spectrum of SN~2017ati at $+148.4$~d ($6100-6600$~\AA).}
    \label{fig:NeO}
\end{figure}

During the nebular phase, SN~2017ati exhibits a nearly symmetric double-peaked profile in [\Oi]~$\lambda\lambda6300,6364$, distinguishing it from SNe~1993J and 2011dh. Double-peaked [\Oi] emission lines are relatively common in late-time spectra of SE-SNe, as seen for example in SNe~2008ax and 2011fu, and are often interpreted as arising from oxygen distributed in a ring-like configuration, with the line of sight approximately aligned with the plane of the ring \citep{Mazzali2005Sci...308.1284M, Mazzali2008Sci...321.1185M, Tanaka2009ApJ...700.1680T, Maurer2010MNRAS.402..161M}. An alternative explanation attributes the apparent double peaks to the intrinsic [\Oi]~$\lambda\lambda6300,6364$ doublet, whose line ratio evolves from unity at early times to 3:1 in the optically thin regime at late epochs \citep{Li1992ApJ...387..309L, Taubenberger2009MNRAS.397..677T, Milisavljevic2010ApJ...709.1343M}. Based on the subsequent analysis of the nebular spectrum of SN~2017ati, this interpretation is favoured.

We performed a continuum-subtracted spectral fit of SN~2017ati at $+148.4$~d over $6100-6600$~\AA, modelling the [\Oi] doublet with two Gaussians and the small red-side bump with two additional Gaussians (see Fig.~\ref{fig:NeO}). In this spectrum, SN~2017ati exhibits a distinct red-side excess similar to that seen in SN~1993J, which cannot be reproduced by the [\Oi] doublet alone, indicating additional line contributions. 
Old nebular-phase modelling showed that non-thermal excitation by radioactive decay products can sustain H$\alpha$ emission at late epochs \citep{Kozma1998ApJ...497..431K, Houck1996ApJ...456..811H}. Modern nebular spectral synthesis models for SE-SNe further suggest that the [\Oi] region can be contaminated by H$\alpha$, \Hei, and [\Nii]~$\lambda\lambda6548,6583$, particularly in Type~IIb events where residual hydrogen is present \citep{Jerkstrand2015MNRAS.448.2482J, Dessart2021A&A...656A..61D}. Comparative studies indicate that such red-side excess emission is common among SE-SNe and does not uniquely trace large-scale ejecta asymmetries, but is consistent with compositional mixing between hydrogen-rich and metal-rich layers \citep{Taubenberger2009MNRAS.397..677T}.

The apparent symmetry and the significant blueshift observed ($\sim 2240$ km/s) in the [\Oi] doublet of SN~2017ati can largely be attributed to line blending and non-thermal excitation effects, similar to those identified in SN~1993J. 
Nevertheless, a non-spherical explosion leading to an asymmetric oxygen distribution cannot be entirely excluded, analogous to the ejecta geometries inferred for SN~2008ax \citep{Mazzali2005Sci...308.1284M, Maeda2007ApJ...666.1069M, Valenti2011MNRAS.416.3138V, Kumar2022ApJ...927...61K}. 
The fitted [\Oi] line ratio close to 3:1 in the nebular spectra of SN~2017ati indicates optically thin conditions for the emission, but this constraint alone does not provide a direct information on the ejecta geometry. 
Overall, the [\Oi] line profile of SN~2017ati likely reflects a combination of line blending, non-thermal excitation, residual contributions from hydrogen and metal-rich material, and a possible underlying asymmetry in the ejecta, encompassing phenomena similar to those observed in SNe~1993J and 2008ax.

\subsubsection{Oxygen mass}
\label{Sect:Omass}

The intensity of the [\Oi]~$\lambda\lambda6300,6364$ feature provides an indication of the oxygen mass synthesised in the stellar core, which is closely linked to the progenitor ZAMS mass \citep{Woosley1995ApJS..101..181W,Thielemann1996ApJ...460..408T}. 
A quantitative determination of the O mass additionally requires the knowledge of the ejecta temperature and the optical depth of the relevant transitions. 
In this work, the O mass of SN~2017ati is estimated following the method of \citet{Jerkstrand2014MNRAS.439.3694J}, where the \Oi\ temperature is inferred from the line ratio of [\Oi]~$\lambda5577$ to [\Oi]~$\lambda\lambda6300,6364$, as both lines are collisionally excited and their ratio is sensitive to the temperature.

Measuring the flux of the [\Oi]~$\lambda\lambda6300,6364$ doublet is non-trivial because the red wing is affected by strong and broad emission, likely produced by a blend of H$\alpha$, He~\textsc{i}, and [\Nii]. As shown in Fig.~\ref{fig:NeO}, the observed profile is modelled with four Gaussian components: two account for the [\Oi]~$\lambda6300$ and $\lambda6364$ transitions, while an additional narrow component and a broad component are introduced to reproduce the blended excess emission (see Sect.~\ref{Nebularsp-O}). The luminosity of [\Oi]  is computed through spectral integration, yielding a ratio of $L_{5577}/L_{6300,6364}=0.33$, with $L_{6300,6364}=2.7 \times 10^{40}\,\mathrm{erg}\,\mathrm{s}^{-1}$.
From the same fitting procedure, the flux ratio of [\Oi]~$\lambda6300$ to [\Oi]~$\lambda6364$ is measured to be $3.2$, indicating that the ejecta are already in an optically thin regime at this epoch. Following \citet{Jerkstrand2014MNRAS.439.3694J}, and adopting escape probabilities from \citet{Sobolev1957SvA.....1..678S} and \citet{Fransson1989ApJ...343..323F}, we take $\beta_{6300,6364}\approx0.5$ and $\beta_{5577/6300,6364}\approx1-2$. Using the relation
\begin{equation}
\frac{L_{5577}}{L_{6300,6364}} =
38  \,\times\,
\exp\!\left(\frac{-25790\,\mathrm{K}}{T}\right) \beta_{5577/6300,6364},
\label{eq:OI_temp}
\end{equation}
the oxygen temperature is estimated to lie in the range
$T \approx 4740-5430$~K. The oxygen mass is then derived from
\begin{equation}
M_{\mathrm{O}} =
\frac{L_{6300,6364} \,/\, \beta_{6300,6364}}
     {9.7 \times 10^{41}\,\mathrm{erg}\,\mathrm{s}^{-1}}
\,\times\,
\exp\!\left(\frac{22720\,\mathrm{K}}{T}\right),
\label{eq:OI_mass}
\end{equation}
resulting in an oxygen mass of
$M_{\mathrm{O}} \approx 1.82-3.34\,\msun$.
The inferred oxygen mass exceeds typical values reported for normal SNe~II and IIb, including SNe~1993J, 2004et, 2011dh,  and 2012aw \citep{Jerkstrand2012A&A...546A..28J, Jerkstrand2014MNRAS.439.3694J, Jerkstrand2015A&A...573A..12J}, while being more comparable to those derived for SNe~2015bs and 2018gk \citep{Anderson2018NatAs...2..574A,  Bose2021MNRAS.503.3472B}. 

Stellar evolution models predict a monotonic increase in the oxygen yield with the ZAMS progenitor mass \citep{Nomoto1997NuPhA.616...79N,Rauscher2002ApJ...576..323R,Limongi2003ApJ...592..404L,Sukhbold2016ApJ...821...38S, van2025Galax..13...72V}. Within this framework, the measured oxygen mass implies a progenitor initial mass of $M_{\mathrm{ZAMS}}\approx19-26\,\msun$.
However, the star rotation can have a significant impact on the oxygen yield \citep{Chieffi2013ApJ...764...21C}.
For non-rotating progenitors, oxygen masses of $0.50$, $1.26$, $2.35$, and $3.80\,\msun$correspond to initial masses of 15, 20, 25, and 30\,$\msun$, respectively. For rapidly rotating progenitors ($v_{rot}=300$~km\,s$^{-1}$), the yields are substantially higher, reaching $1.48$, $2.72$, $3.79$, and $5.75\,\msun$ for the same masses. Consequently, a rotating progenitor can produce $M_{\mathrm{O}} \approx 1.82\,\msun$from an initial mass of only $\sim17\,\msun$, which aligns with the lower-limit progenitor mass estimates discussed later in this paper.
Since the oxygen is predominantly synthesised during hydrostatic burning, variations in the explosion physics are expected to have only a limited impact on the relationship between the oxygen mass and the progenitor ZAMS mass.
The main sources of uncertainty in the inferred oxygen mass stem from the flux measurements of the [\Oi]~$\lambda5577$ and [\Oi]~$\lambda\lambda6300,6364$ lines.

\subsubsection{Modeling spectra in the nebular phase} 
\label{sec_subsubnebmdl}

\begin{figure}
    \centering
    \includegraphics[width=\columnwidth]{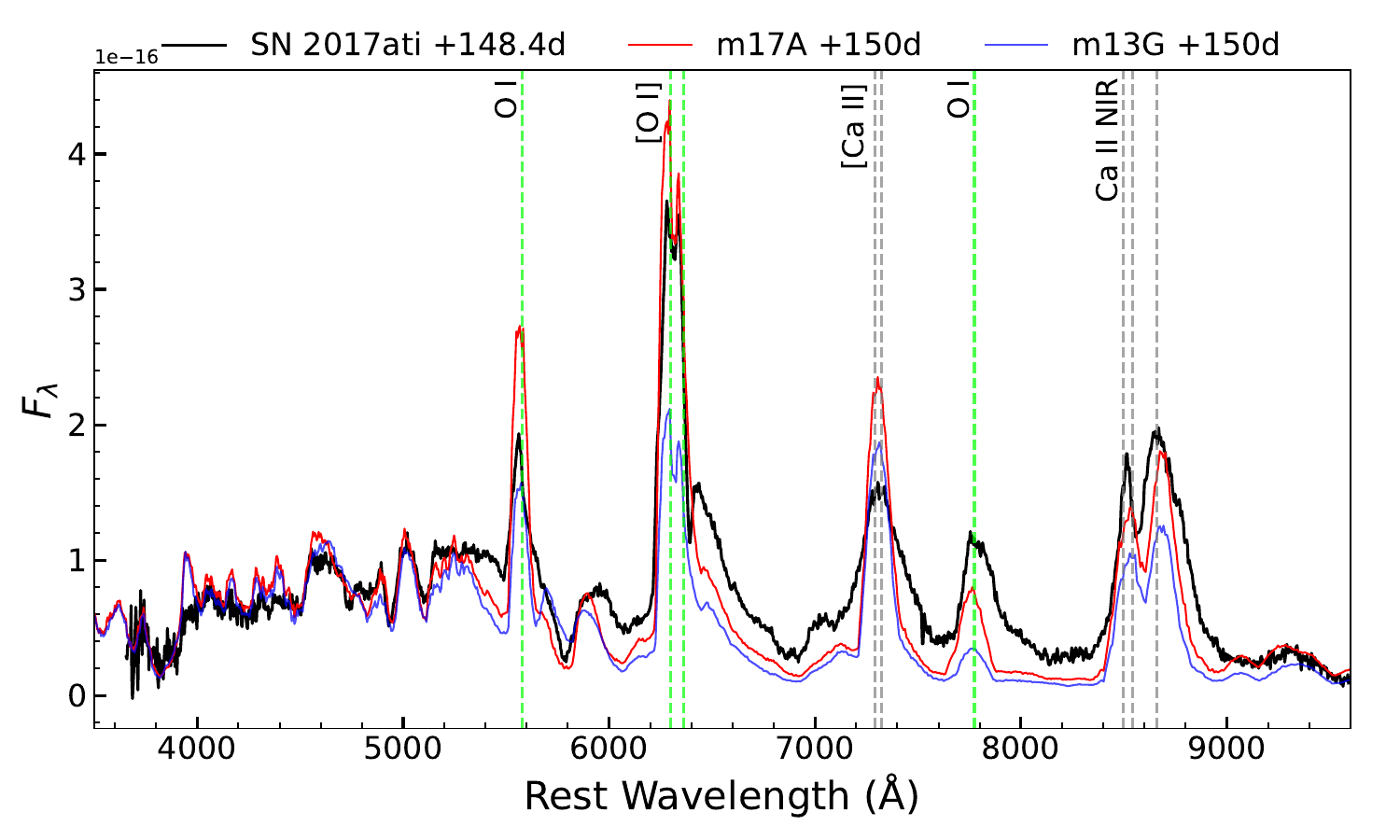}
    \vspace{0.3cm}
    \includegraphics[width=\columnwidth]{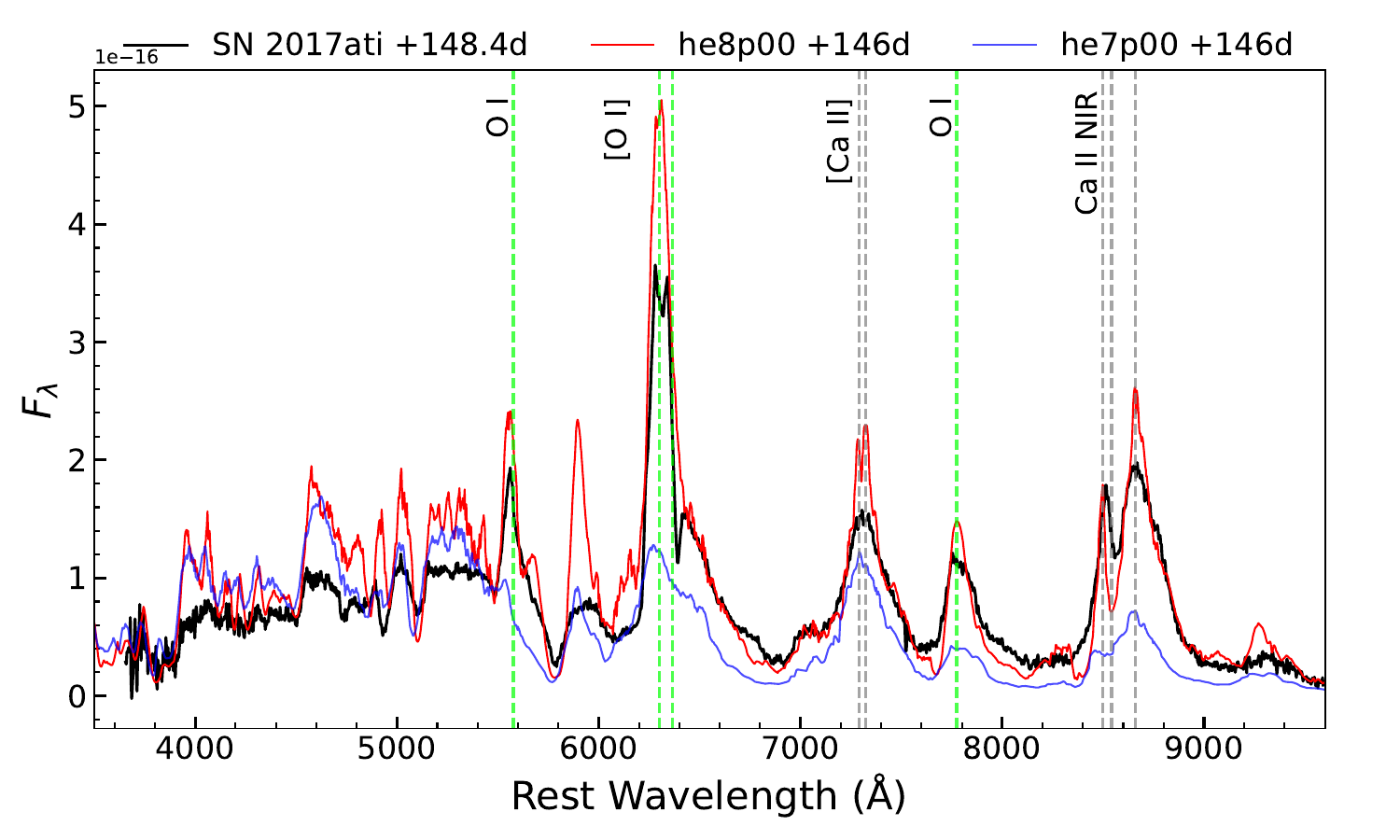}
    \caption{
    Nebular-phase spectrum of SN~2017ati at $+148.4$~d post-explosion (black line), compared with synthetic spectra from different progenitor models.
    \emph{Upper panel}: Comparison with Type~IIb SN models from \citet{Jerkstrand2015A&A...573A..12J}. The blue line represents a model with a progenitor $M_{\mathrm{ZAMS}} = 13\,\mathrm{M_\odot}$, while the red line represents a model with $M_{\mathrm{ZAMS}} = 17\,\mathrm{M_\odot}$.
    \emph{Lower panel}: Comparison with the \texttt{he7p00} and \texttt{he8p00} models presented by \citet{Dessart2023A&A...677A...7D}. The blue line denotes the \texttt{he7p00} model with $M_{\mathrm{preSN}} = 5.04\,\mathrm{M_\odot}$, while the red line corresponds to the \texttt{he8p00} model with $M_{\mathrm{preSN}} = 5.63\,\mathrm{M_\odot}$.
    In both panels, the model spectra are scaled to the same distance and the same $^{56}\mathrm{Ni}$ mass as SN~2017ati.    }
    \label{spec_cmp_neb_models}
\end{figure}

\cite{Fransson1989ApJ...343..323F} proposed that the flux ratio between the [\Caii]\ $\lambda\lambda7291,7323$ and [\Oi]\ $\lambda\lambda6300,6364$ doublets can be used as a semi-quantitative indicator of the ZAMS mass \citep{Fang2018ApJ...864...47F, Fang2019NatAs...3..434F}. This diagnostic is motivated by the different nucleosynthetic origins of the two elements: oxygen is predominantly produced during hydrostatic burning stages, whereas calcium is mainly synthesised during explosive nucleosynthesis. As a consequence, the calcium yield shows only a weak dependence on the prior stellar evolution. A larger [\Caii]/[\Oi] ratio therefore points to a lower initial stellar mass. Furthermore, \cite{Elmhamdi2004A&A} showed that this ratio approaches a stable value at late phases, typically beyond $\sim150$~d after the explosion, and remains nearly constant thereafter. 

For SN~2017ati, the [\Caii]/[\Oi] flux ratio measured at $+148.4$~d post-explosion, after the subtraction of the underlying continuum, is approximately $0.50$. As discussed in Sect.~\ref{Nebularsp-O}, this value may be influenced by residual emission from H$\alpha$, \Hei, and [\Nii]~$\lambda\lambda6548,6583$, and should therefore be considered as a lower limit. This ratio is comparable to those reported for SN~1993J and SN~2020acat, which are also near $0.5$, implying a progenitor ZAMS mass of roughly $14-18~\msun$. However, caution is warranted in interpreting this estimate, as the assumptions underlying the [\Caii]/[\Oi] diagnostic may not be universally applicable to all SNe~IIb. 
For instance, SN~2008ax exhibits a high [\Caii]/[\Oi] ratio despite a relatively massive progenitor \citep{Taubenberger2011MNRAS.413.2140T}, which is noteworthy given the close similarity in spectral evolution between SN~2017ati and SN~2008ax (see Fig.~\ref{fig:sp_com}).

The nebular-phase spectrum of SN~2017ati obtained at $+148.4$~d was compared with synthetic spectra calculated using the \texttt{SUMO} radiative-transfer code \citep{Jerkstrand2015A&A...573A..12J}, as shown in the upper panel of Fig.~\ref{spec_cmp_neb_models}. The models, computed for an epoch of 150~d, are based on the stellar evolution calculations of \cite{Woosley2007PhR}, considering progenitors with ZAMS mass of $13\,\msun$ (model 13G, originally computed for SN~2008ax and selected here for comparison due to its close spectral similarity with SNe~2017ati) and $17\,\msun$ (model 17A).
For reference, the initial model parameters are summarized as follows: for model 13G, the total ejecta mass is $2.1\,\msun$, the hydrogen envelope mass is $0.054\,\msun$, the helium envelope mass is $1.1\,\msun$, and the initial $^{56}$Ni mass is $0.075\,\msun$; for model 17A, the corresponding values are $3.5\,\msun$, $0.054\,\msun$, $1.2\,\msun$, and $0.075\,\msun$, respectively. To ensure a consistent comparison, all model spectra were scaled to the same distance and flux scale as SN~2017ati, adopting a $^{56}\mathrm{Ni}$ mass of $0.21\,\msun$ as inferred in Sect.~\ref{sec:MOSFiT} with the radioactive-decay plus magnetar energy-input model.
Comparison with nebular spectral models, specifically the 17A model, shows that the [\Oi]~$\lambda\lambda6300,6364$ emission of SN~2017ati is well reproduced, indicating an oxygen mass consistent with a progenitor ZAMS mass of approximately $17\,\msun$. The relative weakness of the [\Oi]\,$\lambda5577$ line can be explained by a comparatively low nebular temperature or by additional cooling via molecular formation, and does not necessitate a revision of the inferred progenitor mass.

By contrast, the relatively weak [\Caii]~$\lambda\lambda7291,7324$ emission, together with the prominent \Caii\ NIR triplet, reflects physical conditions in the Ca-emitting region rather than a deficiency in calcium mass. Radiative-transfer calculations indicate that the strengths of calcium lines are sensitive to the local density structure, mixing, and the spatial distribution of the radioactive power deposition \citep{Dessart2020A&A...642A..33D}. Elevated densities or clumping can suppress the forbidden [\Caii] transitions while enhancing the permitted \Caii\ emission. Statistical analyses of nebular spectra presented by \citet{Prentice2022MNRAS.514.5686P} reveal systematic trends in the [\Caii]/[\Oi] ratio and in the strength of the \Caii\ NIR emission across different progenitor environments, in agreement with this interpretation \citep{Ergon2022A&A...666A.104E}.
As discussed in Sect.~\ref{Nebularsp-O}, the measured fluxes may also be influenced by residual emission from H$\alpha$, \Hei, and [\Nii]~$\lambda\lambda6548,6583$, implying that the [\Oi]~$\lambda\lambda6300,6364$ flux could be moderately overestimated.

For Type~Ib/Ic SNe arising from completely stripped helium stars, intense stellar winds act to substantially decrease the final pre-SN mass. By contrast, in SNe~IIb the survival of a low-mass hydrogen envelope suggests that mass removal from the helium core was relatively modest. This scenario is in line with the conclusions of \citet{Dessart2023A&A...677A...7D}, who showed that nebular spectra are primarily governed by the pre-SN mass, with only a weak dependence on the initial stellar mass. We then compare the nebular-phase spectrum of SN~2017ati at $+148.4$~d with the \texttt{he7p00} and \texttt{he8p00} models presented by \citet{Dessart2023A&A...677A...7D}, as shown in the lower panel of Fig.~\ref{spec_cmp_neb_models}. 
The model spectra adopted for this comparison are non-local thermodynamic equilibrium radiative-transfer simulations computed with \texttt{CMFGEN} \citep{Hillier2012MNRAS.424..252H}, and both models assume a distance of 1~kpc. 
For the \texttt{he7p00} model, the adopted parameters are a kinetic energy of $E_{\mathrm{kin}} = 1.38 \times 10^{51}$~erg, an ejecta mass of $M_{\mathrm{ej}} = 3.33\,M_{\odot}$, a $^{56}Ni$ mass of $M_{\mathrm{Ni}} = 0.102\,M_{\odot}$, a pre-explosion mass of $M_{\mathrm{preSN}} = 5.04\,M_{\odot}$, and  $M_{\mathrm{ZAMS}} = 25.68\,M_{\odot}$. 
The \texttt{he8p00} model instead adopts $E_{\mathrm{kin}} = 0.71 \times 10^{51}$~erg, $M_{\mathrm{ej}} = 3.95\,M_{\odot}$, $M_{\mathrm{Ni}} = 0.0546\,M_{\odot}$, $M_{\mathrm{preSN}} = 5.63\,M_{\odot}$, and $M_{\mathrm{ZAMS}} = 27.91\,M_{\odot}$.
The synthetic spectra from the \texttt{he7p00} and \texttt{he8p00} models both show a high degree of similarity to the observed nebular spectrum of SN~2017ati. The most pronounced differences appear in the strengths of the \Oi\,$\lambda\lambda6300,6364$ and \Caii\,$\lambda\lambda7291,7324$ emission features, which are stronger than those predicted by the \texttt{he7p00} model. 
In this respect, the \texttt{he8p00} model provides a closer overall match, reproducing both the relative line strengths and the general spectral morphology more accurately. 
This level of agreement indicates a progenitor with a relatively large helium-core mass, consistent with the oxygen mass inferred from the nebular diagnostics and supporting a higher-mass progenitor scenario for SN~2017ati. 
On this basis, SN~2017ati is more plausibly associated with a pre-SN mass in the range $M_{\mathrm{preSN}} = 5.04-5.63\,M_{\odot}$. Combining the evidence from the [\Caii]/[\Oi] flux ratio, the [\Oi] emission strength, and the comparison with both \texttt{SUMO} and \texttt{CMFGEN} nebular spectral models, we conclude that the progenitor of SN~2017ati likely had a ZAMS mass of $M_{\mathrm{ZAMS}} \geq 17\,\msun$.

\section{Discussion}
\label{section:sum}

\subsection{The nature of light curve evolution}
\label{secti:LCsum}
A small group of SNe~IIb displaying unusual photometric behaviour has been identified, including SNe~2017ckj, 2018gk, and 2017ati. These objects reach peak luminosities $\approx1-2$~mag brighter in the $r$ band than those of typical SNe~IIb (Fig.~\ref{fig:Absolute_magnitude}).
Their light-curve evolution also differs from that of typical SNe~IIb, showing behaviour that is not well matched by either events with a prominent early shock-cooling tail or those without such a feature.
If SN~2017ati is assumed to follow a standard SN~IIb luminosity evolution powered purely by $^{56}$Ni decay, the modelling in Sect.~\ref{sec:MOSFiT} using \texttt{MOSFiT} yields a $^{56}$Ni mass of $\sim0.37\,\msun$, while the poor fit to the early-time light curves further indicates that the luminosity is not solely powered by $^{56}$Ni.
This value is substantially higher than that commonly inferred for normal SNe~IIb, for which a mean $^{56}$Ni mass of $0.066\pm0.006\,\msun$ has been reported in \cite{Osmar2023ApJ...955...71R}.
Such a high luminosity suggests an additional energy source, such as ejecta-CSM interaction or magnetar spin-down \citep{Bose2021MNRAS.503.3472B}. 

\citet{Wang2017ApJ...837..128W} report that magnetar input dominates the light curve mainly around the maximum light ($t_{\rm exp}<50$~d) and again at very late epochs ($t_{\rm exp}>300$~d), whereas the decline after peak ($50$~d < $t_{\rm exp}$ < $300$~d) is largely controlled by the radioactive $^{56}$Ni decay.
This behaviour closely resembles the light curves for SN~2017ati, which show a relatively high peak luminosity and a rapid decline during the first $\sim50$~days after the explosion. 
Beyond $\sim50$~days, the light curves of SN~2017ati exhibit a markedly slower fading, with a decline rate in the $Bg$ bands close to that expected from $^{56}$Co radioactive decay ($\sim0.98$~mag~$100\,\mathrm{d}^{-1}$, see Tab.~\ref{tab:decline_rate}). 
At very late epochs, however, SN~2017ati lacks sufficient observational coverage, preventing a direct comparison in this phase.

While the magnetar-powered model reduces the inferred nickel mass of SN~2017ati to $M_{\mathrm{^{56}Ni}} = 0.21^{+0.08}_{-0.12}\,\mathrm{M_{\odot}}$, it remains high relative to the typical range of $0.03-0.28\,\mathrm{M_{\odot}}$ reported for SNe~IIb \citep{Anderson2019A&A...628A...7A, Meza2020A&A...641A.177M}. This result underscores the exceptional energetics of SN~2017ati compared with more canonical events such as SN~1993J and SN~2011dh, consistent with the relatively high peak luminosity. During the radioactive tail phase, SN~2017ati also remains $\sim1-2$~mag brighter than other SNe~IIb.

An alternative explanation for the light curve of SN~2017ati involves additional energy input from ejecta--CSM interaction. 
Such interaction is classically observed in SNe, which exhibit narrow spectral emission features during the first week after explosion \citep{Fraser2020RSOS....700467F}. 
Unfortunately, such early spectroscopic observations are unavailable for SN~2017ati; high-cadence, very early spectroscopy would have been required to break the degeneracy between these two models.
Based on the spectra obtained at later phases, a dominant contribution from ejecta--CSM interaction appears unlikely, since such interaction is expected to produce distinctive line-profile morphologies \citep{Luc2022} or a prominent broad H$\alpha$ emission similar to that observed in SN~1993J during phases clearly influenced by interaction. 
Nevertheless, SN~2017ati shows notable spectral similarities to SNe~1993J and 2008ax. 
In particular, the broad emission feature on the red side of [\Oi]~$\lambda\lambda6300,6364$ may plausibly be associated with H$\alpha$, as has been suggested for SN~1993J during a transitional phase towards interaction-dominated emission \citep{Matheson2000AJ....120.1499M}. However, the possibility raised by \citet{Jerkstrand2015A&A...573A..12J}, that this feature is instead produced by [\Nii]~$\lambda\lambda6548,6583$ rather than hydrogen, cannot be excluded.
By analogy, this makes possible that signatures of ejecta--CSM interaction could emerge at later epochs in SN~2017ati, even if such effects are not prominent in the currently available data.

\subsection{Constraints on the progenitor and the explosion}
\label{sec_prog}

We apply multiple independent methods to constrain the progenitor mass in Sect.~\ref{Nebularsp}. 
These methods include nebular-phase spectral diagnostics, based on the luminosities and profiles of oxygen emission lines, together with detailed comparisons to theoretical explosion models. 
Taken together, these constraints favour a progenitor with a relatively high initial mass ($M_{\mathrm{ZAMS}} \geq 17\,\msun$) and a compact pre-explosion configuration. 
This interpretation is further supported by the absence of clearly identifiable shock-cooling signatures in the early-time light curves, which argues against a progenitor possessing an extended hydrogen-rich envelope. 
Moreover, the results of \citet{Barmentloo2024MNRAS} indicate that helium stars with higher masses are generally associated with smaller stellar radii. 
While the available evidence points towards a compact progenitor retaining only a low-mass hydrogen-rich envelope, alternative configurations cannot be entirely ruled out given the remaining observational and modelling uncertainties.

Within this framework, different evolutionary channels may in principle give rise to progenitors with similar compact structures. 
As discussed by \citet{Crockett2008MNRAS.391L...5C}, these channels include a relatively massive single star that lost most of the hydrogen-rich envelope prior to explosion, as well as a lower-mass star that experienced substantial envelope stripping through binary interaction. 
For SN~2007ati, the nebular-phase [\Caii]/[\Oi]$\sim 0.5$ ratio provides an additional constraint that is sensitive to the progenitor mass (see Sect.~\ref{sec_subsubnebmdl}), with the observed value being more naturally reproduced by higher-mass progenitor models.
The nebular spectra of SN~2017ati show notable similarities to those of SNe~2011fu, 2015as, and 2020acat, characterised by strong [\Oi] emission and relatively weak \Caii emission. 
Given the pronounced spectral similarity of SN~2017ati to SNe~1993J and 2008ax, a progenitor that underwent substantial envelope stripping via binary interaction is favoured, though a massive single-star origin cannot be completely ruled out based on the available observations. 

\section{Conclusions} 
\label{section:conclusion}
SN~2017ati represents an unusual member of the Type~IIb SN population.
Compared to typical SNe~IIb, it exhibits a significantly higher peak luminosity, reaching $M_{\mathrm{r}} = -18.48 \pm 0.16\,\mathrm{mag}$, approximately $1-2$~mag brighter than the typical range. 
Despite this enhanced peak brightness, the post-peak decline rate beyond $\sim50$~d after the explosion is broadly consistent with that of ordinary SNe~IIb.
The main conclusions of this work are summarised as follows:

\begin{itemize}
\item Light-curve modelling of SN~2017ati using \texttt{MOSFiT} indicates that a pure $^{56}$Ni decay model ($M_{\mathrm{^{56}Ni}} = 0.37^{+0.07}_{-0.10}\,\mathrm{M_{\odot}}$), as well as a $^{56}$Ni decay model combined with CSM interaction, fails to adequately reproduce the observed photometric evolution of SN~2017ati.
The inclusion of an additional energy  contribution from a magnetar yields a notably improved fit, suggesting that such a process may contribute to the luminosity budget.

\item The observed light curve of SN~2017ati is better reproduced by a magnetar-powered model ($B = 13.2^{+4.6}_{-4.3} \times 10^{14}~\mathrm{G}$) implemented in \texttt{MOSFiT}, compared to alternative models.
In this context, the photometric evolution shows similarities to that of SN~1998bw around peak brightness, which have previously been proposed within a magnetar framework. 
However, the available data do not uniquely distinguish between magnetar input and alternative additional power sources.

\item The $^{56}$Ni mass, $M_{\mathrm{^{56}Ni}} = 0.21^{+0.08}_{-0.12}\,\mathrm{M_{\odot}}$, inferred from the magnetar-based modelling lies near the upper end of the range typically found for SNe~IIb, consistent with the relatively high peak luminosity and the elevated luminosity during the radioactive tail phase of SN~2017ati.

\item The optical spectra of SN~2017ati display striking similarities to those of SN~2008ax, with the primary difference being the strength of \Caii\ emission. 
This resemblance suggests a comparable explosion configuration and favours a progenitor that experienced substantial envelope stripping through binary interaction, although alternative evolutionary channels cannot be entirely excluded.

\item Nebular-phase spectral diagnostics using the luminosities and profiles of oxygen emission lines, combined with comparisons to theoretical explosion models, indicate a ZAMS mass of  $M_{\mathrm{ZAMS}} \geq 17\,\msun$ for the progenitor of SN~2017ati.
\end{itemize}

Upcoming observational facilities, including the Chinese Space Station Telescope \citep[CSST;][]{CSST2025} and the Vera C.\ Rubin Observatory \citep{Hambleton2023PASP..135j5002H}, will enable time-domain surveys of Type~II SNe with substantially improved cadence and sensitivity. 
Such datasets will enhance the detection and characterisation of rapidly evolving transients, place tighter constraints on theoretical modelling, and contribute to a deeper physical understanding of this complex subclass of CCSNe.

\section*{Data availability}
Optical photometric measurements of SN 2017ati are available at the CDS via \url{https://cdsarc.cds.unistra.fr/viz-bin/cat/J/A+A/xxx/xxx}.
Our observations of the specta are available via the Weizmann Interactive SN Data Repository (WISeREP; \citealt{Yaron2012PASP..124..668Y}).

\bibliographystyle{aa}
\bibliography{aa59331-26} 

\begin{thebibliography}{139}
\expandafter\ifx\csname natexlab\endcsname\relax\def\natexlab#1{#1}\fi

\bibitem[{{Anderson}(2019)}]{Anderson2019A&A...628A...7A}
{Anderson}, J.~P. 2019, \aap, 628, A7

\bibitem[{{Anderson} {et~al.}(2018){Anderson}, {Dessart}, {Guti{\'e}rrez},
  {Kr{\"u}hler}, {Galbany}, {Jerkstrand}, {Smartt}, {Contreras}, {Morrell},
  {Phillips}, {Stritzinger}, {Hsiao}, {Gonz{\'a}lez-Gait{\'a}n}, {Agliozzo},
  {Castell{\'o}n}, {Chambers}, {Chen}, {Flewelling}, {Gonzalez},
  {Hosseinzadeh}, {Huber}, {Fraser}, {Inserra}, {Kankare}, {Mattila},
  {Magnier}, {Maguire}, {Lowe}, {Sollerman}, {Sullivan}, {Young}, \&
  {Valenti}}]{Anderson2018NatAs...2..574A}
{Anderson}, J.~P., {Dessart}, L., {Guti{\'e}rrez}, C.~P., {et~al.} 2018, Nature
  Astronomy, 2, 574

\bibitem[{{Arcavi} {et~al.}(2017){Arcavi}, {Hosseinzadeh}, {Brown}, {Smartt},
  {Valenti}, {Tartaglia}, {Piro}, {Sanchez}, {Nicholls}, {Monard}, {Howell},
  {McCully}, {Sand}, {Tonry}, {Denneau}, {Stalder}, {Heinze}, {Rest}, {Smith},
  \& {Bishop}}]{Arcavi2017ApJ...837L...2A}
{Arcavi}, I., {Hosseinzadeh}, G., {Brown}, P.~J., {et~al.} 2017, \apjl, 837, L2

\bibitem[{Arnett(1996)}]{Arnett1996SupernovaeAN}
Arnett, D.~C. 1996, in Supernovae and Nucleosynthesis

\bibitem[{{Arnett}(1982)}]{Arnett1982ApJ...253..785A}
{Arnett}, W.~D. 1982, \apj, 253, 785

\bibitem[{{Balakina} {et~al.}(2019){Balakina}, {Pruzhinskaya}, {Moskvitin}, \&
  {Blinnikov}}]{Balakina2019mmag.conf...32B}
{Balakina}, E.~A., {Pruzhinskaya}, M.~V., {Moskvitin}, A.~S., \& {Blinnikov},
  S.~I. 2019, in The Multi-Messenger Astronomy: Gamma-Ray Bursts, Search for
  Electromagnetic Counterparts to Neutrino Events and Gravitational Waves,
  32--36

\bibitem[{{Barbon} {et~al.}(1995){Barbon}, {Benetti}, {Cappellaro}, {Patat},
  {Turatto}, \& {Iijima}}]{Barbon1995A&AS..110..513B}
{Barbon}, R., {Benetti}, S., {Cappellaro}, E., {et~al.} 1995, \aaps, 110, 513

\bibitem[{{Barmentloo} {et~al.}(2024){Barmentloo}, {Jerkstrand}, {Iwamoto},
  {Hachisu}, {Nomoto}, {Sollerman}, \& {Woosley}}]{Barmentloo2024MNRAS}
{Barmentloo}, S., {Jerkstrand}, A., {Iwamoto}, K., {et~al.} 2024, \mnras, 533,
  1251

\bibitem[{{Benetti}(2017)}]{Benetti2017TNSCR.258....1B}
{Benetti}, S. 2017, Transient Name Server Classification Report, 2017-258, 1

\bibitem[{{Beniamini} {et~al.}(2019){Beniamini}, {Hotokezaka}, {van der Horst},
  \& {Kouveliotou}}]{Beniamini2019MNRAS.487.1426B}
{Beniamini}, P., {Hotokezaka}, K., {van der Horst}, A., \& {Kouveliotou}, C.
  2019, \mnras, 487, 1426

\bibitem[{{Bersten} {et~al.}(2012){Bersten}, {Benvenuto}, {Nomoto}, {Ergon},
  {Folatelli}, {Sollerman}, {Benetti}, {Botticella}, {Fraser}, {Kotak},
  {Maeda}, {Ochner}, \& {Tomasella}}]{Bersten2012ApJ...757...31B}
{Bersten}, M.~C., {Benvenuto}, O.~G., {Nomoto}, K., {et~al.} 2012, \apj, 757,
  31

\bibitem[{{Bersten} {et~al.}(2018){Bersten}, {Folatelli}, {Garc{\'\i}a}, {van
  Dyk}, {Benvenuto}, {Orellana}, {Buso}, {S{\'a}nchez}, {Tanaka}, {Maeda},
  {Filippenko}, {Zheng}, {Brink}, {Cenko}, {de Jaeger}, {Kumar}, {Moriya},
  {Nomoto}, {Perley}, {Shivvers}, \& {Smith}}]{Bersten2018Natur.554..497B}
{Bersten}, M.~C., {Folatelli}, G., {Garc{\'\i}a}, F., {et~al.} 2018, \nat, 554,
  497

\bibitem[{{Bose} {et~al.}(2021){Bose}, {Dong}, {Kochanek}, {Stritzinger},
  {Ashall}, {Benetti}, {Falco}, {Filippenko}, {Pastorello}, {Prieto}, {Somero},
  {Sukhbold}, {Zhang}, {Auchettl}, {Brink}, {Brown}, {Chen}, {Fiore}, {Grupe},
  {Holoien}, {Lundqvist}, {Mattila}, {Mutel}, {Pooley}, {Post}, {Reddy},
  {Reynolds}, {Shappee}, {Stanek}, {Thompson}, {Villanueva}, \&
  {Zheng}}]{Bose2021MNRAS.503.3472B}
{Bose}, S., {Dong}, S., {Kochanek}, C.~S., {et~al.} 2021, \mnras, 503, 3472

\bibitem[{{Branch} {et~al.}(2002){Branch}, {Benetti}, {Kasen}, {Baron},
  {Jeffery}, {Hatano}, {Stathakis}, {Filippenko}, {Matheson}, {Pastorello},
  {Altavilla}, {Cappellaro}, {Rizzi}, {Turatto}, {Li}, {Leonard}, \&
  {Shields}}]{Branch2002ApJ...566.1005B}
{Branch}, D., {Benetti}, S., {Kasen}, D., {et~al.} 2002, \apj, 566, 1005

\bibitem[{{Cai} {et~al.}(2018){Cai}, {Pastorello}, {Fraser}, {Botticella},
  {Gall}, {Arcavi}, {Benetti}, {Cappellaro}, {Elias-Rosa}, {Harmanen},
  {Hosseinzadeh}, {Howell}, {Isern}, {Kangas}, {Kankare}, {Kuncarayakti},
  {Lundqvist}, {Mattila}, {McCully}, {Reynolds}, {Somero}, {Stritzinger}, \&
  {Terreran}}]{Cai2018MNRAS.480.3424C}
{Cai}, Y.~Z., {Pastorello}, A., {Fraser}, M., {et~al.} 2018, \mnras, 480, 3424

\bibitem[{{Cardelli} {et~al.}(1989){Cardelli}, {Clayton}, \&
  {Mathis}}]{Cardelli1989ApJ...345..245C}
{Cardelli}, J.~A., {Clayton}, G.~C., \& {Mathis}, J.~S. 1989, \apj, 345, 245

\bibitem[{{Charalampopoulos} {et~al.}(2025){Charalampopoulos}, {Kotak},
  {Sollerman}, {Guti{\'e}rrez}, {Pursiainen}, {Killestein}, {Schulze}, {Pessi},
  {Maeda}, {Kangas}, {Cai}, {Fremling}, {Hinds}, {Jegou du Laz}, {Kankare},
  {Kasliwal}, {Kuncarayakti}, {Lundqvist}, {Masci}, {Mattila}, {Perley},
  {Reguitti}, {Reynolds}, {Stritzinger}, {Tartaglia}, {Van Roestel}, \&
  {Wold}}]{Charalampopoulos2025A&A...700A.138C}
{Charalampopoulos}, P., {Kotak}, R., {Sollerman}, J., {et~al.} 2025, \aap, 700,
  A138

\bibitem[{{Chatzopoulos} {et~al.}(2013){Chatzopoulos}, {Wheeler}, {Vinko},
  {Horvath}, \& {Nagy}}]{Chatzopoulos2013ApJ...773...76C}
{Chatzopoulos}, E., {Wheeler}, J.~C., {Vinko}, J., {Horvath}, Z.~L., \& {Nagy},
  A. 2013, \apj, 773, 76

\bibitem[{{Chen} {et~al.}(2025){Chen}, {Wang}, {Wu}, {Andrews}, {Farah},
  {Ochner}, {Reguitti}, {Brink}, {Zhang}, {Song}, {Liu}, {Filippenko}, {Sand},
  {Albanese}, {Alexander}, {Andrews}, {Bostroem}, {Cai}, {Christy}, {Esamdin},
  {Farina}, {Franz}, {Howell}, {Hsu}, {Hu}, {Iskandar}, {Li}, {Li}, {Li}, {Li},
  {Liu}, {McCully}, {Newsome}, {Ni}, {Pastorello}, {Padilla Gonzalez},
  {Pearson}, {Peng}, {Ransome}, {Shrestha}, {Smith}, {Subrayan}, {Terreran},
  {Valerin}, {Vink{\'o}}, {Vasylyev}, {Wang}, {Wang}, {Wang}, {Wheeler},
  {Wynn}, {Xiang}, {Yan}, {Yuan}, {Zhang}, {Zheng}, \&
  {Zhang}}]{Chen2025arXiv251022997C}
{Chen}, L., {Wang}, X., {Wu}, Q., {et~al.} 2025, arXiv e-prints,
  arXiv:2510.22997

\bibitem[{{Chevalier}(1982)}]{Chevalier1982ApJ...259..302C}
{Chevalier}, R.~A. 1982, \apj, 259, 302

\bibitem[{{Chieffi} \& {Limongi}(2013)}]{Chieffi2013ApJ...764...21C}
{Chieffi}, A. \& {Limongi}, M. 2013, \apj, 764, 21

\bibitem[{{Chugai}(1991)}]{Chugai1991MNRAS.250..513C}
{Chugai}, N.~N. 1991, \mnras, 250, 513

\bibitem[{{Clocchiatti} \& {Wheeler}(1997)}]{Clocchiatti1997ApJ...491..375C}
{Clocchiatti}, A. \& {Wheeler}, J.~C. 1997, \apj, 491, 375

\bibitem[{{Clocchiatti} {et~al.}(1996){Clocchiatti}, {Wheeler}, {Benetti}, \&
  {Frueh}}]{Clocchiatti1996ApJ...459..547C}
{Clocchiatti}, A., {Wheeler}, J.~C., {Benetti}, S., \& {Frueh}, M. 1996, \apj,
  459, 547

\bibitem[{{Crockett} {et~al.}(2008){Crockett}, {Eldridge}, {Smartt},
  {Pastorello}, {Gal-Yam}, {Fox}, {Leonard}, {Kasliwal}, {Mattila}, {Maund},
  {Stephens}, \& {Danziger}}]{Crockett2008MNRAS.391L...5C}
{Crockett}, R.~M., {Eldridge}, J.~J., {Smartt}, S.~J., {et~al.} 2008, \mnras,
  391, L5

\bibitem[{{CSST Collaboration} {et~al.}(2026){CSST Collaboration}, {Gong},
  {Miao}, {Zhan}, {Li}, {Shangguan}, {Li}, {Liu}, {Chen}, {Yuan}, {Zhou},
  {Liu}, {Yu}, {Ji}, {Qi}, {Liu}, {Dai}, {Wang}, {Zheng}, {Hao}, {Dou}, {Ao},
  {Lin}, {Zhang}, {Wang}, {Sun}, {Li}, {Li}, {Xu}, {Li}, {Li}, {Wu}, {Zhang},
  {Wang}, {Bai}, {Cai}, {Cai}, {Cao}, {Chan}, {Chang}, {Chen}, {Chen}, {Chen},
  {Chen}, {Cui}, {Dong}, {Du}, {Duan}, {Fan}, {Fan}, {Fan}, {Fan}, {Fang},
  {Fu}, {Fu}, {Fu}, {Gao}, {Gu}, {Gu}, {Guo}, {Han}, {Hu}, {Huang}, {Ho},
  {Jiang}, {Jiang}, {Jing}, {Kang}, {Kong}, {Li}, {Li}, {Li}, {Li}, {Li}, {Li},
  {Liao}, {Lin}, {Liu}, {Liu}, {Liu}, {Liu}, {Mao}, {Mao}, {Meng}, {Pang},
  {Peng}, {Peng}, {Shan}, {Shen}, {Shen}, {Shen}, {Shi}, {Shi}, {Tan}, {Tian},
  {Wang}, {Wang}, {Wang}, {Wang}, {Wu}, {Wu}, {Wu}, {Xu}, {Xue}, {Xue}, {Yang},
  {Yang}, {Yao}, {Yuan}, {Yuan}, {Zhang}, {Zhang}, {Zhang}, {Zhang}, {Zhang},
  {Zhao}, {Zhao}, {Zhong}, {Zhong}, {Zhou}, {Zhu}, \& {Zu}}]{CSST2025}
{CSST Collaboration}, {Gong}, Y., {Miao}, H., {et~al.} 2026, Science China
  Physics, Mechanics, and Astronomy, 69, 239501

\bibitem[{{Delgado} {et~al.}(2017){Delgado}, {Harrison}, {Hodgkin}, {Leeuwen},
  {Rixon}, \& {Yoldas}}]{Delgado2017TNSTR.184....1D}
{Delgado}, A., {Harrison}, D., {Hodgkin}, S., {et~al.} 2017, Transient Name
  Server Discovery Report, 2017-184, 1

\bibitem[{{Desai} {et~al.}(2023){Desai}, {Ashall}, {Shappee}, {Morrell},
  {Galbany}, {Burns}, {DerKacy}, {Hinkle}, {Hsiao}, {Kumar}, {Lu}, {Phillips},
  {Shahbandeh}, {Stritzinger}, {Baron}, {Bersten}, {Brown}, {de Jaeger},
  {Elias-Rosa}, {Folatelli}, {Huber}, {Mazzali}, {M{\"u}ller-Bravo}, {Piro},
  {Polin}, {Suntzeff}, {Anderson}, {Chambers}, {Chen}, {de Boer}, {Fulton},
  {Gao}, {Gromadzki}, {Inserra}, {Magnier}, {Nicholl}, {Ragosta}, {Wainscoat},
  \& {Young}}]{Desai2023MNRAS.524..767D}
{Desai}, D.~D., {Ashall}, C., {Shappee}, B.~J., {et~al.} 2023, \mnras, 524, 767

\bibitem[{{Dessart} \& {Hillier}(2005)}]{Dessart2005AA}
{Dessart}, L. \& {Hillier}, D.~J. 2005, \aap, 437, 667

\bibitem[{{Dessart} \& {Hillier}(2020)}]{Dessart2020A&A...642A..33D}
{Dessart}, L. \& {Hillier}, D.~J. 2020, \aap, 642, A33

\bibitem[{{Dessart} \& {Hillier}(2022)}]{Luc2022}
{Dessart}, L. \& {Hillier}, D.~J. 2022, \aap, 660, L9

\bibitem[{{Dessart} {et~al.}(2021){Dessart}, {Hillier}, {Sukhbold}, {Woosley},
  \& {Janka}}]{Dessart2021A&A...656A..61D}
{Dessart}, L., {Hillier}, D.~J., {Sukhbold}, T., {Woosley}, S.~E., \& {Janka},
  H.-T. 2021, \aap, 656, A61

\bibitem[{{Dessart} {et~al.}(2016){Dessart}, {Hillier}, {Woosley}, {Livne},
  {Waldman}, {Yoon}, \& {Langer}}]{Dessart2016MNRAS.458.1618D}
{Dessart}, L., {Hillier}, D.~J., {Woosley}, S., {et~al.} 2016, \mnras, 458,
  1618

\bibitem[{{Dessart} {et~al.}(2023){Dessart}, {Hillier}, {Woosley}, \&
  {Kuncarayakti}}]{Dessart2023A&A...677A...7D}
{Dessart}, L., {Hillier}, D.~J., {Woosley}, S.~E., \& {Kuncarayakti}, H. 2023,
  \aap, 677, A7

\bibitem[{{Dessart} {et~al.}(2017){Dessart}, {Hillier}, {Yoon}, {Waldman}, \&
  {Livne}}]{Dessart2017A&A...603A..51D}
{Dessart}, L., {Hillier}, D.~J., {Yoon}, S.-C., {Waldman}, R., \& {Livne}, E.
  2017, \aap, 603, A51

\bibitem[{{Elmhamdi} {et~al.}(2004){Elmhamdi}, {Danziger}, {Cappellaro}, {Della
  Valle}, {Gouiffes}, {Phillips}, \& {Turatto}}]{Elmhamdi2004A&A}
{Elmhamdi}, A., {Danziger}, I.~J., {Cappellaro}, E., {et~al.} 2004, \aap, 426,
  963

\bibitem[{{Ergon} \& {Fransson}(2022)}]{Ergon2022A&A...666A.104E}
{Ergon}, M. \& {Fransson}, C. 2022, \aap, 666, A104

\bibitem[{{Ergon} {et~al.}(2014){Ergon}, {Sollerman}, {Fraser}, {Pastorello},
  {Taubenberger}, {Elias-Rosa}, {Bersten}, {Jerkstrand}, {Benetti},
  {Botticella}, {Fransson}, {Harutyunyan}, {Kotak}, {Smartt}, {Valenti},
  {Bufano}, {Cappellaro}, {Fiaschi}, {Howell}, {Kankare}, {Magill}, {Mattila},
  {Maund}, {Naves}, {Ochner}, {Ruiz}, {Smith}, {Tomasella}, \&
  {Turatto}}]{Ergon2014A&A...562A..17E}
{Ergon}, M., {Sollerman}, J., {Fraser}, M., {et~al.} 2014, \aap, 562, A17

\bibitem[{{Eyles-Ferris} {et~al.}(2025){Eyles-Ferris}, {Jonker}, {Levan},
  {Malesani}, {Sarin}, {Fryer}, {Rastinejad}, {Burns}, {Tanvir}, {O'Brien},
  {Fong}, {Mandel}, {Gompertz}, {Kilpatrick}, {Bloemen}, {Bright},
  {Carotenuto}, {Corcoran}, {Cotter}, {Groot}, {Izzo}, {Laskar},
  {Martin-Carrillo}, {Palmerio}, {Ravasio}, {van Roestel}, {Saccardi},
  {Starling}, {Thakur}, {Vergani}, {Vreeswijk}, {Bauer}, {Campana},
  {Chac{\'o}n}, {Chrimes}, {Covino}, {van Dalen}, {D'Elia}, {De Pasquale},
  {Habeeb}, {Hartmann}, {van Hoof}, {Jakobsson}, {Julakanti}, {Leloudas}, {Mata
  S{\'a}nchez}, {Nixon}, {Pieterse}, {Pugliese}, {Quirola-V{\'a}squez},
  {Rayson}, {Salvaterra}, {Schneider}, {Torres}, \&
  {Zafar}}]{Eyles2025ApJ...988L..14E}
{Eyles-Ferris}, R. A.~J., {Jonker}, P.~G., {Levan}, A.~J., {et~al.} 2025,
  \apjl, 988, L14

\bibitem[{{Fang} \& {Maeda}(2018)}]{Fang2018ApJ...864...47F}
{Fang}, Q. \& {Maeda}, K. 2018, \apj, 864, 47

\bibitem[{{Fang} {et~al.}(2024){Fang}, {Maeda}, {Kuncarayakti}, \&
  {Nagao}}]{Fang2024NatAs...8..111F}
{Fang}, Q., {Maeda}, K., {Kuncarayakti}, H., \& {Nagao}, T. 2024, Nature
  Astronomy, 8, 111

\bibitem[{{Fang} {et~al.}(2019){Fang}, {Maeda}, {Kuncarayakti}, {Sun}, \&
  {Gal-Yam}}]{Fang2019NatAs...3..434F}
{Fang}, Q., {Maeda}, K., {Kuncarayakti}, H., {Sun}, F., \& {Gal-Yam}, A. 2019,
  Nature Astronomy, 3, 434

\bibitem[{{Fang} {et~al.}(2022){Fang}, {Maeda}, {Kuncarayakti}, {Tanaka},
  {Kawabata}, {Hattori}, {Aoki}, {Moriya}, \&
  {Yamanaka}}]{Fang2022ApJ...928..151F}
{Fang}, Q., {Maeda}, K., {Kuncarayakti}, H., {et~al.} 2022, \apj, 928, 151

\bibitem[{{Filippenko}(1997)}]{Filippenko1997ARA&A..35..309F}
{Filippenko}, A.~V. 1997, \araa, 35, 309

\bibitem[{{Fransson} \& {Chevalier}(1989)}]{Fransson1989ApJ...343..323F}
{Fransson}, C. \& {Chevalier}, R.~A. 1989, \apj, 343, 323

\bibitem[{{Fraser}(2020)}]{Fraser2020RSOS....700467F}
{Fraser}, M. 2020, Royal Society Open Science, 7, 200467

\bibitem[{{Gal-Yam}(2017)}]{Gal-Yam2017hsn..book..195G}
{Gal-Yam}, A. 2017, in Handbook of Supernovae, ed. A.~W. {Alsabti} \&
  P.~{Murdin}, 195

\bibitem[{{Gal-Yam}(2019)}]{Gal2019ARA&A..57..305G}
{Gal-Yam}, A. 2019, \araa, 57, 305

\bibitem[{{Gangopadhyay} {et~al.}(2018){Gangopadhyay}, {Misra}, {Pastorello},
  {Sahu}, {Tomasella}, {Tartaglia}, {Singh}, {Dastidar}, {Srivastav}, {Ochner},
  {Brown}, {Anupama}, {Benetti}, {Cappellaro}, {Kumar}, {Kumar}, \&
  {Pandey}}]{Gangopadhyay2018MNRAS.476.3611G}
{Gangopadhyay}, A., {Misra}, K., {Pastorello}, A., {et~al.} 2018, \mnras, 476,
  3611

\bibitem[{{Gangopadhyay} \& {Pessi}(2026)}]{Gangopadhyay2025arXiv251204010G}
{Gangopadhyay}, A. \& {Pessi}, P.~J. 2026, Frontiers in Astronomy and Space
  Sciences, 12, 1708372

\bibitem[{{Goldwasser} {et~al.}(2022){Goldwasser}, {Yaron}, {Sass}, {Irani},
  {Gal-Yam}, \& {Howell}}]{Goldwasser2022TNSAN.191....1G}
{Goldwasser}, S., {Yaron}, O., {Sass}, A., {et~al.} 2022, Transient Name Server
  AstroNote, 191, 1

\bibitem[{{Gomez} {et~al.}(2022){Gomez}, {Berger}, {Nicholl}, {Blanchard}, \&
  {Hosseinzadeh}}]{Gomez2022ApJ...941..107G}
{Gomez}, S., {Berger}, E., {Nicholl}, M., {Blanchard}, P.~K., \&
  {Hosseinzadeh}, G. 2022, \apj, 941, 107

\bibitem[{{Gonz{\'a}lez-Gait{\'a}n} {et~al.}(2015){Gonz{\'a}lez-Gait{\'a}n},
  {Tominaga}, {Molina}, {Galbany}, {Bufano}, {Anderson}, {Gutierrez},
  {F{\"o}rster}, {Pignata}, {Bersten}, {Howell}, {Sullivan}, {Carlberg}, {de
  Jaeger}, {Hamuy}, {Baklanov}, \&
  {Blinnikov}}]{Gonzalez-Gaitan2015MNRAS.451.2212G}
{Gonz{\'a}lez-Gait{\'a}n}, S., {Tominaga}, N., {Molina}, J., {et~al.} 2015,
  \mnras, 451, 2212

\bibitem[{{Grayling} {et~al.}(2021){Grayling}, {Guti{\'e}rrez}, {Sullivan},
  {Wiseman}, {Vincenzi}, {Gonz{\'a}lez-Gait{\'a}n}, {Tucker}, {Galbany},
  {Kelsey}, {Lidman}, {Swann}, {Smith}, {Frohmaier}, {Carollo}, {Glazebrook},
  {Lewis}, {M{\"o}ller}, {Hinton}, {Uddin}, {Abbott}, {Aguena}, {Avila},
  {Bertin}, {Bhargava}, {Brooks}, {Carnero Rosell}, {Carrasco Kind},
  {Carretero}, {Costanzi}, {da Costa}, {De Vicente}, {Desai}, {Diehl}, {Doel},
  {Everett}, {Ferrero}, {Fosalba}, {Frieman}, {Garc{\'\i}a-Bellido},
  {Gaztanaga}, {Gruen}, {Gruendl}, {Gschwend}, {Gutierrez}, {Hoyle}, {Kuehn},
  {Kuropatkin}, {Lima}, {MacCrann}, {Marshall}, {Martini}, {Miquel}, {Morgan},
  {Palmese}, {Paz-Chinch{\'o}n}, {Plazas}, {Romer}, {S{\'a}nchez}, {Sanchez},
  {Scarpine}, {Serrano}, {Sevilla-Noarbe}, {Soares-Santos}, {Suchyta}, {Tarle},
  {Thomas}, {To}, {Varga}, {Walker}, {Wilkinson}, \& {DES
  Collaboration}}]{Grayling2021MNRAS.505.3950G}
{Grayling}, M., {Guti{\'e}rrez}, C.~P., {Sullivan}, M., {et~al.} 2021, \mnras,
  505, 3950

\bibitem[{{Guillochon} {et~al.}(2018){Guillochon}, {Nicholl}, {Villar},
  {Mockler}, {Narayan}, {Mandel}, {Berger}, \&
  {Williams}}]{Guillochon2018ApJS..236....6G}
{Guillochon}, J., {Nicholl}, M., {Villar}, V.~A., {et~al.} 2018, \apjs, 236, 6

\bibitem[{{Hambleton} {et~al.}(2023){Hambleton}, {Bianco}, {Street}, {Bell},
  {Buckley}, {Graham}, {Hernitschek}, {Lund}, {Mason}, {Pepper}, {Pr{\v{s}}a},
  {Rabus}, {Raiteri}, {Szab{\'o}}, {Szkody}, {Andreoni}, {Antoniucci},
  {Balmaverde}, {Bellm}, {Bonito}, {Bono}, {Botticella}, {Brocato},
  {Bu{\v{c}}ar Bricman}, {Cappellaro}, {Carnerero}, {Chornock}, {Clarke},
  {Cowperthwaite}, {Cucchiara}, {D'Ammando}, {Dage}, {Dall'Ora}, {Davenport},
  {de Martino}, {de Somma}, {Di Criscienzo}, {Di Stefano}, {Drout}, {Fabrizio},
  {Fiorentino}, {Gandhi}, {Garofalo}, {Giannini}, {Gomboc}, {Greggio},
  {Hartigan}, {Hundertmark}, {Johnson}, {Johnson}, {Jurkic}, {Khakpash},
  {Leccia}, {Li}, {Magurno}, {Malanchev}, {Marconi}, {Margutti}, {Marinoni},
  {Mauron}, {Molinaro}, {M{\"o}ller}, {Moniez}, {Muraveva}, {Musella}, {Ngeow},
  {Pastorello}, {Petrecca}, {Piranomonte}, {Ragosta}, {Reguitti}, {Righi},
  {Ripepi}, {Rivera Sandoval}, {Stassun}, {Stroh}, {Terreran}, {Trimble},
  {Tsapras}, {van Velzen}, {Venuti}, \& {Vink}}]{Hambleton2023PASP..135j5002H}
{Hambleton}, K.~M., {Bianco}, F.~B., {Street}, R., {et~al.} 2023, \pasp, 135,
  105002

\bibitem[{{Hart} {et~al.}(2023){Hart}, {Shappee}, {Hey}, {Kochanek}, {Stanek},
  {Lim}, {Dobbs}, {Tucker}, {Jayasinghe}, {Beacom}, {Boright}, {Holoien},
  {Ong}, {Prieto}, {Thompson}, \& {Will}}]{Hart2023arXiv230403791H}
{Hart}, K., {Shappee}, B.~J., {Hey}, D., {et~al.} 2023, arXiv e-prints,
  arXiv:2304.03791

\bibitem[{{Hillier} \& {Dessart}(2012)}]{Hillier2012MNRAS.424..252H}
{Hillier}, D.~J. \& {Dessart}, L. 2012, \mnras, 424, 252

\bibitem[{{Houck} \& {Fransson}(1996)}]{Houck1996ApJ...456..811H}
{Houck}, J.~C. \& {Fransson}, C. 1996, \apj, 456, 811

\bibitem[{{Jerkstrand} {et~al.}(2015{\natexlab{a}}){Jerkstrand}, {Ergon},
  {Smartt}, {Fransson}, {Sollerman}, {Taubenberger}, {Bersten}, \&
  {Spyromilio}}]{Jerkstrand2015A&A...573A..12J}
{Jerkstrand}, A., {Ergon}, M., {Smartt}, S.~J., {et~al.} 2015{\natexlab{a}},
  \aap, 573, A12

\bibitem[{{Jerkstrand} {et~al.}(2012){Jerkstrand}, {Fransson}, {Maguire},
  {Smartt}, {Ergon}, \& {Spyromilio}}]{Jerkstrand2012A&A...546A..28J}
{Jerkstrand}, A., {Fransson}, C., {Maguire}, K., {et~al.} 2012, \aap, 546, A28

\bibitem[{{Jerkstrand} {et~al.}(2014){Jerkstrand}, {Smartt}, {Fraser},
  {Fransson}, {Sollerman}, {Taddia}, \&
  {Kotak}}]{Jerkstrand2014MNRAS.439.3694J}
{Jerkstrand}, A., {Smartt}, S.~J., {Fraser}, M., {et~al.} 2014, \mnras, 439,
  3694

\bibitem[{{Jerkstrand} {et~al.}(2015{\natexlab{b}}){Jerkstrand}, {Smartt},
  {Sollerman}, {Inserra}, {Fraser}, {Spyromilio}, {Fransson}, {Chen},
  {Barbarino}, {Dall'Ora}, {Botticella}, {Della Valle}, {Gal-Yam}, {Valenti},
  {Maguire}, {Mazzali}, \& {Tomasella}}]{Jerkstrand2015MNRAS.448.2482J}
{Jerkstrand}, A., {Smartt}, S.~J., {Sollerman}, J., {et~al.}
  2015{\natexlab{b}}, \mnras, 448, 2482

\bibitem[{{Kasen} \& {Bildsten}(2010)}]{Kasen2010ApJ...717..245K}
{Kasen}, D. \& {Bildsten}, L. 2010, \apj, 717, 245

\bibitem[{{Khatami} \& {Kasen}(2019)}]{Khatami2019ApJ...878...56K}
{Khatami}, D.~K. \& {Kasen}, D.~N. 2019, \apj, 878, 56

\bibitem[{{Kozma} \& {Fransson}(1998)}]{Kozma1998ApJ...497..431K}
{Kozma}, C. \& {Fransson}, C. 1998, \apj, 497, 431

\bibitem[{{Kumar} {et~al.}(2022){Kumar}, {Singh}, {Sahu}, \&
  {Anupama}}]{Kumar2022ApJ...927...61K}
{Kumar}, B., {Singh}, A., {Sahu}, D.~K., \& {Anupama}, G.~C. 2022, \apj, 927,
  61

\bibitem[{{Li} \& {McCray}(1992)}]{Li1992ApJ...387..309L}
{Li}, H. \& {McCray}, R. 1992, \apj, 387, 309

\bibitem[{{Li} {et~al.}(2025{\natexlab{a}}){Li}, {Benetti}, {Cai}, {Wang},
  {Pastorello}, {Elias-Rosa}, {Reguitti}, {Borsato}, {Cappellaro}, {Fiore},
  {Fraser}, {Gromadzki}, {Harmanen}, {Isern}, {Kangas}, {Kankare}, {Lundqvist},
  {Mattila}, {Ochner}, {Peng}, {Reynolds}, {Salmaso}, {Srivastav},
  {Stritzinger}, {Tomasella}, {Valerin}, {Wang}, {Zhang}, \&
  {Wu}}]{Li2025A&A...704A.233L}
{Li}, L.-H., {Benetti}, S., {Cai}, Y.-Z., {et~al.} 2025{\natexlab{a}}, \aap,
  704, A233

\bibitem[{{Li} {et~al.}(2025{\natexlab{b}}){Li}, {Zhu}, {Zou}, {Geng}, {Liu},
  {Wang}, {Li}, {Xu}, {Sun}, {Wang}, {Yu}, {Zhang}, {Wu}, {Yang}, {Filippenko},
  {Liu}, {Yuan}, {Aguado}, {An}, {An}, {Buckley}, {Castro-Tirado}, {Fu},
  {Fynbo}, {Howell}, {Hu}, {Jiang}, {Kumar}, {Mao}, {Maund}, {Liu}, {Mockler},
  {Moskvitin}, {Andrews}, {Bom}, {Brink}, {Chatterjee}, {Chen}, {Cheng},
  {Cooke}, {Dai}, {Du}, {Erasmus}, {Fang}, {Farah}, {Goranskij}, {Gritsevich},
  {Gu}, {Guo}, {Hsiao}, {Hu}, {Hua}, {Jacobson-Gal{\'a}n}, {Jia}, {Jin},
  {Kasliwal}, {Kilpatrick}, {Kumar}, {Lei}, {Li}, {Li}, {Li}, {Ling}, {Liu},
  {Liu}, {Liu}, {L{\'o}pez-Oramas}, {Maslennikova}, {McCully}, {Monageng},
  {Newsone}, {Padilla Gonzalez}, {Pan}, {Peng}, {Pignata}, {Poidevin},
  {Potter}, {P{\'e}rez-Fournon}, {Santana-Silva}, {Santos}, {Song}, {Song},
  {Spiridonova}, {Sun}, {Sun}, {Terreran}, {Wang}, {Wang}, {Wang}, {Wang},
  {Wu}, {Xiang}, {Xiao}, {Xu}, {Xue}, {Yan}, {Yang}, {Yu}, {Zhang}, {Zhang},
  {Zhang}, {Zhang}, {Zhang}, {Zheng}, \& {Zou}}]{Li2025arXiv250417034L}
{Li}, W.~X., {Zhu}, Z.~P., {Zou}, X.~Z., {et~al.} 2025{\natexlab{b}}, arXiv
  e-prints, arXiv:2504.17034

\bibitem[{{Limongi} \& {Chieffi}(2003)}]{Limongi2003ApJ...592..404L}
{Limongi}, M. \& {Chieffi}, A. 2003, \apj, 592, 404

\bibitem[{{Ma} {et~al.}(2025{\natexlab{a}}){Ma}, {Wang}, {Mo}, {Andrew Howell},
  {Pellegrino}, {Zhang}, {Wu}, {Yan}, {Liu}, {Arcavi}, {Chen}, {Farah},
  {Padilla Gonzalez}, {Guo}, {Hiramatsu}, {Li}, {Lin}, {Liu}, {McCully},
  {Newsome}, {Sai}, {Terreran}, {Xiang}, \& {Zhang}}]{2025A&A...698A.306M}
{Ma}, X., {Wang}, X., {Mo}, J., {et~al.} 2025{\natexlab{a}}, \aap, 698, A306

\bibitem[{{Ma} {et~al.}(2025{\natexlab{b}}){Ma}, {Wang}, {Mo}, {Howell},
  {Pellegrino}, {Zhang}, {Yan}, {Arcavi}, {Chen}, {Farah}, {Padilla Gonzalez},
  {Guo}, {Hiramatsu}, {Li}, {Lin}, {Liu}, {McCully}, {Newsome}, {Sai},
  {Terreran}, {Xiang}, {Zhang}, \& {Zhang}}]{2025A&A...698A.305M}
{Ma}, X., {Wang}, X., {Mo}, J., {et~al.} 2025{\natexlab{b}}, \aap, 698, A305

\bibitem[{{Maeda} {et~al.}(2007){Maeda}, {Tanaka}, {Nomoto}, {Tominaga},
  {Kawabata}, {Mazzali}, {Umeda}, {Suzuki}, \&
  {Hattori}}]{Maeda2007ApJ...666.1069M}
{Maeda}, K., {Tanaka}, M., {Nomoto}, K., {et~al.} 2007, \apj, 666, 1069

\bibitem[{{Marinoni} {et~al.}(1998){Marinoni}, {Monaco}, {Giuricin}, \&
  {Costantini}}]{Marinoni1998ApJ...505..484M}
{Marinoni}, C., {Monaco}, P., {Giuricin}, G., \& {Costantini}, B. 1998, \apj,
  505, 484

\bibitem[{{Matheson} {et~al.}(2000){Matheson}, {Filippenko}, {Ho}, {Barth}, \&
  {Leonard}}]{Matheson2000AJ....120.1499M}
{Matheson}, T., {Filippenko}, A.~V., {Ho}, L.~C., {Barth}, A.~J., \& {Leonard},
  D.~C. 2000, \aj, 120, 1499

\bibitem[{{Matheson} {et~al.}(2001){Matheson}, {Filippenko}, {Li}, {Leonard},
  \& {Shields}}]{Matheson2001AJ....121.1648M}
{Matheson}, T., {Filippenko}, A.~V., {Li}, W., {Leonard}, D.~C., \& {Shields},
  J.~C. 2001, \aj, 121, 1648

\bibitem[{{Maurer} {et~al.}(2010){Maurer}, {Mazzali}, {Deng}, {Filippenko},
  {Hamuy}, {Kirshner}, {Matheson}, {Modjaz}, {Pian}, {Stritzinger},
  {Taubenberger}, \& {Valenti}}]{Maurer2010MNRAS.402..161M}
{Maurer}, J.~I., {Mazzali}, P.~A., {Deng}, J., {et~al.} 2010, \mnras, 402, 161

\bibitem[{{Mazzali} {et~al.}(2005){Mazzali}, {Kawabata}, {Maeda}, {Nomoto},
  {Filippenko}, {Ramirez-Ruiz}, {Benetti}, {Pian}, {Deng}, {Tominaga},
  {Ohyama}, {Iye}, {Foley}, {Matheson}, {Wang}, \&
  {Gal-Yam}}]{Mazzali2005Sci...308.1284M}
{Mazzali}, P.~A., {Kawabata}, K.~S., {Maeda}, K., {et~al.} 2005, Science, 308,
  1284

\bibitem[{{Mazzali} {et~al.}(2008){Mazzali}, {Valenti}, {Della Valle},
  {Chincarini}, {Sauer}, {Benetti}, {Pian}, {Piran}, {D'Elia}, {Elias-Rosa},
  {Margutti}, {Pasotti}, {Antonelli}, {Bufano}, {Campana}, {Cappellaro},
  {Covino}, {D'Avanzo}, {Fiore}, {Fugazza}, {Gilmozzi}, {Hunter}, {Maguire},
  {Maiorano}, {Marziani}, {Masetti}, {Mirabel}, {Navasardyan}, {Nomoto},
  {Palazzi}, {Pastorello}, {Panagia}, {Pellizza}, {Sari}, {Smartt},
  {Tagliaferri}, {Tanaka}, {Taubenberger}, {Tominaga}, {Trundle}, \&
  {Turatto}}]{Mazzali2008Sci...321.1185M}
{Mazzali}, P.~A., {Valenti}, S., {Della Valle}, M., {et~al.} 2008, Science,
  321, 1185

\bibitem[{{Medler} {et~al.}(2022){Medler}, {Mazzali}, {Teffs}, {Ashall},
  {Anderson}, {Arcavi}, {Benetti}, {Bostroem}, {Burke}, {Cai},
  {Charalampopoulos}, {Elias-Rosa}, {Ergon}, {Galbany}, {Gromadzki},
  {Hiramatsu}, {Howell}, {Inserra}, {Lundqvist}, {McCully}, {M{\"u}ller-Bravo},
  {Newsome}, {Nicholl}, {Padilla Gonzalez}, {Paraskeva}, {Pastorello},
  {Pellegrino}, {Pessi}, {Reguitti}, {Reynolds}, {Roy}, {Terreran},
  {Tomasella}, \& {Young}}]{Medler2022MNRAS.513.5540M}
{Medler}, K., {Mazzali}, P.~A., {Teffs}, J., {et~al.} 2022, \mnras, 513, 5540

\bibitem[{{Meza} \& {Anderson}(2020)}]{Meza2020A&A...641A.177M}
{Meza}, N. \& {Anderson}, J.~P. 2020, \aap, 641, A177

\bibitem[{{Milisavljevic} {et~al.}(2010){Milisavljevic}, {Fesen}, {Gerardy},
  {Kirshner}, \& {Challis}}]{Milisavljevic2010ApJ...709.1343M}
{Milisavljevic}, D., {Fesen}, R.~A., {Gerardy}, C.~L., {Kirshner}, R.~P., \&
  {Challis}, P. 2010, \apj, 709, 1343

\bibitem[{{Modjaz} {et~al.}(2019){Modjaz}, {Guti{\'e}rrez}, \&
  {Arcavi}}]{Modjaz2019NatAs...3..717M}
{Modjaz}, M., {Guti{\'e}rrez}, C.~P., \& {Arcavi}, I. 2019, Nature Astronomy,
  3, 717

\bibitem[{{Morales-Garoffolo} {et~al.}(2014){Morales-Garoffolo}, {Elias-Rosa},
  {Benetti}, {Taubenberger}, {Cappellaro}, {Pastorello}, {Klauser}, {Valenti},
  {Howerton}, {Ochner}, {Schramm}, {Siviero}, {Tartaglia}, \&
  {Tomasella}}]{Morales2014MNRAS.445.1647M}
{Morales-Garoffolo}, A., {Elias-Rosa}, N., {Benetti}, S., {et~al.} 2014,
  \mnras, 445, 1647

\bibitem[{{Morales-Garoffolo} {et~al.}(2015){Morales-Garoffolo}, {Elias-Rosa},
  {Bersten}, {Jerkstrand}, {Taubenberger}, {Benetti}, {Cappellaro}, {Kotak},
  {Pastorello}, {Bufano}, {Dom{\'\i}nguez}, {Ergon}, {Fraser}, {Gao},
  {Garc{\'\i}a}, {Howell}, {Isern}, {Smartt}, {Tomasella}, \&
  {Valenti}}]{Morales2015MNRAS.454...95M}
{Morales-Garoffolo}, A., {Elias-Rosa}, N., {Bersten}, M., {et~al.} 2015,
  \mnras, 454, 95

\bibitem[{{Moriya} {et~al.}(2013){Moriya}, {Maeda}, {Taddia}, {Sollerman},
  {Blinnikov}, \& {Sorokina}}]{Moriya2013MNRAS.435.1520M}
{Moriya}, T.~J., {Maeda}, K., {Taddia}, F., {et~al.} 2013, \mnras, 435, 1520

\bibitem[{{Mould} {et~al.}(2000){Mould}, {Huchra}, {Freedman}, {Kennicutt},
  {Ferrarese}, {Ford}, {Gibson}, {Graham}, {Hughes}, {Illingworth}, {Kelson},
  {Macri}, {Madore}, {Sakai}, {Sebo}, {Silbermann}, \&
  {Stetson}}]{Mould2000ApJ...529..786M}
{Mould}, J.~R., {Huchra}, J.~P., {Freedman}, W.~L., {et~al.} 2000, \apj, 529,
  786

\bibitem[{{Nadyozhin}(1994)}]{Nadyozhin1994ApJS...92..527N}
{Nadyozhin}, D.~K. 1994, \apjs, 92, 527

\bibitem[{{Nagy}(2018)}]{Nagy2018ApJ...862..143N}
{Nagy}, A.~P. 2018, \apj, 862, 143

\bibitem[{{Nakar} \& {Piro}(2014)}]{Nakar2014ApJ...788..193N}
{Nakar}, E. \& {Piro}, A.~L. 2014, \apj, 788, 193

\bibitem[{{Nicholl}(2018)}]{Nicholl_2018}
{Nicholl}, M. 2018, Research Notes of the American Astronomical Society, 2, 230

\bibitem[{{Nicholl} {et~al.}(2017){Nicholl}, {Guillochon}, \&
  {Berger}}]{Nicholl2017ApJ...850...55N}
{Nicholl}, M., {Guillochon}, J., \& {Berger}, E. 2017, \apj, 850, 55

\bibitem[{{Nomoto} {et~al.}(1997){Nomoto}, {Hashimoto}, {Tsujimoto},
  {Thielemann}, {Kishimoto}, {Kubo}, \& {Nakasato}}]{Nomoto1997NuPhA.616...79N}
{Nomoto}, K., {Hashimoto}, M., {Tsujimoto}, T., {et~al.} 1997, \nphysa, 616, 79

\bibitem[{{Pastorello} {et~al.}(2008){Pastorello}, {Kasliwal}, {Crockett},
  {Valenti}, {Arbour}, {Itagaki}, {Kaspi}, {Gal-Yam}, {Smartt}, {Griffith},
  {Maguire}, {Ofek}, {Seymour}, {Stern}, \&
  {Wiethoff}}]{Pastorello2008MNRAS.389..955P}
{Pastorello}, A., {Kasliwal}, M.~M., {Crockett}, R.~M., {et~al.} 2008, \mnras,
  389, 955

\bibitem[{{Patat} {et~al.}(1995){Patat}, {Chugai}, \&
  {Mazzali}}]{Patat1995A&A...299..715P}
{Patat}, F., {Chugai}, N., \& {Mazzali}, P.~A. 1995, \aap, 299, 715

\bibitem[{{Peng} {et~al.}(2026){Peng}, {Benetti}, {Cai}, {Pastorello},
  {Valerin}, {Reguitti}, {Fiore}, {Fang}, {Wang}, {Berton}, {Borsato},
  {Cappellaro}, {Congiu}, {Elias-Rosa}, {Granata}, {Isern}, {La Mura},
  {Ochner}, {Raddi}, {Terreran}, {Tomasella}, {Turatto}, {Yan}, {Pei}, {Wu},
  {Zha}, {Wang}, {Wang}, \& {Pan}}]{Peng2026A&A...705A.104P}
{Peng}, Z.-H., {Benetti}, S., {Cai}, Y.-Z., {et~al.} 2026, \aap, 705, A104

\bibitem[{{Prentice} {et~al.}(2022){Prentice}, {Maguire}, {Siebenaler}, \&
  {Jerkstrand}}]{Prentice2022MNRAS.514.5686P}
{Prentice}, S.~J., {Maguire}, K., {Siebenaler}, L., \& {Jerkstrand}, A. 2022,
  \mnras, 514, 5686

\bibitem[{{Prentice} {et~al.}(2016){Prentice}, {Mazzali}, {Pian}, {Gal-Yam},
  {Kulkarni}, {Rubin}, {Corsi}, {Fremling}, {Sollerman}, {Yaron}, {Arcavi},
  {Zheng}, {Kasliwal}, {Filippenko}, {Cenko}, {Cao}, \&
  {Nugent}}]{Prentice2016MNRAS.458.2973P}
{Prentice}, S.~J., {Mazzali}, P.~A., {Pian}, E., {et~al.} 2016, \mnras, 458,
  2973

\bibitem[{{Rastinejad} {et~al.}(2025){Rastinejad}, {Levan}, {Jonker},
  {Kilpatrick}, {Fryer}, {Sarin}, {Gompertz}, {Liu}, {Eyles-Ferris}, {Fong},
  {Burns}, {Gillanders}, {Mandel}, {Malesani}, {O'Brien}, {Tanvir}, {Ackley},
  {Aryan}, {Bauer}, {Bloemen}, {de Boer}, {Bom}, {Chac{\'o}n}, {Chambers},
  {Chen}, {Chrimes}, {van Dalen}, {D'Elia}, {De Pasquale}, {Fulton}, {Groot},
  {Gupta}, {Hartmann}, {van Hoof}, {Huber}, {Izzo}, {Jacobson-Galan},
  {Jakobsson}, {Kong}, {Laskar}, {Lowe}, {Magnier}, {Maiorano},
  {Martin-Carrillo}, {Mas-Ribas}, {Mata S{\'a}nchez}, {Nicholl}, {Nixon},
  {Oates}, {Paek}, {Palmerio}, {Paris}, {Pieterse}, {Pugliese}, {Quirola
  Vasquez}, {van Roestel}, {Rossi}, {Rouco Escorial}, {Salvaterra},
  {Schneider}, {Smartt}, {Smith}, {Smith}, {Srivastav}, {Torres}, {Ventura},
  {Vreeswijk}, {Wainscoat}, {Yang}, \& {Yang}}]{Rastinejad2025ApJ...988L..13R}
{Rastinejad}, J.~C., {Levan}, A.~J., {Jonker}, P.~G., {et~al.} 2025, \apjl,
  988, L13

\bibitem[{{Rauscher} {et~al.}(2002){Rauscher}, {Heger}, {Hoffman}, \&
  {Woosley}}]{Rauscher2002ApJ...576..323R}
{Rauscher}, T., {Heger}, A., {Hoffman}, R.~D., \& {Woosley}, S.~E. 2002, \apj,
  576, 323

\bibitem[{{Reguitti} {et~al.}(2025){Reguitti}, {Pastorello}, {Smartt},
  {Valerin}, {Pignata}, {Campana}, {Chen}, {Sankar}, {Moran}, {Mazzali},
  {Duarte}, {Salmaso}, {Anderson}, {Ashall}, {Benetti}, {Gromadzki},
  {Guti{\'e}rrez}, {Humina}, {Inserra}, {Kankare}, {Kravtsov}, {Muller-Bravo},
  {Pessi}, {Sollerman}, {Young}, {Chambers}, {de Boer}, {Gao}, {Huber}, {Lin},
  {Lowe}, {Magnier}, {Minguez}, {Smith}, {Smith}, {Srivastav}, {Wainscoat}, \&
  {Benedet}}]{Reguitti2025A&A...698A.129R}
{Reguitti}, A., {Pastorello}, A., {Smartt}, S.~J., {et~al.} 2025, \aap, 698,
  A129

\bibitem[{{Richmond} {et~al.}(1996){Richmond}, {Treffers}, {Filippenko}, \&
  {Paik}}]{Richmond1996AJ....112..732R}
{Richmond}, M.~W., {Treffers}, R.~R., {Filippenko}, A.~V., \& {Paik}, Y. 1996,
  \aj, 112, 732

\bibitem[{{Richmond} {et~al.}(1994){Richmond}, {Treffers}, {Filippenko},
  {Paik}, {Leibundgut}, {Schulman}, \& {Cox}}]{Richmond1994AJ....107.1022R}
{Richmond}, M.~W., {Treffers}, R.~R., {Filippenko}, A.~V., {et~al.} 1994, \aj,
  107, 1022

\bibitem[{{Rodr{\'\i}guez} {et~al.}(2023){Rodr{\'\i}guez}, {Maoz}, \&
  {Nakar}}]{Osmar2023ApJ...955...71R}
{Rodr{\'\i}guez}, {\'O}., {Maoz}, D., \& {Nakar}, E. 2023, \apj, 955, 71

\bibitem[{{Rodr{\'\i}guez} {et~al.}(2024){Rodr{\'\i}guez}, {Nakar}, \&
  {Maoz}}]{Osmar2024Natur.628..733R}
{Rodr{\'\i}guez}, {\'O}., {Nakar}, E., \& {Maoz}, D. 2024, \nat, 628, 733

\bibitem[{{Sahu} {et~al.}(2013){Sahu}, {Anupama}, \&
  {Chakradhari}}]{Sahu2013MNRAS.433....2S}
{Sahu}, D.~K., {Anupama}, G.~C., \& {Chakradhari}, N.~K. 2013, \mnras, 433, 2

\bibitem[{{Sako} {et~al.}(2018){Sako}, {Bassett}, {Becker}, {Brown},
  {Campbell}, {Wolf}, {Cinabro}, {D'Andrea}, {Dawson}, {DeJongh}, {Depoy},
  {Dilday}, {Doi}, {Filippenko}, {Fischer}, {Foley}, {Frieman}, {Galbany},
  {Garnavich}, {Goobar}, {Gupta}, {Hill}, {Hayden}, {Hlozek}, {Holtzman},
  {Hopp}, {Jha}, {Kessler}, {Kollatschny}, {Leloudas}, {Marriner}, {Marshall},
  {Miquel}, {Morokuma}, {Mosher}, {Nichol}, {Nordin}, {Olmstead}, {{\"O}stman},
  {Prieto}, {Richmond}, {Romani}, {Sollerman}, {Stritzinger}, {Schneider},
  {Smith}, {Wheeler}, {Yasuda}, \& {Zheng}}]{Sako2018PASP..130f4002S}
{Sako}, M., {Bassett}, B., {Becker}, A.~C., {et~al.} 2018, \pasp, 130, 064002

\bibitem[{{Sapir} \& {Waxman}(2017)}]{Sapir2017ApJ...838..130S}
{Sapir}, N. \& {Waxman}, E. 2017, \apj, 838, 130

\bibitem[{{Schlafly} \& {Finkbeiner}(2011)}]{Schlafly2011ApJ...737..103S}
{Schlafly}, E.~F. \& {Finkbeiner}, D.~P. 2011, \apj, 737, 103

\bibitem[{{Sharon} \& {Kushnir}(2020)}]{Sharon2020MNRAS.496.4517S}
{Sharon}, A. \& {Kushnir}, D. 2020, \mnras, 496, 4517

\bibitem[{{Shingles} {et~al.}(2021){Shingles}, {Smith}, {Young}, {Smartt},
  {Tonry}, {Denneau}, {Heinze}, {Weiland}, {Flewelling}, {Stalder},
  {Clocchiatti}, {F{\"o}rster}, {Pignata}, {Rest}, {Anderson}, {Stubbs}, \&
  {Erasmus}}]{Shingles2021TNSAN...7....1S}
{Shingles}, L., {Smith}, K.~W., {Young}, D.~R., {et~al.} 2021, Transient Name
  Server AstroNote, 7, 1

\bibitem[{{Sobolev}(1957)}]{Sobolev1957SvA.....1..678S}
{Sobolev}, V.~V. 1957, \sovast, 1, 678

\bibitem[{{Spergel} {et~al.}(2007){Spergel}, {Bean}, {Dor{\'e}}, {Nolta},
  {Bennett}, {Dunkley}, {Hinshaw}, {Jarosik}, {Komatsu}, {Page}, {Peiris},
  {Verde}, {Halpern}, {Hill}, {Kogut}, {Limon}, {Meyer}, {Odegard}, {Tucker},
  {Weiland}, {Wollack}, \& {Wright}}]{Spergel2007ApJS..170..377S}
{Spergel}, D.~N., {Bean}, R., {Dor{\'e}}, O., {et~al.} 2007, \apjs, 170, 377

\bibitem[{{Srinivasaragavan} {et~al.}(2025){Srinivasaragavan}, {Hamidani},
  {Schroeder}, {Sarin}, {Ho}, {Piro}, {Cenko}, {Anand}, {Sollerman}, {Perley},
  {Maeda}, {O'Connor}, {Kuncarayakti}, {Miller}, {Ahumada}, {Vail}, {Duffell},
  {Dastidar}, {Andreoni}, {Bochenek}, {Brennan}, {Carney}, {Chen}, {Freeburn},
  {Gal-Yam}, {Jacobson-Gal{\'a}n}, {Kasliwal}, {Li}, {Li}, {Sravan}, \&
  {Warshofsky}}]{Srinivasaragavan2025ApJ...988L..60S}
{Srinivasaragavan}, G.~P., {Hamidani}, H., {Schroeder}, G., {et~al.} 2025,
  \apjl, 988, L60

\bibitem[{{Stritzinger} {et~al.}(2018{\natexlab{a}}){Stritzinger}, {Anderson},
  {Contreras}, {Heinrich-Josties}, {Morrell}, {Phillips}, {Anais}, {Boldt},
  {Busta}, {Burns}, {Campillay}, {Corco}, {Castellon}, {Folatelli},
  {Gonz{\'a}lez}, {Holmbo}, {Hsiao}, {Krzeminski}, {Salgado}, {Ser{\'o}n},
  {Torres-Robledo}, {Freedman}, {Hamuy}, {Krisciunas}, {Madore}, {Persson},
  {Roth}, {Suntzeff}, {Taddia}, {Li}, \& {Filippenko}}]{Stritzinger2018A&A}
{Stritzinger}, M.~D., {Anderson}, J.~P., {Contreras}, C., {et~al.}
  2018{\natexlab{a}}, \aap, 609, A134

\bibitem[{{Stritzinger} {et~al.}(2018{\natexlab{b}}){Stritzinger}, {Taddia},
  {Burns}, {Phillips}, {Bersten}, {Contreras}, {Folatelli}, {Holmbo}, {Hsiao},
  {Hoeflich}, {Leloudas}, {Morrell}, {Sollerman}, \&
  {Suntzeff}}]{Stritzinger2018A&A...609A.135S}
{Stritzinger}, M.~D., {Taddia}, F., {Burns}, C.~R., {et~al.}
  2018{\natexlab{b}}, \aap, 609, A135

\bibitem[{{Sukhbold} {et~al.}(2016){Sukhbold}, {Ertl}, {Woosley}, {Brown}, \&
  {Janka}}]{Sukhbold2016ApJ...821...38S}
{Sukhbold}, T., {Ertl}, T., {Woosley}, S.~E., {Brown}, J.~M., \& {Janka}, H.-T.
  2016, \apj, 821, 38

\bibitem[{{Taddia} {et~al.}(2018){Taddia}, {Stritzinger}, {Bersten}, {Baron},
  {Burns}, {Contreras}, {Holmbo}, {Hsiao}, {Morrell}, {Phillips}, {Sollerman},
  \& {Suntzeff}}]{Taddia2018A&A...609A.136T}
{Taddia}, F., {Stritzinger}, M.~D., {Bersten}, M., {et~al.} 2018, \aap, 609,
  A136

\bibitem[{{Taggart} \& {Perley}(2021)}]{Taggart2021MNRAS.503.3931T}
{Taggart}, K. \& {Perley}, D.~A. 2021, \mnras, 503, 3931

\bibitem[{{Tanaka} {et~al.}(2009){Tanaka}, {Yamanaka}, {Maeda}, {Kawabata},
  {Hattori}, {Minezaki}, {Valenti}, {Della Valle}, {Sahu}, {Anupama},
  {Tominaga}, {Nomoto}, {Mazzali}, \& {Pian}}]{Tanaka2009ApJ...700.1680T}
{Tanaka}, M., {Yamanaka}, M., {Maeda}, K., {et~al.} 2009, \apj, 700, 1680

\bibitem[{{Taubenberger} {et~al.}(2011){Taubenberger}, {Navasardyan}, {Maurer},
  {Zampieri}, {Chugai}, {Benetti}, {Agnoletto}, {Bufano}, {Elias-Rosa},
  {Turatto}, {Patat}, {Cappellaro}, {Mazzali}, {Iijima}, {Valenti},
  {Harutyunyan}, {Claudi}, \& {Dolci}}]{Taubenberger2011MNRAS.413.2140T}
{Taubenberger}, S., {Navasardyan}, H., {Maurer}, J.~I., {et~al.} 2011, \mnras,
  413, 2140

\bibitem[{{Taubenberger} {et~al.}(2009){Taubenberger}, {Valenti}, {Benetti},
  {Cappellaro}, {Della Valle}, {Elias-Rosa}, {Hachinger}, {Hillebrandt},
  {Maeda}, {Mazzali}, {Pastorello}, {Patat}, {Sim}, \&
  {Turatto}}]{Taubenberger2009MNRAS.397..677T}
{Taubenberger}, S., {Valenti}, S., {Benetti}, S., {et~al.} 2009, \mnras, 397,
  677

\bibitem[{{Thielemann} {et~al.}(1996){Thielemann}, {Nomoto}, \&
  {Hashimoto}}]{Thielemann1996ApJ...460..408T}
{Thielemann}, F.-K., {Nomoto}, K., \& {Hashimoto}, M.-A. 1996, \apj, 460, 408

\bibitem[{{Thomas}(2013)}]{Thomas2013ascl.soft08008T}
{Thomas}, R.~C. 2013, {SYN++: Standalone SN spectrum synthesis}, Astrophysics
  Source Code Library, record ascl:1308.008

\bibitem[{{Thomas} {et~al.}(2011){Thomas}, {Nugent}, \& {Meza}}]{SYN}
{Thomas}, R.~C., {Nugent}, P.~E., \& {Meza}, J.~C. 2011, \pasp, 123, 237

\bibitem[{{Tsvetkov} {et~al.}(2009){Tsvetkov}, {Volkov}, {Baklanov},
  {Blinnikov}, \& {Tuchin}}]{Tsvetkov2009PZ.....29....2T}
{Tsvetkov}, D.~Y., {Volkov}, I.~M., {Baklanov}, P., {Blinnikov}, S., \&
  {Tuchin}, O. 2009, Peremennye Zvezdy, 29, 2

\bibitem[{{Tsvetkov} {et~al.}(2012){Tsvetkov}, {Volkov}, {Sorokina},
  {Blinnikov}, {Pavlyuk}, \& {Borisov}}]{Tsvetkov2012PZ.....32....6T}
{Tsvetkov}, D.~Y., {Volkov}, I.~M., {Sorokina}, E., {et~al.} 2012, Peremennye
  Zvezdy, 32, 6

\bibitem[{{Valenti} {et~al.}(2011){Valenti}, {Fraser}, {Benetti}, {Pignata},
  {Sollerman}, {Inserra}, {Cappellaro}, {Pastorello}, {Smartt}, {Ergon},
  {Botticella}, {Brimacombe}, {Bufano}, {Crockett}, {Eder}, {Fugazza},
  {Haislip}, {Hamuy}, {Harutyunyan}, {Ivarsen}, {Kankare}, {Kotak}, {Lacluyze},
  {Magill}, {Mattila}, {Maza}, {Mazzali}, {Reichart}, {Taubenberger},
  {Turatto}, \& {Zampieri}}]{Valenti2011MNRAS.416.3138V}
{Valenti}, S., {Fraser}, M., {Benetti}, S., {et~al.} 2011, \mnras, 416, 3138

\bibitem[{{van Loon}(2025)}]{van2025Galax..13...72V}
{van Loon}, J.~T. 2025, Galaxies, 13, 72

\bibitem[{{Wang} {et~al.}(2017){Wang}, {Yu}, {Liu}, {Wang}, {Han}, {Xu}, {Dai},
  {Qiu}, \& {Wei}}]{Wang2017ApJ...837..128W}
{Wang}, L.~J., {Yu}, H., {Liu}, L.~D., {et~al.} 2017, \apj, 837, 128

\bibitem[{{Wang} {et~al.}(2024){Wang}, {Pastorello}, {Maeda}, {Reguitti},
  {Cai}, {Andrew Howell}, {Benetti}, {Buckley}, {Cappellaro}, {Carini},
  {Cartier}, {Chen}, {Elias-Rosa}, {Fang}, {Gal-Yam}, {Gangopadhyay},
  {Gromadzki}, {Gan}, {Hiramatsu}, {Hu}, {Inserra}, {McCully}, {Nicholl},
  {Olivares E.}, {Pignata}, {Pineda-Garc{\'\i}a}, {Pursiainen}, {Ragosta},
  {Rau}, {Roy}, {Sollerman}, {Tartaglia}, {Terreran}, {Valerin}, {Wang},
  {Wang}, {Young}, {Aryan}, {Bronikowski}, {Concepcion}, {Galbany}, {Lin},
  {Melandri}, {Petrushevska}, {Ramirez}, {Shi}, {Warwick}, {Zhang}, {Wang},
  {Wang}, \& {Zhu}}]{Wang2024A&A...691A.156W}
{Wang}, Z.~Y., {Pastorello}, A., {Maeda}, K., {et~al.} 2024, \aap, 691, A156

\bibitem[{{Woosley}(2010)}]{Woosley2010ApJ...719L.204W}
{Woosley}, S.~E. 2010, \apjl, 719, L204

\bibitem[{{Woosley} \& {Heger}(2007)}]{Woosley2007PhR}
{Woosley}, S.~E. \& {Heger}, A. 2007, \physrep, 442, 269

\bibitem[{{Woosley} \& {Weaver}(1995)}]{Woosley1995ApJS..101..181W}
{Woosley}, S.~E. \& {Weaver}, T.~A. 1995, \apjs, 101, 181

\bibitem[{{Yaron} \& {Gal-Yam}(2012)}]{Yaron2012PASP..124..668Y}
{Yaron}, O. \& {Gal-Yam}, A. 2012, \pasp, 124, 668

\bibitem[{Young(2020)}]{Young_plot_atlas_fp}
Young, D.~R. 2020, {plot\_atlas\_fp.py}

\bibitem[{{Zhao} {et~al.}(2026){Zhao}, {Benetti}, {Cai}, {Pastorello},
  {Elias-Rosa}, {Reguitti}, {Valerin}, {Wang}, {Cappellaro}, {Feng}, {Fiore},
  {Fitzpatrick}, {Fraser}, {Isern}, {Kankare}, {Kravtsov}, {Kumar},
  {Lundqvist}, {Matilainen}, {Mattila}, {Mazzali}, {Moran}, {Ochner}, {Peng},
  {Reynolds}, {Salmaso}, {Srivastav}, {Stritzinger}, {Taubenberger},
  {Tomasella}, {Vink{\'o}}, {Wheeler}, {Williams}, {Pei}, {Yang}, {Liu}, {Liu},
  \& {Yang}}]{Zhao2025arXiv251209384Z}
{Zhao}, J.-W., {Benetti}, S., {Cai}, Y.-Z., {et~al.} 2026, \aap, 706, A271

\bibitem[{{Zhu} {et~al.}(2025){Zhu}, {Zheng}, \&
  {Zhang}}]{Zhu2025MNRAS.544L.139Z}
{Zhu}, J.-P., {Zheng}, J.-H., \& {Zhang}, B. 2025, \mnras, 544, L139

\end{thebibliography}
\begin{appendix}
\section{Acknowledgements}
\begin{small}
We gratefully thank the anonymous referee for his/her insightful comments and suggestions that improved the paper.
We thank Luc Dessart for his invaluable contributions to the interpretation of  light curves and spectra, as well as his insightful comments and revisions that significantly improved this work.
This work is supported by the National Natural Science Foundation of China (No. 12303054), the National Key Research and Development Program of China (Grant No. 2024YFA1611603), the Yunnan Fundamental Research Projects (Grant Nos. 202401AU070063, 202501AS070078), and the International Centre of Supernovae, Yunnan Key Laboratory (No. 202302AN360001). Y.-Z.C. and A.R. acknowledge financial support from the SOXS project (PI S. Campana).
AP, AR, SB, EC, NER and  LT acknowledge support from the PRIN-INAF 2022, "Shedding light on the nature of gap transients: from the observations to the models".
EC acknowledges support from MIUR, PRIN 2020 (METE, grant 2020KB33TP).
N.E.R. also acknowledges support from the Spanish Ministerio de Ciencia e Innovaci\'on (MCIN) and the Agencia Estatal de Investigaci\'on (AEI) 10.13039/501100011033 under the program Unidad de Excelencia Mar\'ia de Maeztu CEX2020-001058-M.
T.K. acknowledges support from the Research Council of Finland project 360274.
T.M.R. is part of the Cosmic Dawn Center (DAWN), which is funded by the Danish National Research Foundation under grant DNRF140. 
T.M.R and S. Mattila acknowledge support from the Research Council of Finland project 350458.
M.D.S. is funded by the Independent Research Fund Denmark (IRFD, grant number  10.46540/2032-00022B).
S.-P. Pei is supported by the  Science and Technology Foundation of Guizhou Province (QKHJCMS[2026]752). 
Y.-J. Yang is supported by the National Natural Science Foundation of China (Grants No. 12305066).
We thank Luhan Li for providing some public datasets of comparison objects. 

Based on observations obtained with the Cima Ekar 1.82 m Telescopio Copernico, installed at the INAF (Istituto Nazionale di Astrofisica) - Astronomical Observatory of Padova, Italy.
Based on observations made with the Nordic Optical Telescope (NOT), owned in collaboration by the University of Turku and Aarhus University, and operated jointly by Aarhus University, the University of Turku, and the University of Oslo, representing Denmark, Finland, and Norway, the University of Iceland, and Stockholm University at the Observatorio del Roque de los Muchachos, La Palma, Spain, of the Instituto de Astrofisica de Canarias.

We acknowledge ESA Gaia, DPAC and the Photometric Science Alerts Team (\url{http://gsaweb.ast.cam.ac.uk/alerts}).
This work has made use of data from the Asteroid Terrestrial-impact Last Alert System (ATLAS) project. ATLAS is primarily funded to search for near-Earth objects through NASA grants NN12AR55G, 80NSSC18K0284, and 80NSSC18K1575; byproducts of the NEO search include images and catalogs from the survey area. The ATLAS science products have been made possible through the contributions of the University of Hawaii Institute for Astronomy, the Queen's University Belfast, STScI, and the South African Astronomical Observatory, and The Millennium Institute of Astrophysics (MAS), Chile.
We thank Las Cumbres Observatory and its staff for their continued support of ASAS-SN. ASAS-SN is funded in part by the Gordon and Betty Moore Foundation through grants GBMF5490 and GBMF10501 to the Ohio State University, and also funded in part by the Alfred P. Sloan Foundation grant G-2021-14192. Development of ASAS-SN has been supported by NSF grant AST-0908816, the Mt. Cuba Astronomical Foundation, the Center for Cosmology and AstroParticle Physics at the Ohio State University, the Chinese Academy of Sciences South America Center for Astronomy (CAS-SACA), and the Villum Foundation.
This research has made use of the NASA/IPAC Extragalactic Database (NED), which is operated by the Jet Propulsion Laboratory, California Institute of Technology, under contract with the National Aeronautics and Space Administration.
{\sc iraf} was distributed by the National Optical Astronomy Observatory, which was managed by the Association of Universities for Research in Astronomy (AURA), Inc., under a cooperative agreement with the U.S. NSF.

\section{Observations and data reductions}
\label{section:data}
\subsection{Photometric data}

We performed optical photometric follow-up of SN~2017ati in multiple bands, covering the Sloan $ugriz$ and Johnson--Bessell $BV$ systems, starting shortly after its discovery. The observations were conducted using several ground-based facilities. The 1.82 m Copernico Telescope, fitted with the Asiago Faint Object Spectrograph and Camera (AFOSC), is operated by the INAF Padova Astronomical Observatory at the Asiago Observatory in Italy. The 2.56 m Nordic Optical Telescope (NOT) was employed with the Alhambra Faint Object Spectrograph and Camera (ALFOSC).

The optical photometry from the ground-based facilities was reduced using the {\sl ecsnoopy}\footnote{{\sl ecsnoopy} is a software for SN photometry employing PSF fitting and/or template subtraction, developed by E. Cappellaro. Documentation is available at \url{https://sngroup.oapd.inaf.it/ecsnoopy.html}.} 
pipeline, adopting the data-processing methodology outlined by \citet[][]{Cai2018MNRAS.480.3424C}.
In addition to our own observations, we collected publicly available archival photometric data from several surveys, including the Asteroid Terrestrial-impact Last Alert System (ATLAS), \textit{Gaia}\footnote{\url{http://gsaweb.ast.cam.ac.uk/alerts}}, and the All-Sky Automated Survey for Supernovae (ASAS-SN). The ATLAS orange ($o$)- and cyan ($c$)-band light curves were extracted using the ATLAS Forced Photometry service\footnote{\url{https://fallingstar-data.com/forcedphot/}} \citep{Shingles2021TNSAN...7....1S}. 

The ATLAS photometric measurements were further processed using a script described by \citet{Young_plot_atlas_fp}, which applies a rolling-window algorithm to identify and remove spurious points and bins the remaining data into 1-d intervals. Additional $V$-band photometry was retrieved from the ASAS-SN Sky Patrol\footnote{\url{https://asas-sn.osu.edu}} \citep{Hart2023arXiv230403791H}.
The final calibrated optical photometry of SN~2017ati has been made publicly available through the Strasbourg Astronomical Data Centre (CDS).

\subsection{Spectroscopic data}

Spectroscopic follow-up observations of SN~2017ati were obtained with the 1.82m Copernico Telescope equipped with AFOSC and the 2.56m NOT using ALFOSC. The spectra acquired with both Copernico/AFOSC and NOT/ALFOSC were reduced with the dedicated \textsc{Foscgui}\footnote{{\sc Foscgui} is a graphic user interface aimed at extracting SN spectroscopy and photometry obtained with FOSC-like instruments. It was developed by E. Cappellaro. A package description can be found at \url{http://sngroup.oapd.inaf.it/foscgui.html}.} pipeline.

All raw spectroscopic frames were reduced following standard procedures using \textsc{iraf}. The basic reduction steps included bias subtraction, overscan correction, flat-field correction, and image trimming, consistent with those applied to the imaging data. One-dimensional spectra were then extracted from the two-dimensional frames.
Wavelength calibration was performed using arc-lamp exposures obtained during the same observing nights, while flux calibration relied on observations of spectrophotometric standard stars. Telluric absorption features, primarily due to O$_2$ and H$_2$O, were corrected using the standard-star spectra. The reliability of the flux calibration was verified by comparing synthetic photometry derived from the spectra with contemporaneous broadband photometric measurements.
A summary of the instruments and observational setups employed for the spectroscopic data is provided in Table~\ref{tab:spec} in Appendix~\ref{SpecInfo}.

\end{small}
\section{Supplementary tables}
\label{SpecInfo}

\begin{table*}[ht]
\caption{Log of spectroscopic observations of SN 2017ati.}
\label{tab:spec}
\centering

\begin{tabular}{cccccccc}
\toprule
\hline  
Date &MJD & Phase$^a$&Telescope+Instrument & Grism/Grating+Slit & Spectral range & Resolution & Exp.time  \\
 &  & [days]& &  & [\AA]    & [\AA]           & [s] \\
\hline
2017-02-27  & 57811.8 & +27.3 &Copernico1.82m+AFOSC & GR04+1.69" & $3430-8110$ & $13.7$ & $1800$ \\
2017-03-09  & 57826.0 & +41.5 &NOT+ALFOSC      & gm4+1.3"&  $3300-8860$   & 17.7  & $1800$ \\
2017-03-17  & 57829.8 & +45.3 &Copernico1.82m+AFOSC & VPH7+1.69" &  $3600-9150$   &  14.6   & $1800$ \\
2017-03-25  & 57837.0 & +52.5 &Copernico1.82m+AFOSC & VPH6+1.69" &  $3670-9150$   &  14.4   & $1800$ \\
2017-03-28  & 57840.0 & +55.5 &Copernico1.82m+AFOSC & VPH7+1.69" &  $3600-9140$   &  15.1   & $2400$ \\
2017-04-06  & 57849.1 & +64.6 &NOT+ALFOSC      & gm4+1.3"&  $3560-9500$  &  17.7  & $1800$ \\
2017-04-21  & 57864.9 & +80.4 &NOT+ALFOSC      & gm4+1.0"&  $3650-9400$  &  13.7  & $1250$ \\
2017-05-08  & 57881.9 & +97.4 &NOT+ALFOSC      & gm4+1.3"&  $3800-9500$  &  17.9  & $1800$ \\
2017-05-30  & 57903.9 & +119.4&NOT+ALFOSC      & gm4+1.0"&  $3660-9540$  &  13.9  & $1800$ \\
2017-06-28  & 57932.9 & +148.4&NOT+ALFOSC      & gm4+1.0"&  $3640-9540$  &  14.0  & $3200$ \\

\hline\hline 
\multicolumn{8}{l}{{$^a$ Phases are calculated relative to the explosion epoch (MJD = 57784.5) in the reference frame of the observer.}} 
\\

\end{tabular}
\end{table*}

\begin{table*}[ht]
\centering
\caption{Parameters for the comparison sample of SNe IIb.}
\resizebox{\textwidth}{!}{
    \begin{tabular}{ccccccccc}
        \hline\hline
        SN IIb & Explosion Date & Redshift & Distance & $E(B-V)_\mathrm{Gal}$ & $E(B-V)_\mathrm{Host}$& $M^{peak}_{r/R}$$^a$ &$L_{peak}$ $^a$ & References \\
         & [MJD] & $z$ & [Mpc] & [mag] & [mag]& [mag] &[erg s$^{-1}$]& \\
        \hline
        1993J  & 49072.0 & -0.000113 & 2.9 & 0.069 & 0.11&$-17.73 \pm 0.33$ &$1.43 \times 10^{42}$& 1 \\
        2008ax & 54528.8 & 0.00456 & 9.6$\pm$1.3 & 0.022 & 0.278 &$-17.28 \pm 0.46$ &$1.30 \times 10^{42}$& 2 \\
        2011dh & 55712.5 & 0.001638 & 8.03$\pm$0.77 & 0.03 & 0.04&$-17.50 \pm 0.22$  &$8.93 \times 10^{41}$& 3 \\
        2011fu & 55824.5 & 0.001845 & 74.5$\pm$5.2 & 0.068 & 0.035 &$-17.89 \pm 0.12$ &$1.89 \times 10^{42}$& 4 \\
        2013df & 56449.5 & 0.00239   & 21.38 $\pm$ 2.95&0.017 & 0.081&$-17.00 \pm 0.08$ &$7.51 \times 10^{41}$  &5\\
        2015as & 57332.0 & 0.0036 & 19.2$\pm1.4$ & 0.008 & 0 &$\sim-17.28$ &$8.49 \times 10^{41}$& 6 \\ 
        2016gkg & 57651.2 & 0.0049 & 21.8 & 0.0166 & 0.09 &$\sim-17.53$ &$1.05 \times 10^{42}$& 7 \\
        2017ckj & 57836.6 & 0.037 & 158.1$\pm11.1$ & 0.013 & 0 &$-18.46 \pm 0.07$ &$6.59 \times 10^{42}$&  8\\
        2018gk & 58130.1 & 0.031010 & 140.5$\pm$2.3 & 0.0086 & 0 &$-19.64 \pm 0.24$ &\(  1.43 \times 10^{43}\)& 9\\
        2020acat & 59192.0 & 0.007932 & 35.3$\pm$4.4 & 0.0207 & 0 &$-16.55 \pm 0.35$ &$1.48 \times 10^{42}$& 10 \\
        2021bxu & 59246.3 & 0.0178 & 72$\pm5$ & 0.014 & 0 &$\sim 15.93$ &$  3.79 \times 10^{41}$& 11\\
        2022ngb& 59749.9&0.00965&32.2$\pm2.8$&0.085&0.085&$-16.55 \pm 0.35$ &$  7.85 \times 10^{41}$&12\\
        2024abfo&60628.3 & 0.003512 & 10.85$\pm0.53$ &0.0097 &0&$-16.50 \pm 0.10$ &$  7.23 \times 10^{41}$& 13\\
        2017ati&57784.5& 0.01696&70.80$\pm5.20$&0.104&0&$-18.48 \pm 0.16$ &$3.00 \times 10^{42}$&14\\
        \hline\hline
    \end{tabular}}
    \begin{minipage}{\textwidth}
\small
$^a$ Peak magnitudes and luminosities refer to the maximum. 
\end{minipage}
\begin{flushleft}
    References: 
    1= \citet{Richmond1994AJ....107.1022R}, \citet{Barbon1995A&AS..110..513B}, \citet{Richmond1996AJ....112..732R}, 2= \citet{Pastorello2008MNRAS.389..955P}, \citet{Tsvetkov2009PZ.....29....2T}, \citet{Taubenberger2011MNRAS.413.2140T}, 3= \citet{Tsvetkov2012PZ.....32....6T}, \citet{Sahu2013MNRAS.433....2S}, \citet{Ergon2014A&A...562A..17E}, 4= \citet{Morales2015MNRAS.454...95M}, 5=\citet{Morales2014MNRAS.445.1647M}, 6=\citet{Gangopadhyay2018MNRAS.476.3611G}, 7= \citet{Arcavi2017ApJ...837L...2A}, \citet{Bersten2018Natur.554..497B}, 8=\citet{Li2025A&A...704A.233L}, 9= \citet{Bose2021MNRAS.503.3472B},  10= \citet{Medler2022MNRAS.513.5540M}, 11= \citet{Desai2023MNRAS.524..767D}, 12=\citet{Zhao2025arXiv251209384Z} , 13=\citet{Reguitti2025A&A...698A.129R}, 14=This Work.
    \end{flushleft}
    \label{tab:SNe_IIb}
\end{table*}

\begin{table}[htbp]
\centering
\caption{Estimated \({}^{56}\mathrm{Ni}\) mass from different methods.}
\label{tab:Ni}
\resizebox{\columnwidth}{!}{
\begin{tabular}{lc}
\hline\hline
Method & $M_{\mathrm{Ni}}$ ($M_\odot$) \\
\hline
Arnett, peak, pseudo-bol & $0.40 \pm 0.09$ \\
\citet{Clocchiatti1997ApJ...491..375C}, peak, pseudo-bol & $0.29^{+0.05}_{-0.04}$ \\
\citet{Osmar2023ApJ...955...71R}, peak, pseudo-bol$^a$& $0.18^{+0.05}_{-0.04}$\\
1987A, tail, pseudo-bol & $0.47 \pm 0.13$ \\
\texttt{MOSFiT}, Pure $^{56}$Ni & $0.37^{+0.07}_{-0.10}$ \\
\texttt{MOSFiT}, $^{56}$Ni + CSM & $0.49_{-0.12}^{+0.19}$ \\
\texttt{MOSFiT}, $^{56}$Ni + Magnetar & $0.21^{+0.08}_{-0.12}$ \\
\hline\hline
\end{tabular}}
\begin{flushleft}
$^a$ $\log\left(L_{\mathrm{peak}}/M_{\mathrm{Ni}}\right) - b\,\Delta m_{15}(\mathrm{bol})
= a - \log\left(t_{L}^{\mathrm{peak}}/10\,\mathrm{d}\right)$, \\
$a = 0.571 \pm 0.020$, $b = 0.326 \pm 0.028$. \\
For SN~2017ati, adopting $L_{\mathrm{peak}} = 3\times10^{42}\,\mathrm{erg\,s^{-1}}$ and 
$t_{\mathrm{peak}} = 29.1\,\mathrm{d}$, we obtain 
$\log(M_{\mathrm{Ni}}/M_{\odot}) = -0.745 \pm 0.100$~dex.
\end{flushleft}
\end{table}

\begin{table}
\caption{Marginalised posteriors for the \texttt{MOSFiT} model of SN~2017ati.}
\label{tab:MOSFiT}
\centering
\resizebox{\columnwidth}{!}{
  \begin{tabular}{ccccc}
  \hline\hline
  Parameter &$^{56}$Ni &$^{56}$Ni +CSM& $^{56}$Ni+Magnetar& Units\\
  \hline
$B$ & -- &-- &$13.18^{+4.60}_{-4.27}$&$10^{14} G$\\
$P_{spin}$&-- &-- & $28.18^{+2.72}_{-6.3}$ & ms \\
$n_{\rm H,\mathrm{host}}$ & -- & $10.91^{+1.39}_{-0.68}$ &--& $10^{20}$ cm$^{-2}$ \\
$E_{\rm k}$ & --& $8.71^{+2.68}_{-1.79}$ &--& 10$^{50}$ erg \\
$M_{\rm CSM}$ & --& $0.13^{+0.10}_{-0.05}$ &--& M$_\odot$ \\
$R_0$ & --& $3.82^{+4.65}_{2.15}$ &--& AU \\
$\rho_{0}$ & --& $2.33^{+8.61}_{-1.81}$ &--& $10^{-10}$ g\,cm$^{-3}$ \\
$M_{\rm ej}$ &$1.70^{+0.44}_{-0.38}$ &$2.18^{+0.29}_{-0.57}$ & $1.82^{+0.53}_{-0.59}$ & M$_\odot$ \\
$M_{\mathrm{^{56}Ni}}$&$0.37_{-0.10}^{+0.07}$ &$0.49_{-0.12}^{+0.19}$ &$0.21_{-0.12}^{+0.08}$&M$_{\odot}$ \\
$\kappa_{\rm \gamma}$&$0.06^{+0.03}_{-0.01}$ &$0.27^{+0.11}_{-0.08}$ & $0.33^{+0.12}_{-0.05}$ & cm$^2$\,g$^{-1}$ \\
$T_{\rm min}$ &$4.47^{+0.21}_{-0.20}$ &$5.17^{+0.11}_{-0.13}$ & $4.57^{+0.11}_{-0.10}$ &$10^{3}$ K \\
$\sigma$ &$0.37^{+0.03}_{-0.02}$ &$0.18^{+0.01}_{-0.01}$& $0.17^{+0.01}_{-0.02}$ &     \\
$v_{ej} $ &$3.63^{+0.26}_{-0.24}$&$8.18^{+2.71}_{-1.88}$ & $7.24^{+0.70}_{-0.64}$ &$10^{3}$ km s$^{-1}$\\
$t_{exp}$ &$-33.97^{+1.55}_{-1.03}$ &$-3.62^{+0.46}_{-0.53}$ & $ -7.36^{+1.04}_{-0.99}$ & days  \\
\hline
\hline
\end{tabular}}
\end{table}

\newpage
\section{Supplementary figures}
\label{sect:supfigures}

\begin{figure}
\centering
    \includegraphics[width=0.5\textwidth]{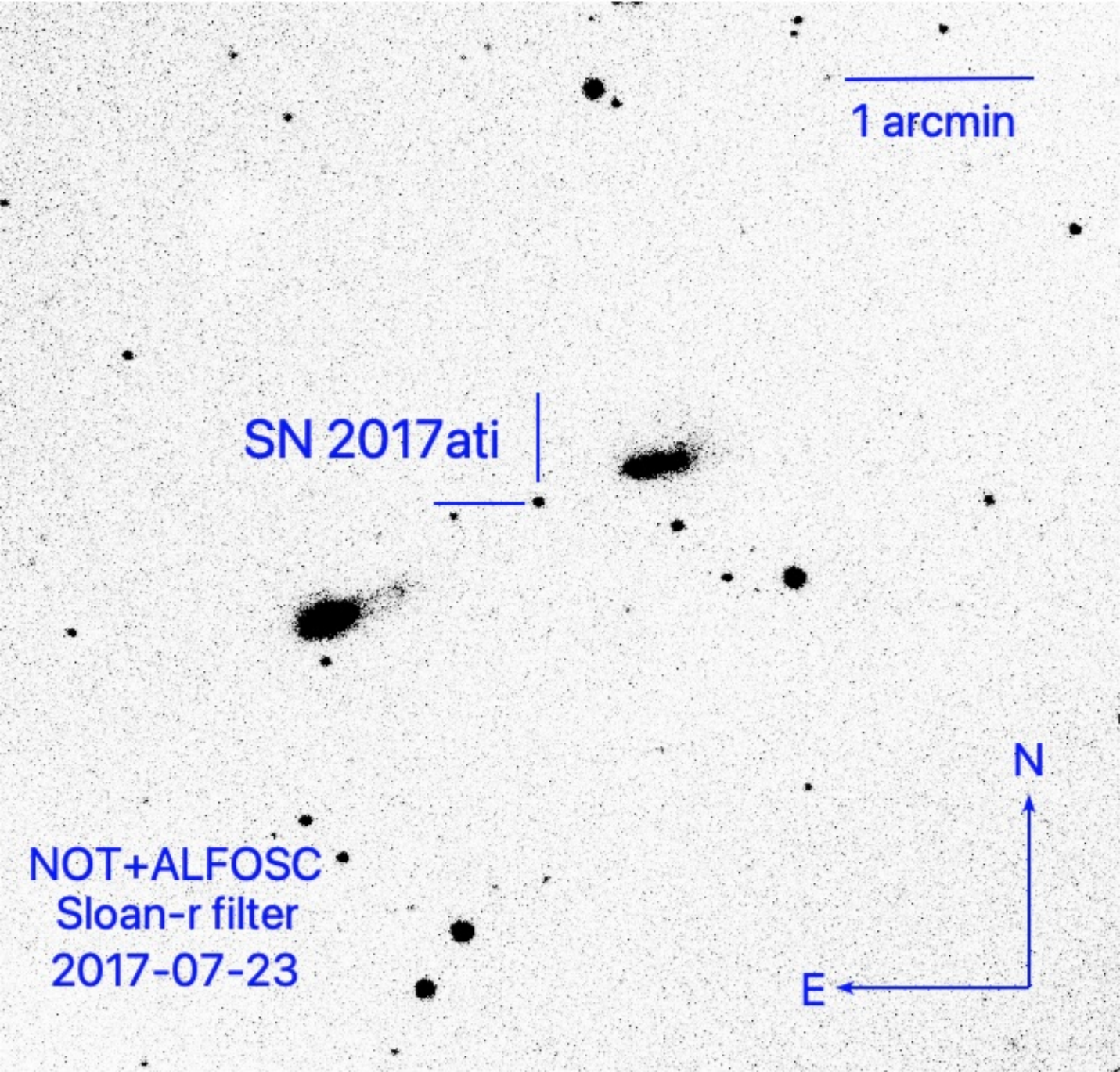}
    \caption{An image showing the location of SN 2017ati, obtained on July 23, 2017, using the $r$-Sloan filter with the NOT/ALFOSC. The orientation and scale are included.}
    \label{fig:location}
\end{figure}

\begin{figure}[htbp]
\centering
\includegraphics[width=0.5\textwidth]{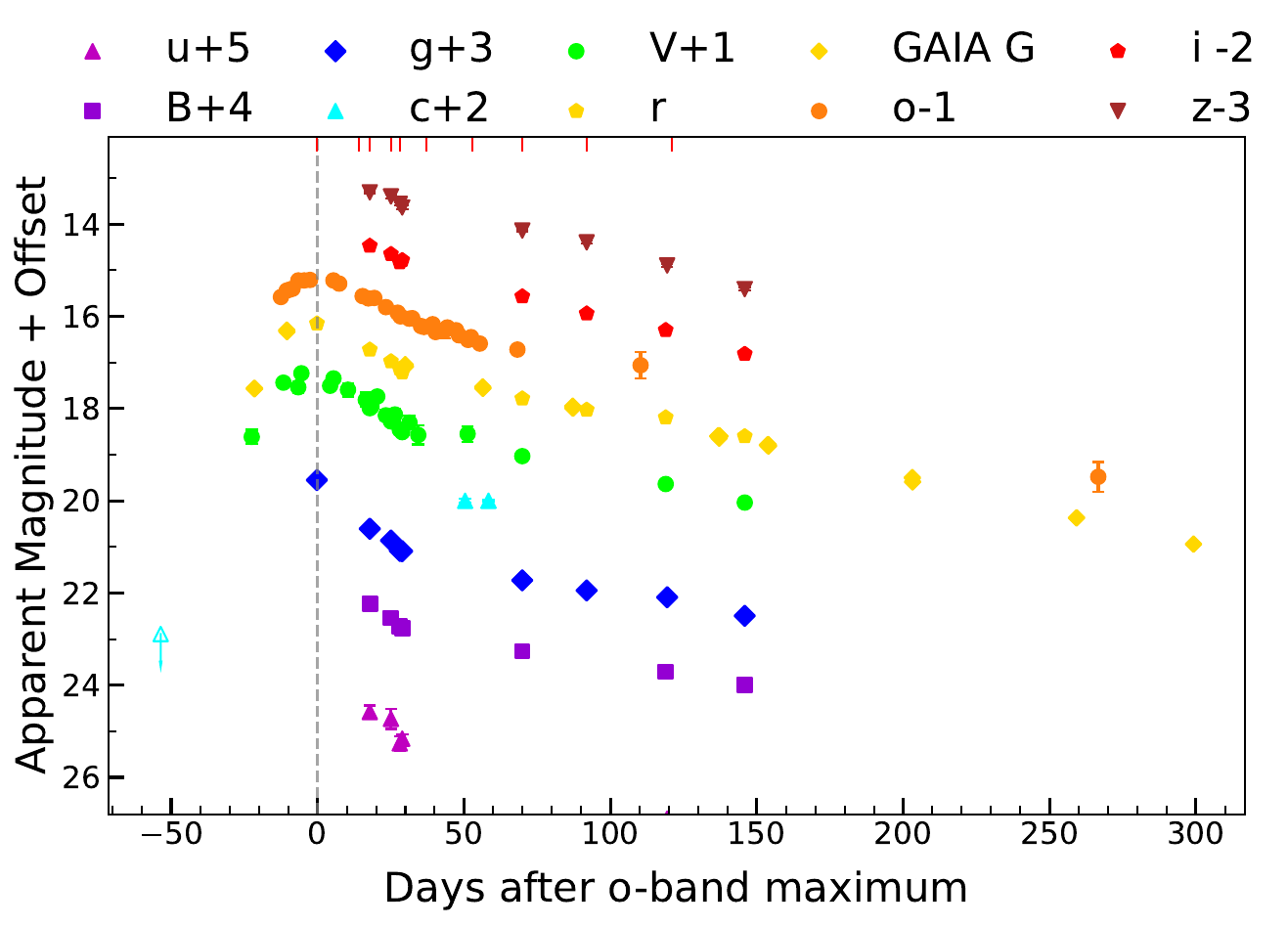}
\caption{Multi-band light curves of SN~2017ati. The dashed vertical line marks the reference epoch, corresponding to the maximum light in the $o$ band. Epochs of our spectroscopic observations are indicated by solid red vertical lines at the top. Upper limits are represented by empty symbols with downward arrows. For clarity, the light curves have been vertically offset by constant values, as indicated in the legend. In most cases, the magnitude uncertainties are smaller than the sizes of the plotted symbols.}
\label{fig:light_curve}
\end{figure}

\begin{figure}[ht]
    \centering
    \includegraphics[width=0.5\textwidth]{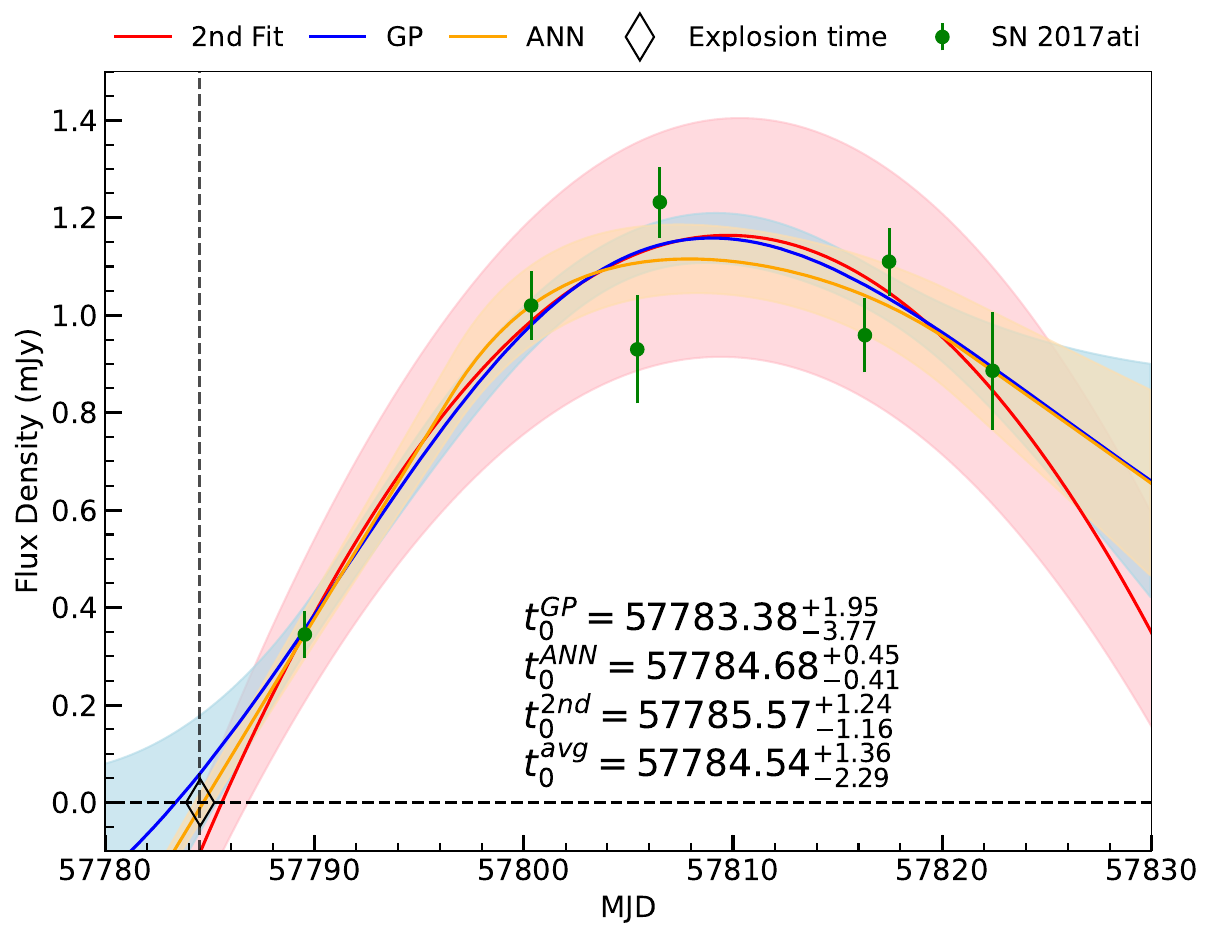} 
    \includegraphics[width=0.5\textwidth]{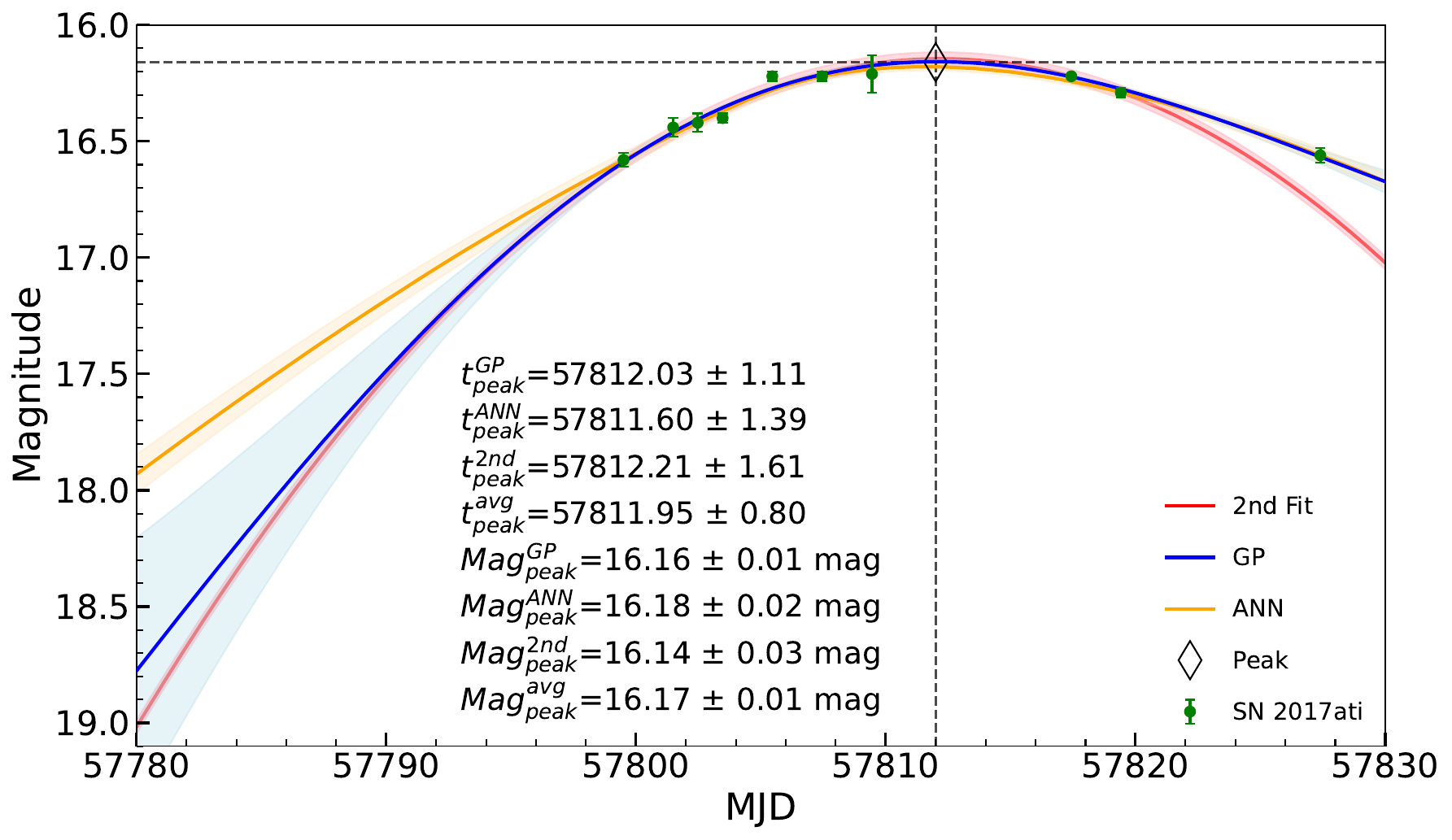}
    \caption{The fitted explosion and peak epochs of SN~2017ati are shown. \emph{Upper panel}: the ASAS-SN $V$-band light curve in flux space (mJy) is fitted with a solid line, and the shaded region represents the $3\sigma$ uncertainty of the explosion epoch. \emph{Lower panel}: the $o$-band light curve is fitted with a solid line, with the shaded region indicating the $3\sigma$ uncertainty of the peak epoch.}
    \label{fig:explosion+peak}
\end{figure}

\begin{figure*}[ht]
    \centering
    \includegraphics[width=1\textwidth]{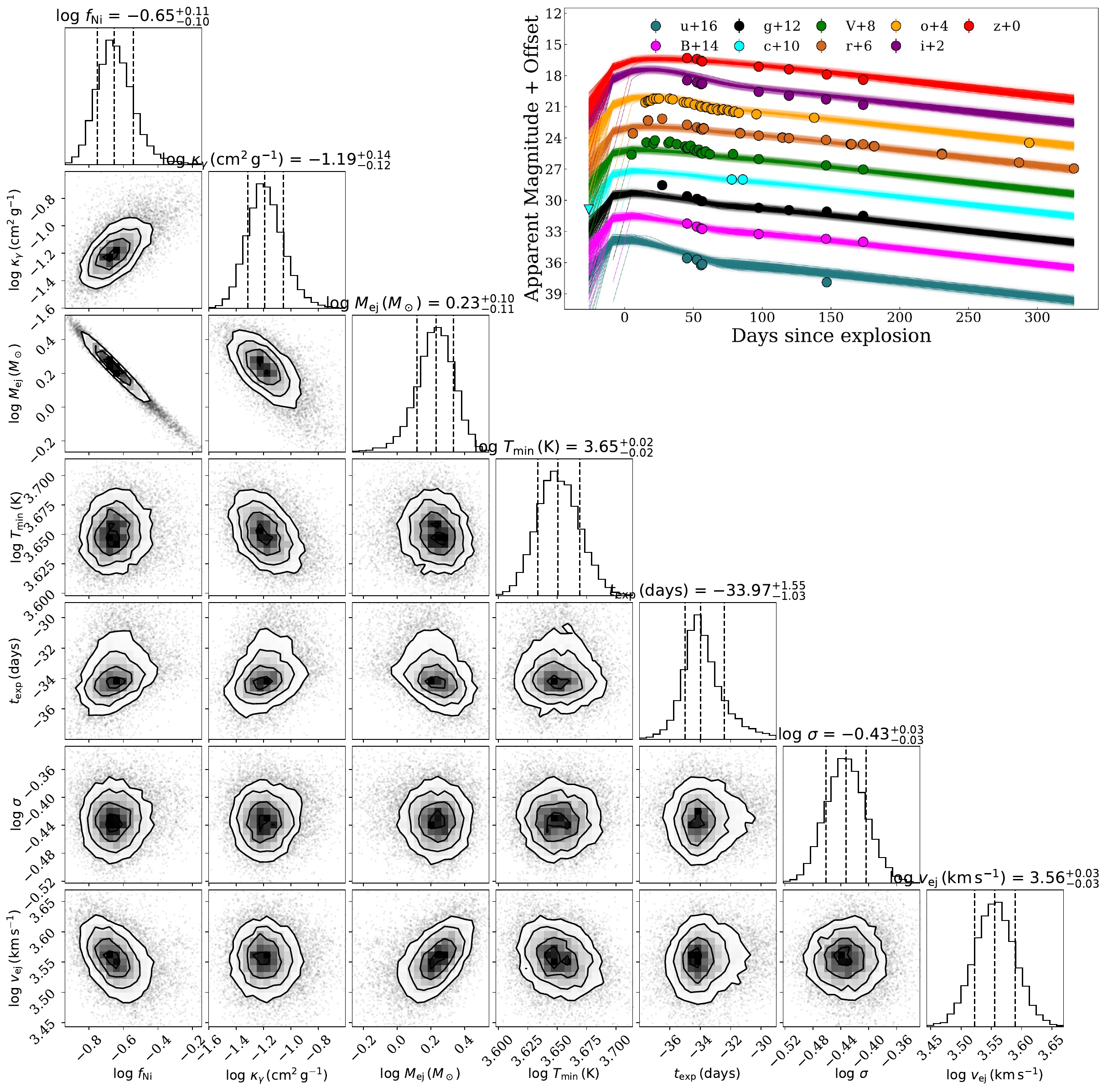}
    \caption{Multi-band light curve of SN~2017ati fitted with the radioactive decay model by \texttt{MOSFiT}, together with the parameter posteriors from the MCMC sampling displayed in the corner plot.The relevant parameters are listed in Table \ref{tab:MOSFiT}.}
    \label{fig:MOSFIT-Ni}
\end{figure*}

\begin{figure*}[ht]
    \centering
    \includegraphics[width=1\textwidth]{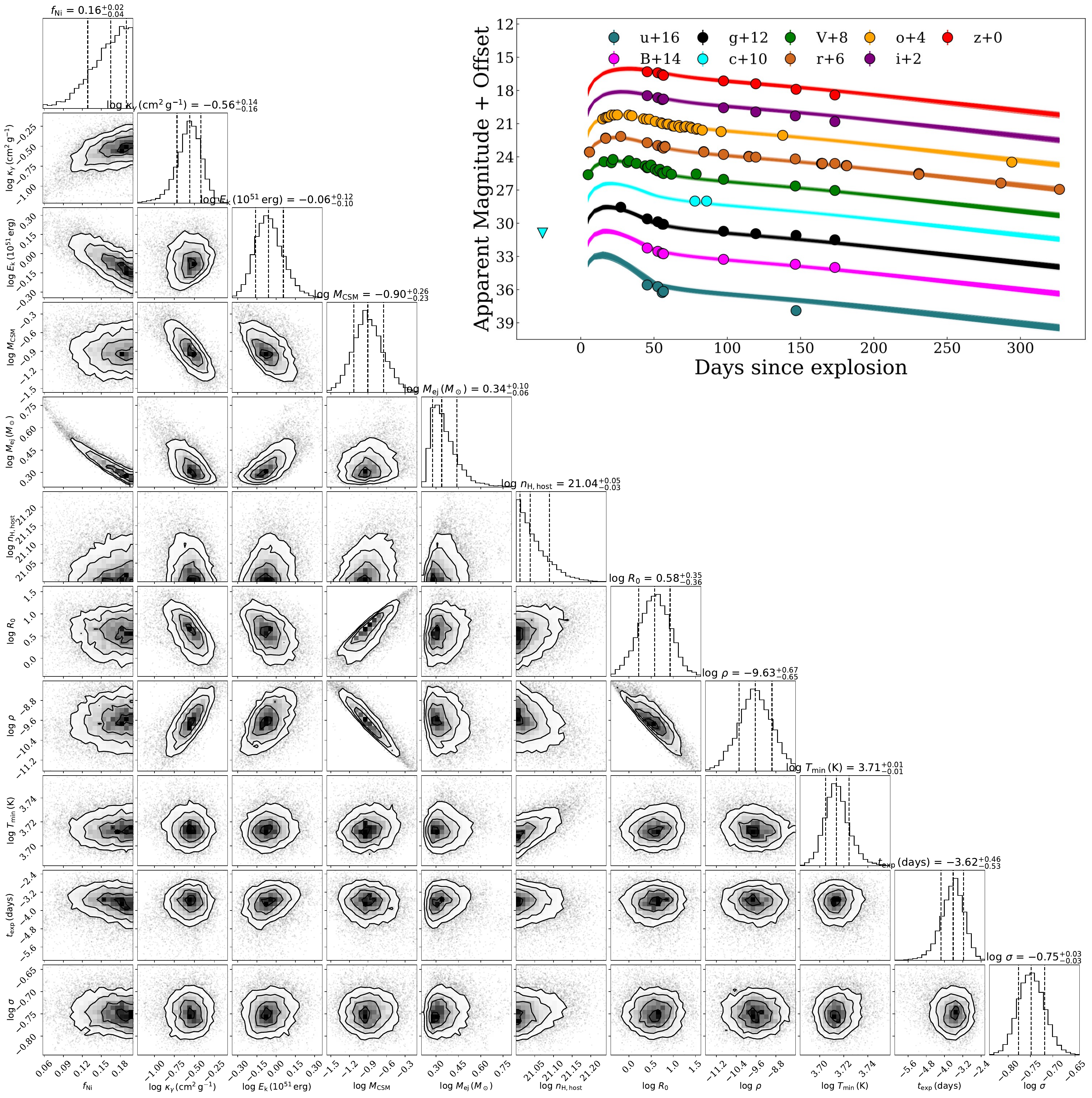}
    \caption{Multi-band light curve of SN~2017ati reproduced with the combined radioactive decay and CSM interaction model in \texttt{MOSFiT}, with the corner plot showing constraints from the MCMC sampling. The relevant parameters are listed in Table \ref{tab:MOSFiT}.}
    \label{fig:CSMNI}
\end{figure*}

\begin{figure*}[ht]
    \centering 
    \includegraphics[width=\textwidth]{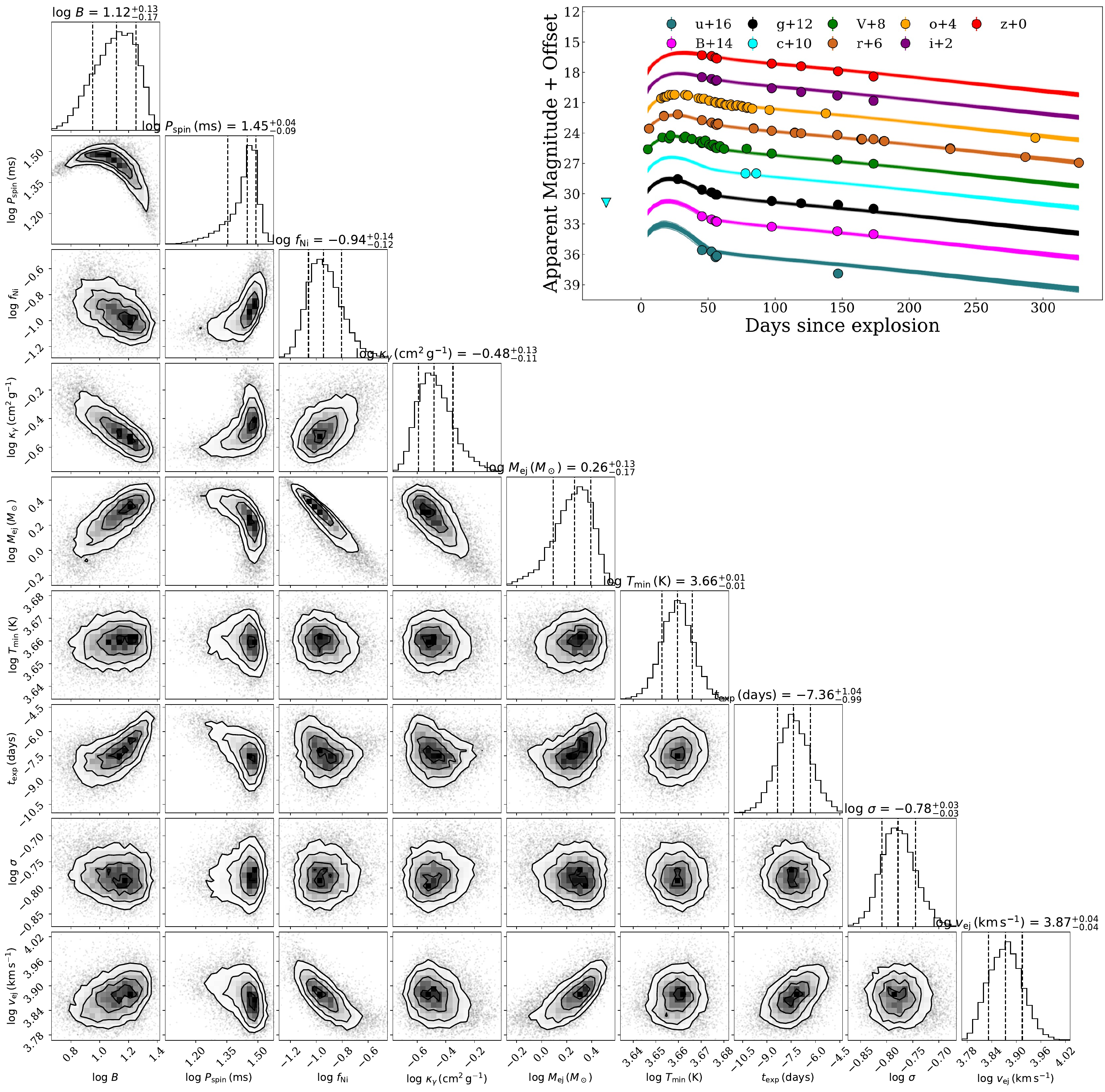}
    \caption{Multi-band light curve of SN~2017ati reproduced with the radioactive decay plus magnetar central-engine model  in \texttt{MOSFiT}, with the corner plot showing constraints from the MCMC sampling. The relevant parameters are listed in Table \ref{tab:MOSFiT}.}
    \label{fig:mosfit}
\end{figure*}

\end{appendix}

\end{document}